# X-RAY DIFFRACTION AND SPECTROSCOPIC TECHNIQUES FROM LIQUID SURFACES AND INTERFACES

**David Vaknin**
*Ames Laboratory and Department of Physics Iowa State University, Ames, Iowa*

## INTRODUCTION

X-ray and neutron scattering techniques are probably the most effective tools when it comes to determining the structure of liquid interfaces on molecular-length scales. These techniques are, in principle, not different from conventional x-ray diffraction techniques that are commonly applied to three-dimensional crystals, liquids, solid surfaces etc. However, special diffractometers and spectrometers that enable scattering from fixed horizontal surfaces are required to carry out the experiments. Indeed, systematic studies of liquid surfaces had not begun until the introduction of a dedicated liquid surface reflectometer (Als-Nielsen and Pershan, 1983).

A basic property of a liquid-vapor interface is the length scale over which the molecular density changes from the bulk value to that of the homogeneous gaseous medium. Inter molecular correlation and capillary waves, which depend on surface tension, gravity, (Buff 1965, Evans, 1979), and instrumental resolution that determines the length scale over which these correlations are averaged (Braslau et al., 1988; Sinha et al., 1988, Sanyal1991), are among the most important factors that define the density profile across the interface and the planar correlations. In some instances the topmost layers of liquids are packed differently than in the bulk, giving rise to layering phenomena at the interface. Monolayers of compounds different than the liquid can be spread at the gas-liquid interface (generally referred to as Langmuir monolayers). (Gaines, 1966). The spread compound might "wet" the liquid surface forming a film of homogeneous thickness or cluster to form an inhomogeneous rough surface. Two major diffraction techniques, complementing one another, are used to determine interfacial structure. One is the x-ray reflectivity (XR) technique which allows one to determine the electron density across such interfaces, from which the molecular density and the total thickness can be extracted. The other is the grazing x-ray diffraction (GIXD also referred to as GID) technique which is commonly used to obtain lateral arrangements and correlations of the topmost layers at interfaces. GIXD is especially efficient in cases where surface crystallization of the liquid or spread monolayers occurs. GIXD has been expanded to small angles (SA-GIXD, or GI-SAXS) for probing the structure of more complicated macromolecular systems, such as membrane proteins (Fukuto et al. 2010), and polymers. Diffuse x-ray scattering (DXS) near the reflectivity line (or specular ridge), which fall into the general GIXD category, is commonly used to evaluate interfacial correlations, due to capillary waves, for instance. Both, XR and GIXD techniques provide structural information that is averaged over macroscopic areas, in contrast to scanning probe microscopies (SPMs), where local arrangements are probed (see, e.g., SCANNING TUNNELING MICROSCOPY). For an inhomogeneous interface, the reflectivity is an incoherent sum of reflectivities, accompanied by strong diffuse scattering, which, in general, is difficult to interpret definitively and often requires complementary techniques to support the x-ray analysis. Therefore, whenever the objective of the experiment is not compromised, preparation of well-defined homogeneous interfaces is a key to a more definitive and straightforward interpretation.

The X-ray fluorescence spectroscopy technique near total reflection from ion-enriched liquid interfaces were developed at the same time as the diffraction techniques (Bloch1985, Yun and Bloch 1990, Daillant 1991) adding insight into ion adsorption and phenomena at liquid interfaces (Shapovalov 2007). The facile tunablility of photon energy at synchrotron x-ray facilities enabled the extension of the fluorescence technique to obtain the near-resonance surface x-ray absorption spectroscopy (XANES) of specific interfacial ions (Bu and Vaknin, 2009). Photon energy tunability also facilitated the application of anomalous diffraction and spectroscopic methods commonly used for bulk materials. Anomalous XR (Vaknin, 2003; Pittler 2006), near edge energy scans at fixed momentum transfers (Bu 2006) have been implemented and used to determine ion distributions and ionic coordination at charged Langmuir monolayers and liquid-liquid interfaces.

### Competitive and Related Techniques

Although modern scanning probe microscopies (SPMs) such as scanning tunneling microscopy (STM; SCANNING TUNNELING MICROSCOPY; Binning and Rohrer, 1983) and atomic force microscopy (AFM; Binning, 1992) rival x-ray scattering techniques in probing atomic arrangements of solid surfaces, they have not yet become suitable techniques for free liquid surfaces (but see X-RAY DIFFRACTION TECHNIQUES FOR LIQUID SURFACES AND MONOMOLECULAR LAYERS). The large fluctuations due to the two-dimensional nature of the liquid interface, high molecular mobility, and the lack of electrical conductivity (which is necessary for STM) are among the main obstacles that make it difficult, practically impossible, to apply these techniques to vapor-liquid interfaces. Furthermore, in dealing with volatile liquids, inadvertent deposition or wetting of the probe by the liquid can occur, which may obscure the measurements. In addition, the relatively strong interaction of the probe with the surface might alter its pristine properties. For similar and other reasons, electron microscopy (see Section 11a) and electron diffraction techniques (LOW-ENERGY ELECTRON DIFFRACTION), which are among the best choices for probing solid surfaces, are not suitable to most liquid surfaces, and in particular to aqueous interfaces. On the other hand, visible light microscopy techniques such as Brewster angle microscopy (BAM; Azzam and Bashara, 1977; Henon and Meunier, 1991; Hönig and Möbius, 1991) or fluorescence microscopy Lösche and Möhwald, 1984) have been used successfully in revealing the morphology of the interface on the micrometer-length scale. This information is complementary to that extracted from x-ray scattering. These techniques are very useful for characterizing inhomogeneous surfaces with two or more distinct domains, for which XR and GIXD results are usually difficult to interpret. However, it is practically impossible to determine the position of the domains with respect to the liquid interface, their thicknesses, or their chemical nature with these light microscopy techniques. Ellipsometry (ELLIPSOMETRY) is another tool that exploits visible light to determine film thickness on a molecular length scale; for its interpretation the refractive indices of the substrate and film (Azzam and Bashara, 1977; Ducharme et al., 1990) have to be guessed. Either of these values might be different from its corresponding bulk value, and therefore is not always reliable. Spectroscopic techniques such as infrared (Mitchel and Dluhy 1988, Blaudez 1993), second-harmonic generation (Heinz 1982), and sum-frequency have been extremely valuable in determining molecular orientation (Guyot-Sionnest 1988), and indirectly ionic species at liquid interfaces (Le Calvez 2001). These techniques are superior in determining properties or phenomena that involve light elements which cannot be easily probed by x-ray scattering techniques. In this regard, it should be noted that neutron reflectometry is an essential probe for studying the structure of light-element based materials at interfaces. Isotope substitution, in particular, substitution of deuterium for hydrogen, has been extensively used to determine structures of surfactants at liquid interfaces (Penfold and Thomas 1990) and Langmuir monolayers (Vaknin1991c, Bu and Vaknin 2008).

In the following sections, theoretical background to the x-ray techniques is presented together with experimental procedures and data analysis concerning liquid surfaces. This unit is intended to provide a basic formulation which can be developed for further specific applications. Several examples of these techniques applied to a variety of problems are presented briefly to demonstrate the strengths and limitations of the techniques. It should be borne in mind that the derivations and procedures described below are mostly general and can be applied to solid surfaces, and, vice versa, many results applicable to solid surfaces can be used for liquid surfaces. X-ray reflectivity from surfaces and GIXD have been treated in a few reviews (Als-Nielsen and Kjaer, 1989; Russell, 1990; Zhou and Chen, 1995, Als-Nielsen and McMorrow 2001). The same experimental methods and theoretical principles have also been employed to study liquid-liquid interfaces (Sclossman 2002, Schlossman 2008). Two books, with different emphases were also dedicated to X-ray scattering from liquid and soft matter surfaces (Daillant 1999, Tolan 1999).

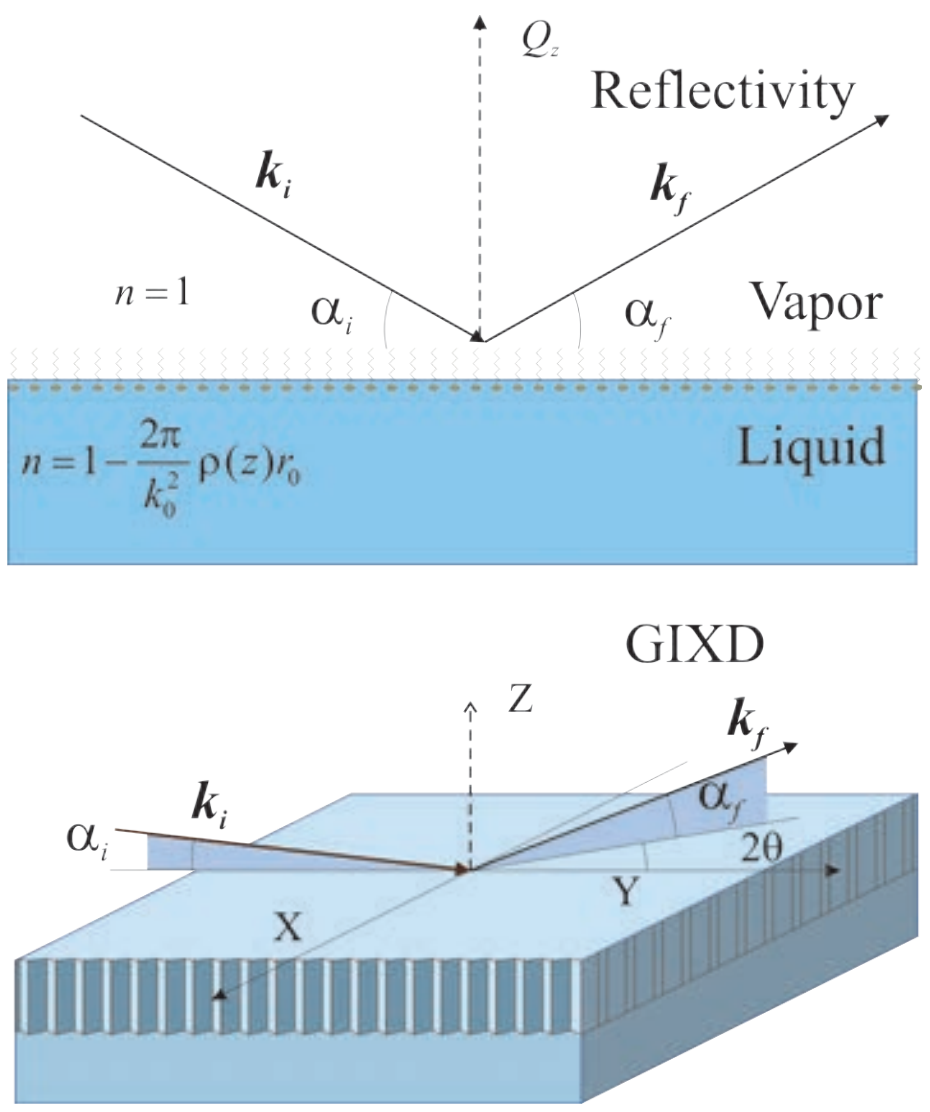

**Figure 1:** The geometry of incident and scattered beam in specular reflectivity (top) and in non-specular diffraction experiments such as GIXD (bottom). The incident beam in GIXD experiments is kept below the critical angle for total reflection to enhance the sensitivity to surface scattering. The coordinate system defined in this figure is used throughout the manuscript.

## PRINCIPLES OF THE METHOD

We assume that a plane harmonic x-ray wave of frequency ω and wave-vector $\mathbf{k}_0$ (with electric field,

$$\mathbf{E} = \mathbf{E}_0 e^{i\omega t - i\mathbf{k_0}\cdot\mathbf{r}}, \tag{1}$$

is scattered from a distribution of free electrons, with a number density $\rho(\mathbf{r})$. Due to the interaction with the electric field of the x-ray wave, each free electron experiences a displacement proportional to the electric field,

$$\mathbf{X} = (-e/m_e\omega^2)\mathbf{E} \quad , \tag{2}$$

where $e$ and $m_e$ are the electric charge and mass of the electron. This displacement gives rise to a polarization distribution vector

$$\mathbf{P}(\mathbf{r}) = \rho(\mathbf{r})e\mathbf{X} = \frac{-\rho(\mathbf{r})e^2}{\omega^2 m_e}\mathbf{E}(\mathbf{r}) = -\frac{4\pi\varepsilon_0 r_0}{k_0^2}\rho(\mathbf{r})\mathbf{E}(\mathbf{r}) \tag{3}$$

where $r_0 = e^2/4\pi\varepsilon_0 mc^2 = 2.82 \times 10^{-13}$ cm, $k_0=|\mathbf{k}_0|=\omega/c$. The electron density (ED) is what we wish to extract from reflectivity and GIXD experiments and relate it to atomic or molecular positions at a liquid interface. The displacement vector **D** can now be constructed from Equation 3, as follows

$$\mathbf{D} = \varepsilon_0\left(1 - \frac{4\pi\rho(\mathbf{r})r_0}{k_0^2}\right)\mathbf{E} = \varepsilon_0\varepsilon_r(\mathbf{r})\mathbf{E} \tag{4}$$

where $\varepsilon_r(\mathbf{r})$ is the relative permittivity, that with the relative permeability $\mu_r$of the medium, defines the refractive index $n = (\mu_r\varepsilon_r)^{1/2}$ . For non magnetic media (i.e., the liquid media we are dealing with here) $\mu_r = 1$, and the the index of refraction is given by

$$n(\mathbf{r}) = \sqrt{1 - 2\delta(\mathbf{r})} \approx 1 - \delta(\mathbf{r}) \tag{5}$$

and we define $\delta = 2\pi\varepsilon_0\rho/k_0^2$.

**Table 1.** Electron Number Density, Critical Momentum Transfers, and Critical Angles Absorption Term for Water, methanol, and Liquid Mercury at λ = 1.5404 Å.

| | $\rho_S$ ($e$/Å$^3$) | $Q_c$ (Å$^{-1}$) | $\alpha_c$ (deg.) | $\delta(\times 10^{-6})$ | $\beta(\times 10^{-8})$ |
|---|---|---|---|---|---|
| $H_2O$ | 0.334 | 0.0217 | 0.152 | 3.555 | 1.215 |
| $CH_3OH$ | 0.2677 | 0.0195 | 0.136 | 2.851 | 0.690 |
| Hg | 3.265 | 0.0680 | 0.477 | 34.583 | 360.9 |

Typical values of the ED ($\rho_s$) for various liquids are listed in Table 1. In the absence of true charges in the scattering medium (i.e., a neutral medium) and under the assumption that the medium is nonmagnetic (magnetic permeability $\mu_r$ = 1) the wave equations that need to be solved to calculate the scattering from a known ED can be derived from the following Maxwell equations (Panofsky and Phillips, 1962)

$$\nabla \bullet \mathbf{D} = 0 \qquad \nabla \bullet \mathbf{B} = 0$$
$$\nabla \times \mathbf{E} = -\partial \mathbf{H}/\partial t \quad \nabla \times \mathbf{H} = \partial \mathbf{D}/\partial t \tag{6}$$

Under the assumption of harmonic plane waves, $\mathbf{E} = \mathbf{E}_0 e^{i\omega t - i\mathbf{k}_0 \cdot \mathbf{r}}$, and the relation $k_0^2 = \mu_0 \varepsilon_0 \omega^2$ the following general equations are obtained from Equations 6

$$\nabla^2 \mathbf{E} + k_0^2 n^2(\mathbf{r})\mathbf{E} = -\nabla(\nabla \ln \varepsilon_r(\mathbf{r}) \bullet \mathbf{E}) \tag{7}$$

$$\nabla^2 \mathbf{H} + k_0^2 n^2(\mathbf{r})\mathbf{H} = -\nabla \ln \varepsilon_r(\mathbf{r}) \times (\nabla \times \mathbf{H}) \tag{8}$$

For most cases in the literature, the right hand side of Equation 7 and Equation 8 are ignored (i.e., approximated to zero), presumably justified because of the weak variation of ln($\varepsilon_r$) term. For some particular cases these terms are exactly zero. For instance, for an incident s-wave, $E$ = ($E_x$,0,0), and a stratified medium, i.e, $\varepsilon_r(\mathbf{r})$ = $\varepsilon_r(z)$ we have $\nabla \ln \varepsilon_r(z) \cdot \mathbf{E} = 0$. In those cases, standard mathematical tools, such as the Born approximation (BA) and the distorted wave Born approximation (DWBA; see DYNAMICAL DIFFRACTION and SURFACE X-RAY DIFFRACTION) can be used (Schiff, 1968).

**X-ray Reflectivity**

In reflectivity experiments a monochromatic x-ray beam of wavelength λ [wavevector $k_0$ = 2π/λ and $\mathbf{k}_i = (0, k_y, k_z)$] is incident at an angle $\alpha_i$ on a liquid surface and is detected at an outgoing angle $\alpha_f$ such that $\alpha_i$ = $\alpha_f$, as shown in Figure 1, with a final wave vector $\mathbf{k}_f$. The momentum transfer for the scattering process is defined in terms of the incident and reflected beam as follows

$$\mathbf{Q} = \mathbf{k}_f - \mathbf{k}_i \tag{9}$$

and in reflectivity **Q** is strictly along the surface normal, with $Q_z$ = $2k_0\sin\alpha = 2k_z$.

**Single, Ideally Sharp Interface: Fresnel Reflectivity.** Solving the scattering problem exactly for the ideally sharp interface, although simple, is very useful for applications of more complicated electron density profiles across interfaces. The wavefunctions are also essential for multiple scattering effects in GIXD and diffuse scattering, for instance. Herein, the s-polarized x-ray beam case is derived (see the Lekner 1987 for the case of the p-polarized electromagnetic beam).

For a stratified medium with an electron density that varies along one direction, $z$, $\rho(\mathbf{r}) = \rho(z)$, assuming no absorption, an s-type polarized x-ray beam with the electrical field parallel to the surface (along the $x$ axis; see Fig. 1) obeys the stationary wave equation as derived from Equation 7

$$\nabla^2 E_x + k_0^2 n^2(z) E_x = 0 \tag{10}$$

The general solution to Equation 10 is then given by

$$E_x = E(z) e^{ik_y y} \tag{11}$$

where the momentum transfer along $y$ is conserved when the wave travels through the medium, leading to the well-known Snell's rule for refraction. Inserting Equation 11 in Equation 10 leads to a one-dimensional wave equation

$$\frac{d^2E}{dz^2} + \kappa^2 E = 0 \tag{12}$$

where κ, the wavevector in the medium is $\kappa^2 = k_0^2(\sin^2\alpha - 2\delta)$. For the ideal interface Equation 12 takes the form,

$$\frac{d^2E}{dz^2} + k_z^2 E_z(z) = 0 \qquad z \geq 0 \text{ in gas}$$
$$\frac{d^2E}{dz^2} + k_s^2 E_z(z) = 0 \qquad z \leq 0 \text{ in liquid}$$

where $k_z$=$k_0$sinα and $k_s^2 = k_z^2 - 2\delta = k_z^2 - 4\pi\rho_s r_0 = k_z^2 - k_c^2$ with the following solutions in terms of the reflectance and the transmittance $r$, and $t$ respectively.

$$E(z) = e^{ik_z z} + re^{-ik_z z} \qquad z \geq 0 \text{ in gas}$$
$$E(z) = te^{ik_s z} \qquad z \geq 0 \text{ in liquid} \tag{13}$$

By applying continuity conditions to the wavefunctions and to their derivatives at $z$ = 0, the Fresnel equations for reflectance, $r$, and transmission, $t$, are obtained

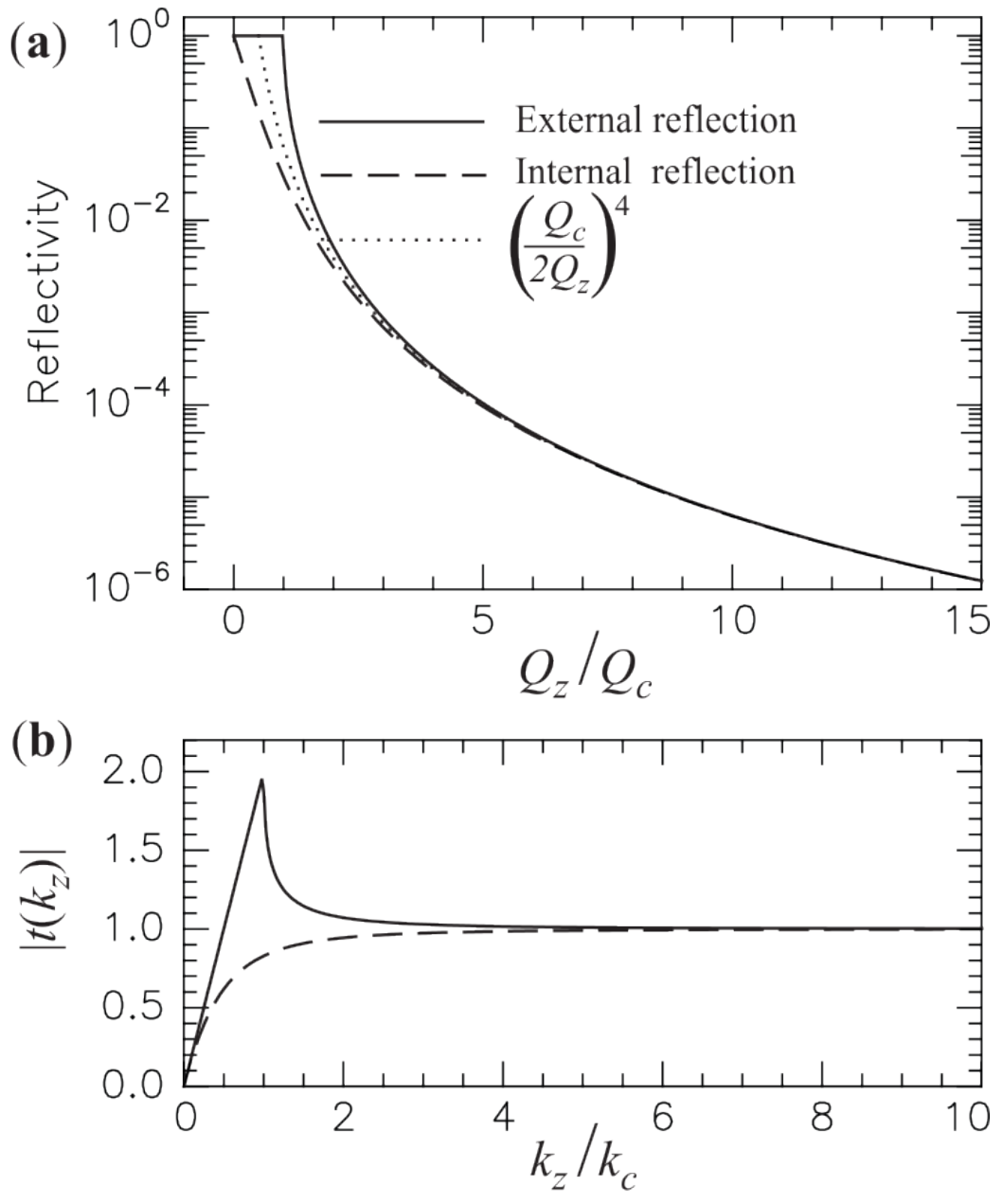

$$1 + r = t; \qquad k_z(1 - r) = k_s t \tag{14}$$

**Figure 2 : (A)** Calculated reflectivity curves for external (solid line) and internal (dashed line) scattering from an ideally flat interface versus momentum transfer given in units of the critical momentum transfer, $Q_c = 4(\pi\rho_s)^{1/2}$. The dotted line is kinematical approximation $(Q_c/2Q_z)^4$. (B) The amplitude of the wave in the medium for external (solid line) and internal reflection (dashed line).

which by algebraic rearrangements yield

$$r = \frac{k_z - k_s}{k_z + k_s}; \qquad t = \frac{2k_z}{k_z + k_s} \tag{15}$$

The measured reflectivity is given by $R_F = |r(k_z)|^2$ is usually displayed as a function of the momentum transfer $Q_z = 2k_z = 2k_0 \sin\alpha$ and is given by

$$R_F(Q_z) = \left| \frac{Q_z - \sqrt{Q_z^2 - Q_c^2}}{Q_z + \sqrt{Q_z^2 - Q_c^2}} \right|^2 \quad ; \quad |t(k_z)|^2 = \left| \frac{2Q_z}{Q_z + \sqrt{Q_z^2 - Q_c^2}} \right|^2 \tag{16}$$

Below a critical momentum transfer,

$$Q_c \equiv 2k_c \equiv 4\sqrt{\pi \rho_s r_0} \tag{17}$$

$k_s$ is an imaginary number and $R_F(Q_Z) = 1$, i.e., total external reflection occurs. Notice that whereas the critical momentum transfer does not depend on the x-ray wavelength, the critical angle for total reflection does, and it is given by $\alpha_c \approx \lambda\sqrt{\rho_s r_0 / \pi}$. Typical values of critical angles for x-rays of wavelength $\lambda_{CuKa}$ = 1.5404 Å are listed in Table 1. For $Q_z$ ? $Q_c$, $R_F(Q_Z)$ can be approximated to a form that is known as the Born approximation

$$R_F(Q_z) \sim \left( \frac{Q_c}{2Q_z} \right)^4 \tag{18}$$

Total reflection does not occur for x-rays traveling from the liquid into the vapor interface, except for the point $Q_z$=0. Calculated external and internal reflectivity curves from an ideally flat surface, $R_F$, displayed versus momentum transfer (in $Q_c$ units) are shown in Figure 2A. Both reflectivities converge at large momentum transfer where they can be both approximated by Equation 18. The dashed line in the same figure shows the approximation $(Q_c/2Q_z)^4$, which fails in describing the reflectivity close to the critical momentum transfer.

To account for absorption we note that the transmitted beam traveling a distance $z/\sin\alpha$ is attenuated by a factor $e^{-\xi z/\sin\alpha}$, where $\xi$ is the linear absorption coefficient (given in tables in Wilson, 1992). We can add this factor to the wave amplitude of the transmitted beam (Eq. 13) and manipulate algebraically as

$$e^{ik_s z - \frac{\xi z}{2\sin\alpha}} = e^{ik_0\left[\sqrt{\sin^2\alpha - k_c^2} + i\frac{\xi}{2k_0 \sin\alpha}\right]z} \cong e^{ik_0\sqrt{\sin^2\alpha - 2\delta + i2\beta z}}, \tag{19}$$

where $\beta = \xi/2k_0$ and high order terms are neglected in the last step. From Equation 19 we extract $\kappa$ for an absorbing medium

$$\kappa^2 = k_0^{\,2}(\sin^2\alpha - 2\delta + i2\beta) \tag{20}$$

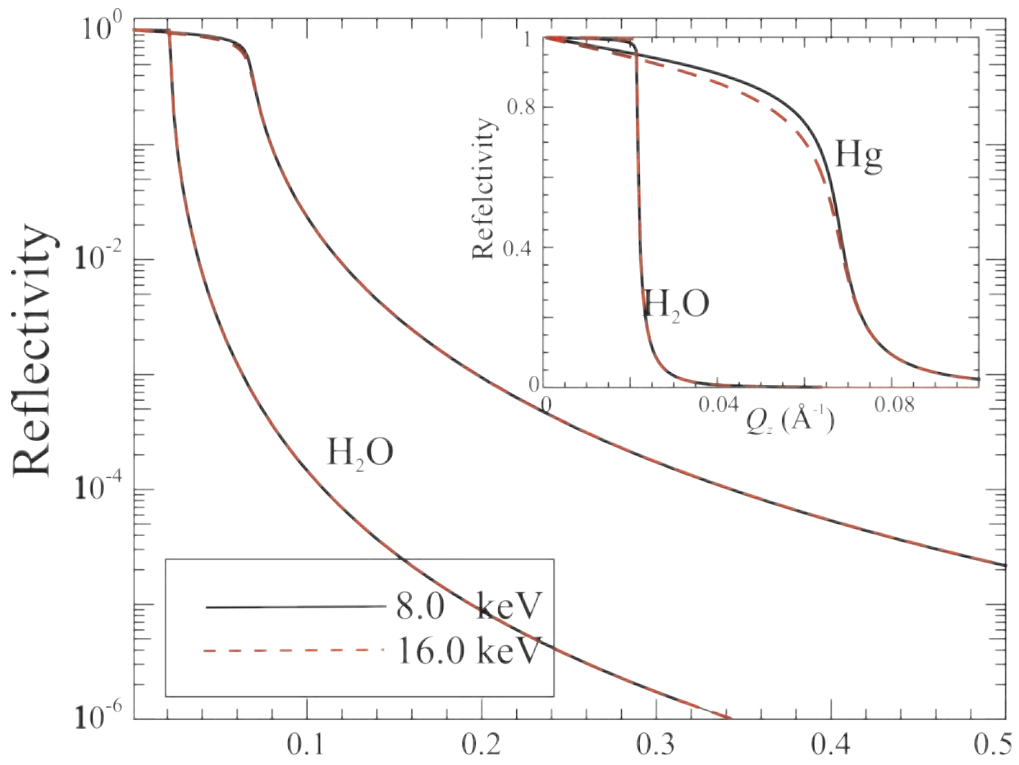

**Figure 3:** Calculated reflectivities for water and mercury including absorption effects at two photon energies as indicated.

With which we generalize a complex refractive index $n^2 = 1 - 2\delta + i2\beta$. Typically for light-element materials, $\delta \sim 10^{-6}$ and $\beta \sim 10^{-8}$ (see Table 1). Defining the real part of the electron density as $\rho = \rho' = k_0^2\delta/(2\pi r_0)$ and similarly an absorptive imaginary function $\rho'' = k_0^2\beta/(2\pi r_0)$ a generalized electron density (complex function) can be constructed as follows

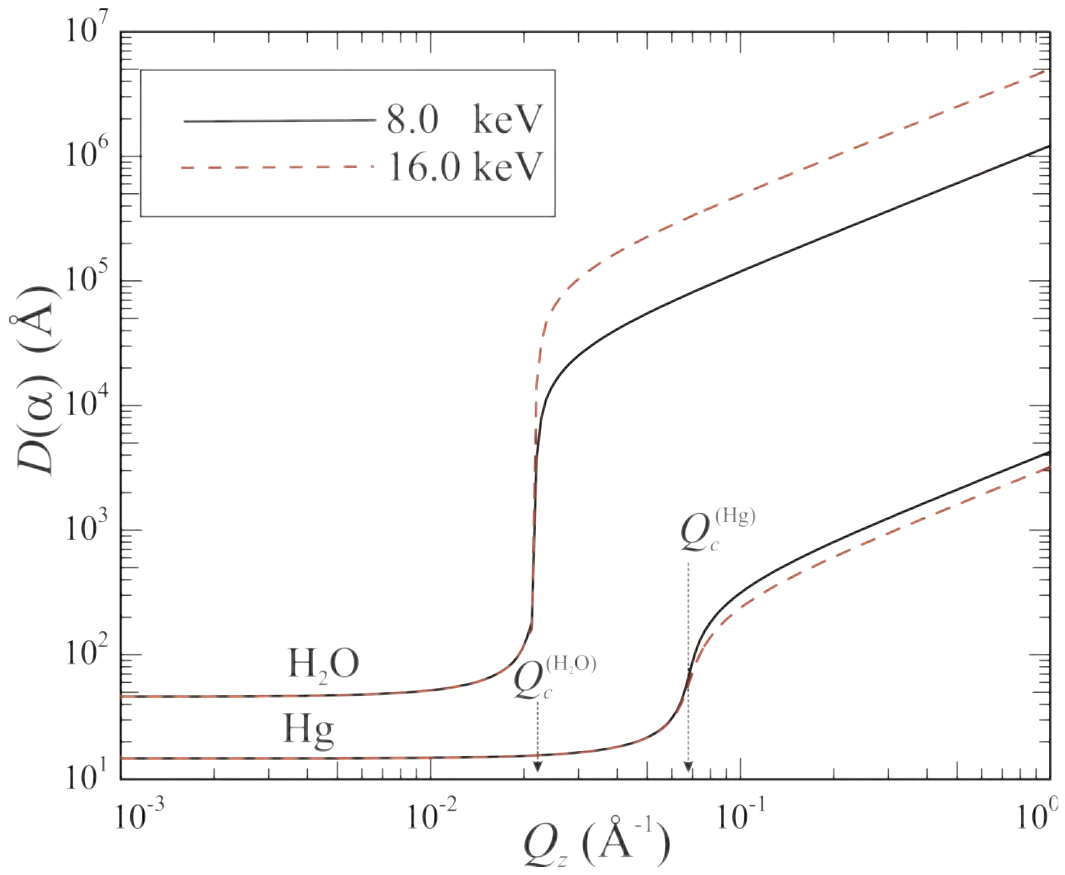

**Figure 4:** Penetration depth of x-rays as a function of $Q_z$ for water and for mercury at two photon energies as indicated.

$$\rho = \rho' - i\rho'' \tag{21}$$

From this point on in this manuscript, $\rho$ is a complex function, and a compact form for the wavevector in the medium is now given by

$$\kappa = \sqrt{k_z^2 - 4\pi\rho r_0} \tag{22}$$

Calculated reflectivities from water and liquid mercury, demonstrating that the effect of absorption is practically insignificant for the former, yet has the strongest influence near the critical angle for the latter, are shown in Figure 3.

Before moving on, let us explore the penetration depth $D(\alpha)$ (or $D(Q_z)$) of the beam as it is attenuated by a factor 1/e. This is a valuable parameter for calculating integrated intensities of scattering events that occur in the beam path (fluorescence, for instance). According to Equation 13 the wave function of the transmitted beam is $te^{ik_s z}$, where $k_s$ is a complex function and the imaginary part of it $\mathrm{Im}(k_s)$ causes beam attenuation. Thus, by definition

$$D(\alpha) = \frac{1}{2\,\mathrm{Im}(k_s)} = \frac{1}{2k_0\,\mathrm{Im}\left[\sqrt{\sin^2\alpha - 2\delta + i2\beta}\right]} \tag{23}$$

Or

$$D(Q_z) = \frac{1}{\mathrm{Im}\left(\sqrt{Q_z^2 - Q_c^2 + i8\beta k_0^2}\right)} \tag{24}$$

The $D(\alpha)$'s for pure water and for Hg at 8 and 16 keV are plotted in Figure 4. Two distinct regions separated by $Q_c$ are clearly observed. Below the critical angle, $D(Q_z) \approx 1/Q_c$, namely the penetration depth is independent of photon energy, and is practically a constant. This is the penetration of the evanescent wave, which for water and Hg is approximately 60 and 15 Å, respectively. At exactly the critical angle $Q_z = Q_c$, $D(\alpha) = (2\beta k_0)^{1/2}$. Above the critical angle, $D(Q_z) \approx Q_z/(4\beta k_0^2) = \sin\alpha/\xi$. The reverse ratios of penetration depth at 8 compared to 16 keV for water and Hg, is simply because, $\xi$(8 keV)/2$\xi$(16 keV) is larger than 1 for water and smaller than one for Hg (that is because Hg has L-edges between 8 and 16 keV).

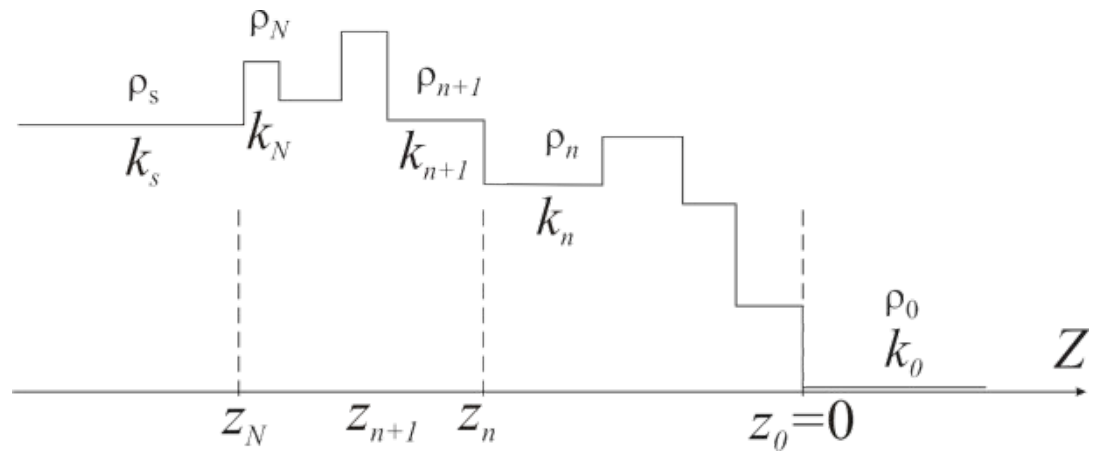

**Figure 5:** Stepwise ED as a function of position and definitions that are used to derive the matrix (or recursive) formulas to calculate the reflectivity.

**Multiple Stepwise and Continuous Interfaces.** On average, the electron density across a liquid interface is a continuously varying function, and is a constant far away on both sides of the interface. The reflectivity for a general function $\rho(z)$ can be then calculated by one of several methods, classified into two major categories: dynamical and kinematical solutions. The dynamical solutions (see DYNAMICAL DIFFRACTION) are in general more exact, and include all the features of the scattering, in particular the low-angle regime, close to the critical angle where multiple scattering processes occur. For a finite number of discrete interfaces (as shown in Fig. 5), exact solutions can be obtained by use of standard recursive (Parratt, 1954) or matrix (Born and Wolf, 1959) methods. These methods can be extended to compute, with very high accuracy, the scattering from any continuous potential by slicing it into a set of finite layers but with a sufficient number of interfaces. On the other hand, the kinematical approach (see KINEMATIC DIFFRACTION OF X-RAYS) neglects multiple scattering effects and fails in describing the scattering at small angles.

1. *The matrix method*. Consider a stepwise complex electron density with N layers each with $\rho_n = \rho'_n - i\rho''_n$ and layer thickness $d_n = z_{n-1} - z_n$ with $N+1$ interfaces (see Figure 5). The z-component wavevector in the n[th] layer is

$$k_n \equiv \sqrt{k_z^2 - 4\pi\rho_n r_0} \tag{25}$$

(for clarity we use the notation $k_n = k_{z,n}$, from here on). At each layer the wave Equation (12) at each position $z = z_n$, gives the following wavefunctions

$$\begin{array}{cc|cc} \rho_{n+1} & & & \rho_n \\ R_{n+1}e^{-ik_{n+1}(z-z_{n+1})} & \rightarrow & \rightarrow & R_n e^{-ik_n(z-z_n)} \\ +T_{n+1}e^{ik_{n+1}(z-z_{n+1})} & \leftarrow & \leftarrow & +T_n e^{ik_n(z-z_n)} \end{array}$$

$$z = z_n$$

for which the continuity conditions of the functions and their derivatives yield

$$\begin{aligned} R_n &= \left(\frac{k_n + k_{n+1}}{2k_n}\right) e^{i\Delta_{n+1}} R_{n+1} + \left(\frac{k_n - k_{n+1}}{2k_n}\right) e^{-i\Delta_{n+1}} T_{n+1} \\ T_n &= \left(\frac{k_n - k_{n+1}}{2k_n}\right) e^{i\Delta_{n+1}} R_{n+1} + \left(\frac{k_n + k_{n+1}}{2k_n}\right) e^{-i\Delta_{n+1}} T_{n+1} \end{aligned} \tag{26}$$

The solution at each interface can be written in terms of a transfer matrix, $M_n$ as,

$$\begin{pmatrix} R_n \\ T_n \end{pmatrix} = \frac{e^{-i\Delta_{n+1}}}{t_n} \begin{pmatrix} e^{i2\Delta_{n+1}} & r_n \\ r_n e^{i2\Delta_{n+1}} & 1 \end{pmatrix} \begin{pmatrix} R_{n+1} \\ T_{n+1} \end{pmatrix} \tag{27}$$

Where $\Delta_{n+1} \equiv k_{n+1} d_{n+1}$ and

$$r_n = \frac{k_n - k_{n+1}}{k_n + k_{n+1}} \;;\; t_n = \frac{2k_n}{k_n + k_{n+1}} \tag{28}$$

are the Fresnel reflectance and transmittance from layer $n$ to layer $n+1$ . The reflectivity is calculated by applying the matrix

$$\begin{pmatrix} R_0 \\ T_0 \end{pmatrix} = (M_1)(M_2)\cdots(M_n)\cdots(M_{N+1})\begin{pmatrix} 0 \\ 1 \end{pmatrix} \tag{29}$$

by noting that beyond the last interface, (i.e., in the bulk) there is only a transmitted beam that we choose to have an arbitrary amplitude, $T_{N+1} = t_{N+1} = 1$, so that $R_{N+1} = r_{N+1} = 0$, and $d_{N+1} = 0$. Without loss of information, the pre-factor of the matrix in Equation (27) can be removed to give a compact matrix

$$M_n = \begin{pmatrix} e^{i2\Delta_{n+1}} & r_n \\ r_n e^{i2\Delta_{n+1}} & 1 \end{pmatrix} \tag{30}$$

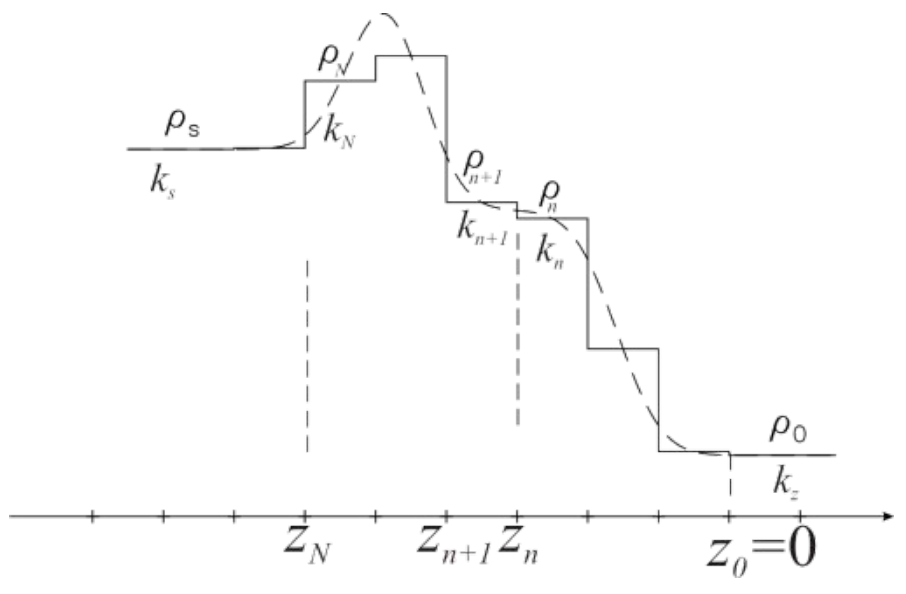

Figure 6: Schematic continuous ED profile (dashed line) sliced into a histogram (solid line) that can be used to calculate the reflectivity to practically any continuous ED function by using the matrix method.

We also note that

$$r_N = \frac{k_N - k_s}{k_N + k_s}; \quad \text{and } M_{N+1} = \begin{pmatrix} 1 & r_N \\ r_N & 1 \end{pmatrix} \tag{30}$$

The reflectivity is then given by the ratio

$$R(Q_z) = \left| \frac{R_0}{T_0} \right|^2 \tag{31}$$

The matrix method is equivalent to the recurisve Parratt formalism, by taking the ratio of the Equations 26

$$\frac{R_n}{T_n} = \frac{(k_n + k_{n+1})e^{i\Delta_{n+1}} R_{n+1} + (k_n - k_{n+1})e^{-i\Delta_{n+1}} T_{n+1}}{(k_n - k_{n+1})e^{i\Delta_{n+1}} R_{n+1} + (k_n + k_{n+1})e^{-i\Delta_{n+1}} T_{n+1}} \tag{32}$$

And by defining $A_n \equiv \frac{R_n}{T_n}$ we obtain

$$A_n = \frac{r_n + A_{n+1} e^{i2\Delta_{n+1}}}{1 + r_n A_{n+1} e^{i2\Delta_{n+1}}} \tag{33}$$

and recursively evaluating $A_0$ the reflectivity $R = |A_0|^2$ is obtained.

Applying this procedure to the *one-box model* of thickness *d, i.e.,* a single step between the gas and liquid interfaces yields

$$R(Q_z \equiv 2k_z) = \left| \frac{r_1 + r_2 e^{i2k_s d}}{1 + r_1 r_2 e^{i2k_s d}} \right|^2 \tag{34}$$

Figure 7 shows the calculated reflectivities from a flat liquid interface with two kinds of films (one box) of the same thickness $d$ but with different scattering length densities, $\rho_1$ and $\rho_2$. The reflectivities are almost indistinguishable when the normalized EDs ($\rho_i/\rho_s$) of the films are complementary to one ($\rho_1/\rho_s + \rho_2/\rho_s = 1$) except for a very minute difference near the first minimum. In the BA method described below, the two potentials shown in Figure 7 yield identical reflectivities (see Pershan 1994).

The matrix method can be used to calculate the exact solution from a finite number of interfaces, and it is most powerful when used with computers to calculate practically any continuous ED profile by slicing it into a histogram as schematically shown in Figure 6. The criteria for determining the optimum number of slices to use is based on the convergence of the calculated reflectivity.

### The Born Approximation

For an s polarized wave, Equation (10) can be written in the form

$$\nabla^2 E_x + \left[k_0^2 - V(\mathbf{r})\right] E_x = 0 \tag{35}$$

where $V(\mathbf{r}) = 4\pi\rho(\mathbf{r}) r_0$ for which the Born-approximation (BA) provides the scattering length amplitude as follows (Schiff, 1968)

$$F(\mathbf{Q}) = r_0 \int \rho(\mathbf{r}) e^{i\mathbf{Q}\cdot\mathbf{r}} d^3\mathbf{r} \tag{36}$$

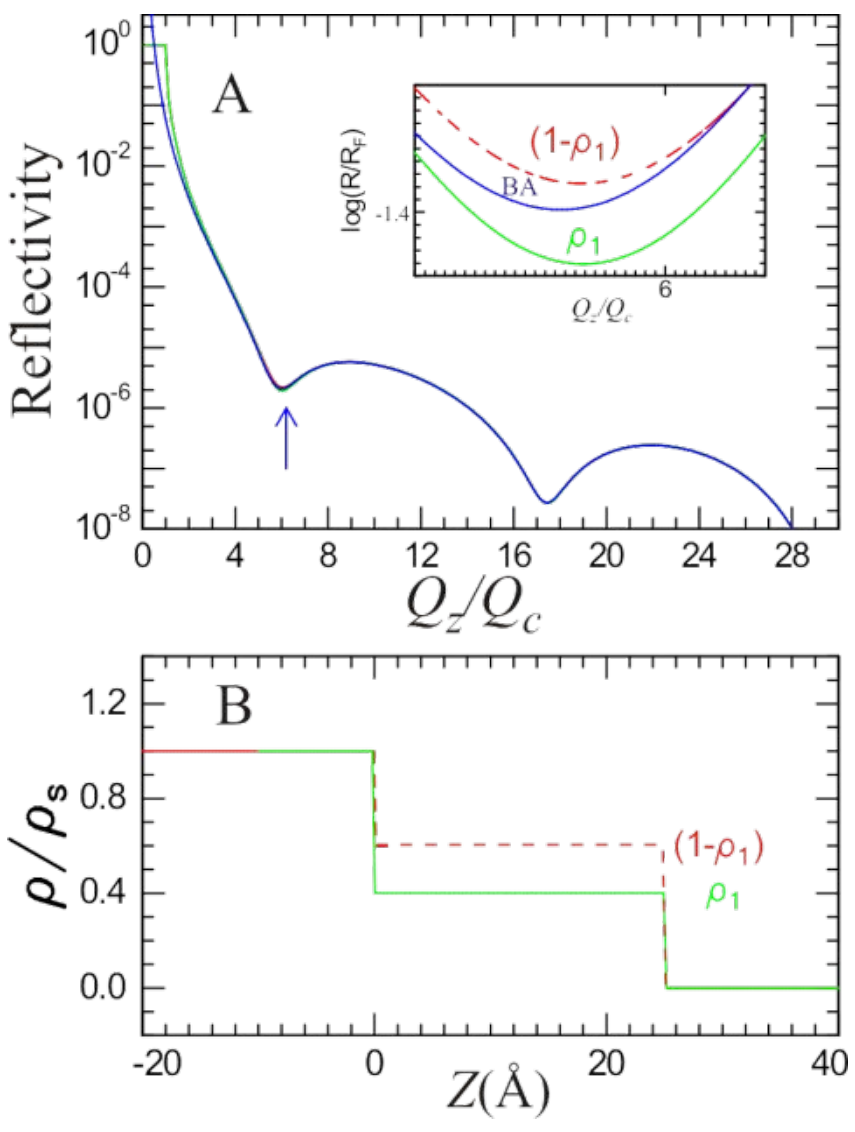

**Figure 7**. (A) Calculated reflectivities for two films with identical thicknesses but with two distinct normalized electron densities $\rho_1$ (solid line) and ($1 - \rho_1$) (dashed line) (B), using Equation 34 ($\sigma = 0$). The two reflectivities are almost identical except for small differences near the first minimum (see arrow in figure). The Born approximation (dotted line) for the two models yields identical reflectivities. The inset shows the normalized reflectivities near the first minimum. As $Q_z$ is increased, the three curves converge. This is the simplest demonstration of the phase problem, i.e., the nonuniqueness of models where two different potentials give almost the same reflectivities .

and the differential cross section is

$$\frac{d\sigma}{d\Omega} = |F(\mathbf{Q})|^2 \tag{37}$$

In x-ray scattering experiments and in particular in reflectivity measurements we record the scattered intensity and normalize it to the incident beam intensity, $I_0$. Using the configuration in our experiment as shown in Figure 1, and noting that the beam flux on the surface is given by $I_0 \sin\alpha_i / L_x L_y$, where $L_y/\sin\alpha_i$ and $L_x$ are the beam footprints on the surface. In addition, in reflectivity experiments, the implied assumption is that the resolution of the spectrometer is relaxed so that we integrate over the entire the scattered beam i.e, over the outgoing solid angle,

$$R(\mathbf{Q}) = \frac{I(\mathbf{Q})}{I_0} = \frac{1}{L_x L_y \sin\alpha} \int \left( \frac{d\sigma}{d\Omega} \right) d\Omega \tag{38}$$

We assume a an ED of a liquid interface that varies conformally along z as follows

$$\rho(\mathbf{r}_{xy}, z) = \begin{cases} 0 & z \geq z(x,y) \\ \rho(z) & z \leq z(x,y) \end{cases} \tag{39}$$

Using Equations (37) and (39) we obtain

$$\frac{d\sigma}{d\Omega} = r_0^2 \left| \int \rho(\mathbf{r}_{xy}, z) e^{i\mathbf{Q}_x\cdot\mathbf{r}_{xy}} d^2\mathbf{r}_{xy} e^{iQ_z z} dz \right|^2 \tag{40}$$

Using the following identity

$$\int \rho(z) e^{iQ_z z} dz = -\frac{1}{iQ_z} \int \frac{d\rho(z)}{dz} e^{iQ_z z} dz \quad (41)$$

noting that

$$d\Omega = \frac{dQ_x dQ_x}{k_0^2 \sin \alpha_f} \quad (42)$$

and using Equation (38) we get

$$R(\mathbf{Q}_{xy}, Q_z) = \frac{4r_0^2}{L_x L_y Q_z^4} \int G(\mathbf{Q}_{xy} - \mathbf{Q}'_{xy}) \times \left| \int \frac{d\rho(\mathbf{r}_{xy}, z)}{dz} e^{i\mathbf{Q}_{xy} \cdot \mathbf{r}_{xy}} d^2\mathbf{r}_{xy} e^{iQ_z z} dz \right|^2 d^2\mathbf{Q}'_{xy} \quad (43)$$

where we used $Q_z = 2k_0 \sin\alpha_i = 2k_0 \sin\alpha_f$ and assumed an ideally perfect instrumental resolution function $G(\mathbf{Q}_{xy})$ such that $\int d^2 \mathbf{Q}_{xy} G(\mathbf{Q}_{xy}) = 1$. If we further simplify the surface so that the ED is uniform in the plane, so that

$$\rho(\mathbf{r}_{xy}, z) = \begin{cases} 0 & z \geq 0 \\ \rho(z) & z \leq 0 \end{cases} \quad (44)$$

we obtain the following

$$R(Q_z) = \frac{r_0^2 16\pi^2 L_x L_y}{L_x L_y Q_z^4} \times \int d^2\mathbf{Q}'_{xy} \delta^2(\mathbf{Q}_{xy}) G(\mathbf{Q}_{xy} - \mathbf{Q}'_{xy}) \left| \int \rho'(z) e^{iQ_z z} dz \right|^2 \quad (45)$$

Where we used the following

$$\left| \int e^{iQ_x x} dx \right|^2 = \iint e^{iQ_x (x-x')} d(x-x') dx = 2\pi L_x \delta(Q_x) \quad (46)$$

The reflectivity is measured at exactly $\mathbf{Q}_{xy} = 0$ and noting that $Q_c = 4(\pi \rho_s r_0)^{1/2}$ we obtain

$$R(Q_z) = \frac{16 r_0^2 \pi^2 \rho_s^2}{Q_z^4 \rho_s^2} \left| \int \frac{d\rho(z)}{dz} e^{iQ_z z} dz \right|^2 \quad (47)$$

To give the

$$R(Q_z) = \left( \frac{Q_c}{2Q_z} \right)^4 \left| \int \frac{1}{\rho_s} \frac{d\rho(z)}{dz} e^{iQ_z} dz \right|^2 = \left( \frac{Q_c}{2Q_z} \right)^4 |\Phi(Q_z)|^2 \quad (48)$$

where, $\Phi(Q_z)$ can be regarded as the generalized structure factor of the interface, analogous to the structure factor of a unit cell in 3-D crystals. Also we recognize the behavior of the Fresnel reflectivity at $Q_z > Q_c$.

A simple way to derive the same result from a continuous ED can be obtained by writing

$$k(z) = \sqrt{k_z^2 - 4\pi r_0 \rho(z)}; \quad \frac{dk}{dz} = -\frac{2\pi}{k(z)} \frac{d\rho}{dz} \quad (49)$$

so the reflectance across an arbitrary point $z$, using Equation 15, is given by

$$r(z) = \frac{k(z+\Delta z) - k(z)}{k(z+\Delta z) + k(z)} \approx -\frac{4\pi r_0 \rho_s}{4k(z)^2 \rho_s} \frac{d\rho}{dz} dz \approx (Q_c / 2Q_z)^2 \frac{1}{\rho_s} \frac{d\rho}{dz} dz \quad (50)$$

In the last step of derivation, the identity $Q_c^2 \equiv 16\pi r_0 \rho_s$ was used. Assuming no multiple scattering, the reflectivity is calculated by integrating over all reflectances at each point, $z$, with a phase factor $e^{iQ_z z}$ as follows

$$\left| \frac{1}{\rho_s} \int r(z) e^{iQ_z z} dz \right|^2 = \left( \frac{Q_c}{2Q_z} \right)^4 \left| \frac{1}{\rho_s} \int \left( \frac{d\rho}{dz} \right) e^{iQ_z z} dz \right|^2 \quad (51)$$

which the same as Equation 48.

**The Distorted Wave Born Approximation (DWBA):**

Due to the weak interaction of the electromagnetic field of x-rays with matter (i.e., electrons), the BA is sufficient for most $Q_z$ values. But, this is not the case when $Q_z$ is of order of $Q_c$. Then, multiple scattering processes become important. The distorted wave Born approximation has been developed for such multiple scattering processes, that involve potentials that consist of two (or more) potentials in proximity for which some knowledge of the wavefunctions of scattered beam from each is known (Rodberg and Thaler, 1967). A wavefunction after scattering from one potential is used to calculate the secondary scattering from the other. All multiple scattering are accounted for to yield a compounded wave function with which to predict scattered intensities. For simplicity, we assume the two potentials are the one from the liquid surface $V_{sub}$ and the other, $V_1$ from an object at some position ($\mathbf{r}_{xy}$, $z$) with respect to the surface ($z = 0$),

$$V(\mathbf{r}) = V_{sub} + V_1 = 4\pi r_0 \rho_s + 4\pi r_0 \rho_1(\mathbf{r}) \quad (52)$$

Let us assume that the wavefunction of a beam scattered directly by the incident from $V_1$ is given by $\psi(\mathbf{Q})$. Let us follow another coherent incident wave from the packet that is reflected from the surface with wave function $r(k_z)$ (Equation 15) and it scatters from $V_1$ with a similar wavefunction $\psi(\mathbf{Q'})$ where $\mathbf{Q} - \mathbf{Q'} = Q_z$, so that for very small $Q_z < Q_c$ we can assume $\psi(\mathbf{Q'}) \cong \psi(\mathbf{Q})$. Under these approximations the intensity of the multiply scattered wave is

$$|\Psi(\mathbf{Q})|^2 \left| \left(1 + r(k_z^i)\right) \right|^2 = |\Psi(\mathbf{Q})|^2 \left| t(k_z^i) \right|^2 = |\Psi(\mathbf{Q})|^2 \left| t(\alpha_i) \right|^2 \quad (53)$$

So, for a hypothetical object of negligible thickness at exactly the interface line i.e., $z = 0$ the scattering intensity will increase from 0 at $Q_z = 0$ and peak at the critical angle as the incident beam angle approaches $Q_c$ and then decay to a constant, with an amplitude as shown in Figure 2 (B). The intensity at $Q_c$, from such an object, is four times the intensity at large $Q_z >> Q_c$. Similarly, the directly scattered wave from $V_1$ has an interference path with its own wave that reflects from the surface so that the scattered wave this time is modulated by $|t(\alpha_f)|^2$. That means that the intensity as a function of the outgoing beam angle has a similar shape to that we described for incident beam with a peak at $\alpha_f = \alpha_c$. It should be noted here, as implied above, that observing a peak at $Q_c$, from a scattering process is not a guarantee that the object is at exactly the interface, $z = 0$. This is discussed later in this chapter in the context of fluorescence from ions as a function of their depth from the interface.

The DWBA is used in the interpretation of inplane GIXD from films at liquid interfaces and for the evaluation of fluorescence of ions as discussed later. It is instructive to extend the DWBA to reflectivity calculated above by the BA. Applying the DWBA to Equation 48 is straight forward yielding

$$R(Q_z) = \left|t(k_z^i)\right|^2 \left|t(k_z^f)\right|^2 \left(\frac{Q_c}{2Q_z}\right)^4 |\Phi(Q_z)|^2 \qquad (54)$$

In reflectivity $\alpha_I = \alpha_f$ and by straightforward algebraic manipulation it is shown below that the DWBA of an ideally pure single interface is equivalent to the exact solution, Equation 16.

$$\left|t(k_z^i)\right|^2 \left|t(k_z^f)\right|^2 \left(\frac{Q_c}{2Q_z}\right)^4 = \left|\frac{Q_c}{Q_z + \sqrt{Q_z^2 - Q_c^2}}\right|^4 =$$
$$\left|\frac{Q_z + \sqrt{Q_z^2 - Q_c^2}}{Q_z + \sqrt{Q_z^2 - Q_c^2}}\right|^2 \left|\frac{Q_z - \sqrt{Q_z^2 - Q_c^2}}{Q_z + \sqrt{Q_z^2 - Q_c^2}}\right|^2 = R_F(Q_z) \qquad (55)$$

In the process we obtained a combined exact/BA reflectivity formula from a film on a subphase which referred to in the literature as the "master formula"

$$R(Q_z) = R_F(Q_z) |\Phi(Q_z)|^2 \qquad (56)$$

As an example of the use of Equation 56 we assume that the ED at a liquid interface can be approximated by a sum of error functions as follows

$$\rho(z) = \frac{\rho_{N+1}}{2} + \sum_{j=1}^{N} \frac{(\rho_j - \rho_{j+1})}{2} \mathrm{erf}\left(\frac{z - z_j}{\sqrt{2}\sigma_j}\right) \qquad (57)$$

where $\rho_0$ is the SLD of the vapor phase and $\rho_N = \rho_s$. Using Equation 32 the reflectivity is given by

$$R(Q_z) = R_{\mathrm{F}}(Q_z) \left|\sum_j \left(\frac{\rho_j - \rho_{j-1}}{\rho_s}\right) e^{-\frac{(Q_z\sigma_j)^2}{2}} e^{iQ_z z_j}\right|^2 \qquad (58)$$

Assuming one interface at $z_1 = 0$ with surface roughness $\sigma_1 = \sigma$, the Fresnel reflectivity, $R_F(Q_z)$, is simply modified by a Debye-Waller-like factor

$$R(Q_z) = R_{\mathrm{F}}(Q_z) e^{-(Q_z\sigma)^2} \qquad (59)$$

Applying Equation 58 to the one-box model discussed above (see Fig. 6), and assuming conformal roughness, $\sigma_j = \sigma$, the calculated reflectivity in terms of ED normalized to $\rho_s$ is

$$R(Q_z) = R_{\mathrm{F}}(Q_z)[(1-\rho_1)^2 + \rho_1^2 + 2\rho_1(1-\rho_1)\cos(Q_z d)] e^{-(Q_z\sigma)^2} \qquad (60)$$

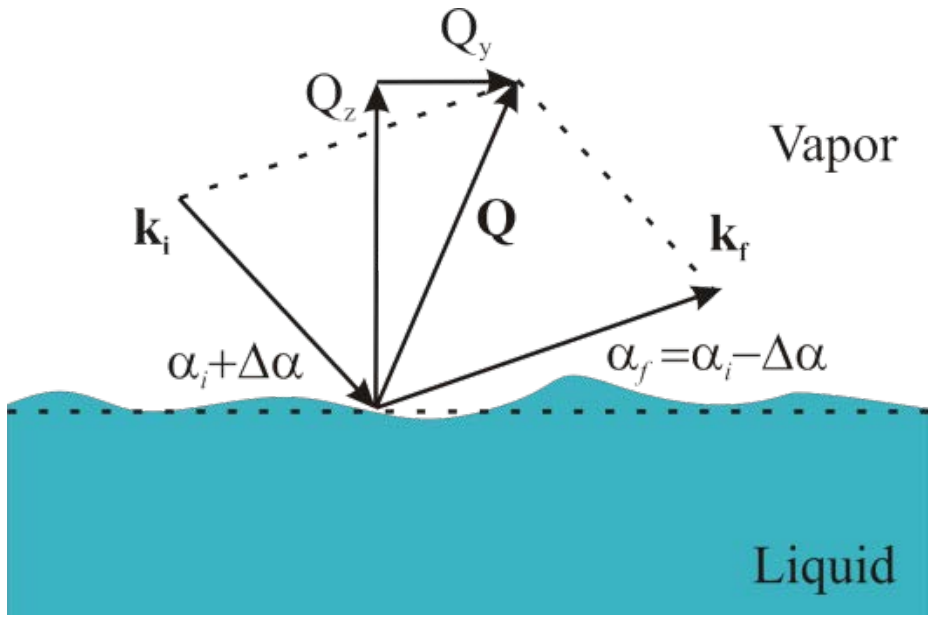

**Figure 8:** Setup used for diffuse x-ray scattering and definitions of geometrical parameters.

In this approximation, the roles of the two ED profiles in Figure 6(B) are equivalent, i.e., both models are mathematically identical. This is the simplest of many examples where two or more distinct ED profiles yield identical reflectivities in the Born approximation. When using the kinematical approximation to invert the reflectivity to ED, there is always a problem of facing a nonunique result. For discussion of ways to distinguish between such models see Data Analysis and Initial Interpretation.

In some instances, the scattering length density can be generated by several step functions representing different moieties of the molecules on the surface, that are further smeared with one Gaussian (conformal roughness $\sigma_j = \sigma$). In this case the reflectivity can be calculated by using a combination of the dynamical and the BA (Als-Nielsen and Kjaer, 1989). First, the exact reflectivity from the step-like functions ($\sigma = 0$) is calculated using the matrix/recursive method, $R_{\mathrm{dyn}}(Q_z)$, and the effect of surface roughness is incorporated by multiplying the calculated reflectivity with a Debye-Waller-like factor as follows (Als-Nielsen and Kjaer, 1989)

$$R(Q_z) = R_{\mathrm{dyn}}(Q_z) e^{-(Q_z\sigma)^2} \qquad (61)$$

**Near-Specular Scattering – rough surfaces**

The geometry for non-specular reflection and GIXD in general is shown in Figure 1B. The scattering from a 2-D system is very weak thus; working near the critical angles of either incident or scattered beam or both to take advantage of intensity enhancements due to multiple scattering processes is desirable. As is shown in Figure 1B the momentum transfer **Q** has a finite component parallel to the liquid surface ($\mathbf{Q}_{xy} \equiv \mathbf{k}^f_{xy} - \mathbf{k}^i_{xy}$), enabling determination of lateral correlations in the 2-D plane. Another geometry commonly employed in DXS experiments, is the one shown in Figure 8, with which long-scale correlations can be measured. The calculation performed below is applicable to the diffuse scattering due to capillary waves, but it can be readily modified to calculate other types of interfacial correlations. The surface density in the form of Equation 39 is assumed, and we apply the DWBA to Equation 43, assuming that $\alpha_i \neq \alpha_f$

$$R(\mathbf{Q}_{xy}, Q_z) = \frac{\left|t(k_z^i)\right|^2 \left|t(k_z^f)\right|^2 r_0^2}{L_x L_y Q_z^2 k_0^2 \sin\alpha_i \sin\alpha_f} \int G(\mathbf{Q}_{xy} - \mathbf{Q}'_{xy}) \times$$
$$\left|\int \frac{d\rho(\mathbf{r}_{xy}, z)}{dz} e^{i\mathbf{Q}'_{xy}\cdot\mathbf{r}_{xy}} d^2\mathbf{r}_{xy} e^{iQ_z z} dz\right|^2 d^2\mathbf{Q}'_{xy} \qquad (62)$$

The second line of this equation can be written as follows

$$\left|\int \frac{d\rho(z)}{dz} e^{iQ_z z} dz\right|^2 \iint e^{iQ_z(z-z')} e^{i\mathbf{Q}_{xy}\cdot(\mathbf{r}_{xy} - \mathbf{r}'_{xy})} d^2\mathbf{r}_{xy} d^2\mathbf{r}'$$
$$= L_x L_y \left|\Phi(Q_z)\right|^2 \int e^{iQ_z(z-z')} e^{i\mathbf{Q}_{xy}\cdot(\mathbf{r}_{xy} - \mathbf{r}'_{xy})} d^2\mathbf{r}_{xy} \qquad (63)$$

Assuming that the $z - z'$ are correlated but this correlation is on average independent of position, $\mathbf{r}_{xy}$, so that in general $z - z' = z(0) - z(\mathbf{r}_{xy})$ further simplify Equation (62)

$$R(\mathbf{Q}_{xy},Q_z)=\frac{\left|t(k_z^i)\right|^2\left|t(k_z^f)\right|^2 Q_c^4}{(16\pi)^2 Q_z^2 k_0^2 \sin\alpha_i \sin\alpha_f}\left|\Phi(Q_z)\right|^2 F(\mathbf{Q}_{xy})$$

$$F(\mathbf{Q}_{xy})\equiv\int G(\mathbf{Q}_{xy}-\mathbf{Q}'_{xy})\int e^{iQ_z\left[z(0)-z(\mathbf{r}_{xy})\right]}e^{i\mathbf{Q}'_{xy}\cdot\mathbf{r}_{xy}}d^2\mathbf{r}_{xy}d^2\mathbf{Q}'_{xy} \quad (64)$$

Employing the Baker-Hausdorff theorem identity for the average

$$\left\langle e^{iQ_z\left[z(0)-z(r_{xy})\right]}\right\rangle=\exp\left(-\frac{Q_z^2}{2}\left\langle\left[z(0)-z(r_{xy})\right]^2\right\rangle\right) \quad (65)$$

Where the term in the exponent is the height-height correlation function, $h(r_{xy})$.

$$h(r_{xy})=\left\langle\left[z(0)-z(r_{xy})\right]^2\right\rangle \quad (66)$$

Capillary waves theory predicts for a liquid with surface tension $\gamma$ at temperature $T$ (Buff net al. 1965),

$$h(r_{xy})\approx\frac{k_BT}{\pi\gamma}\ln(Q_{\max}r_{xy})=\sigma_{CW}^2(r_{xy}) \quad (67)$$

where an upper limit wavenumber cutoff of capillary waves, $Q_{\max}$, typically on the order of a few molecular units, depends on the liquid. The integral in Equation 64 above can be now written as

$$\int e^{iQ_z\left[z(0)-z(r_{xy})\right]}e^{i\mathbf{Q}'_{xy}\cdot\mathbf{r}_{xy}}d^2\mathbf{r}_{xy}=\int e^{-Q_z^2\sigma_{CW}^2(r_{xy})/2+iQ'_{xy}r_{xy}\cos\theta}d\theta r_{xy}dr_{xy} \quad (68)$$

Which by defining $\eta=k_BTQ_z^2/(2\pi\gamma)$ can be integrated analytically to give

$$\begin{aligned}&2\pi Q_{\max}^{-\eta}Q_{xy}^{\eta-2}\int_0^\infty(r_{xy}Q_{xy})^{1-\eta}J_0(r_{xy}Q_{xy})d(r_{xy}Q_{xy})\\&=2\pi 2^{1-\eta}\pi Q_{\max}^{-\eta}Q_{xy}^{\eta-2}\frac{\Gamma(1-\eta/2)}{\Gamma(\eta/2)}\end{aligned} \quad (69)$$

This shows that non-specular diffuse scattering obeys a power law with respect to the inplane momentum transfer, $(Q_{xy})^{\eta-2}$. This power law, can in principle be confirmed experimentally if delta function-like resolution functions both in $Q_x$ and $Q_y$ could be practically achieved in the laboratory. Therefore, the actual resolution function of the spectrometer has to be determined and be folded with this result to compare with measurements. As discussed later (in Experimental Setup section), the resolution function in synchrotron experiments, for DXS and XR, is dominated by the divergence in the acceptance angle of the outgoing beam, so that the FWHM of the resolution function in the $Q_y$ direction, $\Delta Q_y \approx k_0\sin\alpha_f\Delta\alpha_f$ . As a simple and practical example we assume the resolution along $Q_y$ is a square-like function

$$G(Q_y)=\frac{1}{\Delta Q_y}\begin{cases}1 & -\Delta Q_y/2<Q_y<\Delta Q_y/2\\0 & \text{otherwise}\end{cases} \quad (70)$$

On the other hand, in these experiments the resolution in the $Q_x$ direction is so relaxed that one can assume the detector collects all scattering in this direction. Below, the integration in both directions is performed, first over $Q_x$ and subsequently over the $Q_y$ direction.

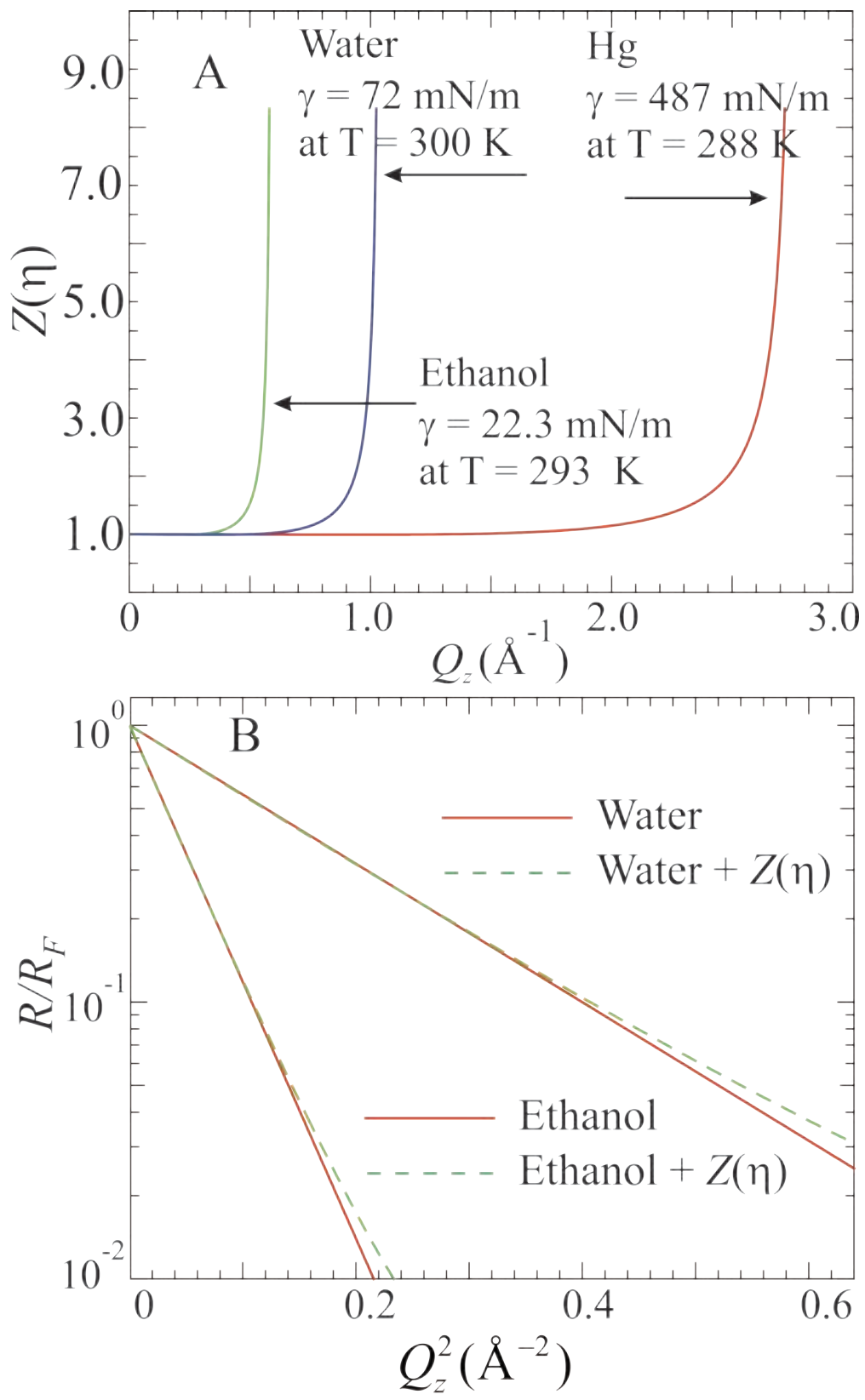

**Figure 9:** (A) Calculated $Z(\eta)$ functions defined in Equation 73 for various liquids with different surface tensions. (B) Calculated normalized reflectivity for water and ethanol using Equation (74) versus the square of $Q_z$ using $Z(\eta)$ = 1 (solid line) and $Z(\eta)$ as shown in (A) (dashed line).

$$\begin{aligned}F(Q_{xy})&=\frac{2\pi 2^{1-\eta}\Gamma(1-\eta/2)}{\Gamma(\eta/2)Q_{\max}^{\eta}}\int_{Q_y-\Delta Q_y/2}^{Q_y-\Delta Q_y/2}\int_{-\infty}^{\infty}(Q_x'^2+Q_y'^2)^{(\eta-2)/2}dQ_x'dQ_y'\\&=\frac{2\pi 2^{1-\eta}\Gamma(1-\eta/2)}{\Gamma(\eta/2)Q_{\max}^{\eta}}\int_{Q_y-\Delta Q_y/2}^{Q_y-\Delta Q_y/2}Q_y'^{\eta-1}\int_{-\infty}^{\infty}(1+t)^{(\eta-2)/2}dtdQ_y'\end{aligned}$$

$$\begin{aligned}F(Q_{xy})&=\frac{2\pi 2^{1-\eta}\sqrt{\pi}\Gamma(1/2-\eta/2)}{\Gamma(\eta/2)Q_{\max}^{\eta}}\int_{Q_y-\Delta Q_y/2}^{Q_y-\Delta Q_y/2}Q_x'^{\eta-1}dQ_y'\\&\left[\int_{-\infty}^{\infty}(1+x)^{\alpha}dx=\frac{\sqrt{\pi}\Gamma(-\alpha-1/2)}{\Gamma(-\alpha)}\right]\\&=\frac{2\pi 2^{1-2\eta}\sqrt{\pi}\Gamma(1/2-\eta/2)}{\eta\Gamma(\eta/2)Q_{\max}^{\eta}}D(Q_y)\end{aligned} \quad (71)$$

and defining

$$D(Q_y) \equiv (\Delta Q_y - 2Q_y)\left|\Delta Q_y - 2Q_y\right|^{\eta-1} + (\Delta Q_y + 2Q_y)\left|\Delta Q_y + 2Q_y\right|^{\eta-1} \tag{72}$$

We now examine the reflectivity by setting $Q_y = 0$

$$R(Q_z) = \left|t(k_z^i)\right|^2 \left|t(k_z^f)\right|^2 \left(\frac{Q_c}{2Q_z}\right)^4 \left|\Phi(Q_z)\right|^2 Z(\eta)\left(\frac{\Delta Q_y}{Q_{\max}}\right)^\eta$$
$$Z(\eta) \equiv \frac{2^{1-2\eta}\Gamma(1/2-\eta/2)}{\sqrt{\pi}\eta\Gamma(\eta/2)} \tag{73}$$

Where we used Equation (55) to identify $R_F(Q_z)$. Thus the final result for the reflectivty from a liquid surface with capillary waves is

$$R(Q_z) = R_F(Q_z)Z(\eta)\left|\Phi(Q_z)\right|^2\left(\frac{Q_{\min}}{Q_{\max}}\right)^\eta \tag{74}$$

The function $Z(\eta)$ in Equation 74 is plotted in Figure 9(A) for different liquids with different surface tension and temperatures, as indicated. In the region of X-ray reflectivity $Z(\eta)$ is practically unity, but it diverges at $\eta=1$ (Pershan 2000). This liquid dependent divergence can be understood as the point where the scattering from the bulk liquid cannot be distinguished from surface scattering. Indeed, for water at RT this point is ~1 Å$^{-1}$. In reality, the height-height correlation function (Equation 66) has another term independent of $r_{xy}$, namely an uncorrelated random component (due to atomic discreteness, for instance). Incorporating such a random term, and following procedure similar to the one from Equation (64) to (66) it can be readily shown that such uncorrelated height distribution gives rise to an intrinsic roughness, $\sigma_I$, that can be incorporated in the reflectivity Equation 74 in the form

$$R(Q_z) = R_F(Q_z)Z(\eta)\exp\left(-Q_z^2\sigma_{eff}^2\right), \tag{75}$$

where, the effective roughness is introduced

$$\sigma_{eff}^2 = \sigma_I^2 + \sigma_{CW}^2 = \sigma_I^2 + \frac{k_BT}{2\pi\gamma}\ln\left(\frac{Q_{\max}}{Q_{\min}}\right) \tag{76}$$

and $Q_{min} = \Delta Q_y(Q_y{=}0) = Q_z\Delta\alpha_f/2$. Equations (75) and (76) show that the reflectivity is $Q_z$ dependent due to the logarithmic term in $\sigma_{CW}$ and also at the singularity of the function $Z(\eta)$ as mentioned above. Figure 9B shows calculated reflectivities ($R/R_F$ versus $Q_z^2$), for water and ethanol using Equation 75 with $Z(\eta)$ =1 and $Z(\eta)$ as shown in Figure 9A. It shows that the two calculations are different only at relatively large $Q_z$ values where the experimental errors are generally larger than the difference between two. Likewise, the logarithmic dependence on $Q_z$ through $Q_{\min}$ of the resolution can be to a good approximation ignored in synchrotron studies (Pershan 2000). However, some correction has to be made when performing XR experiments on in house instruments using common rotating anode or sealed tube type generators. These experiments require continuous opening of incident and scattered beam slits to maintain a reasonable XR signal. One phenomenological way to incorporate a $Q_z$ dependence is to assume that to a first approximation $\sigma_{eff} \sim\sigma_0(1+a\,Q_z)$ (see Tolan 1999 for more details).

**Non-specular diffraction**

As discussed above, the actual ED of a liquid or its interface is made not uniform in the ($x$,$y$) plane by capillary waves, intrinsic roughness, restructuring, or by a deposited film. Such non-uniform in-plane density gives rise to a diffraction pattern in the plane or scattering at finite $Q_x$ and $Q_y$, not necessarily easy to interpret. Capillary waves give rise to non-patterned diffuse scattering, whereas in-plane ordering in a deposited film at the liquid/vapor interface yields Bragg reflections with 2-D-like features (i.e., reflection are rod-like shaped along $Q_z$). For instance, a monolayer of amphiphilic molecules that are spread on a liquid surface self organize and order in the two dimensional (2-D) surface plane. The hydrocarbon chains of these molecules, if compressed to sufficiently high surface pressure, become densely packed as cylinders with a structure that is either hexagonal or a distorted hexagonal (Kjaer 1987, Dutta 1987). Surface crystallization of *n*-alkane molecules on molten alkane has been also reported (Wu et al., 1993b; Ocko et al., 1997).

For the setup of a liquid spectrometer, as illustrated in Figure 1, the components of the general momentum transfer, using Equation 9, are given by

$$\begin{aligned} Q_x &= -k_0\cos\alpha_f\sin 2\theta \\ Q_y &= k_0(\cos\alpha_f\cos 2\theta - \cos\alpha_i) \\ Q_z &= k_0(\sin\alpha_i + \sin\alpha_f) \end{aligned} \tag{77}$$

In most cases, the 2-D order on liquid surfaces is powder-like, and the lateral scans are averaged over $Q_{xy}$ given by

$$Q_{xy} = k_0\sqrt{\cos^2\alpha_i + \cos^2\alpha_f - 2\cos\alpha_i\cos\alpha_f\cos 2\theta} \tag{78}$$

**Diffuse X-ray Scattering (DXS)**

Diffuse scattering from a liquid interface establishes the type of height-height correlation function applicable to the surface and can help determine the type of mechanisms that govern the form of the interface. In principle, with sufficiently tight resolution in 2θ (i.e. approximately $Q_x$ in the configuration shown in Fig. 1) one can measure the effect of these correlations(Fradin 2000, Mechler 2010). But the geometry shown in Figure 8 makes it much more practical to collect the diffuse scattering in the $Q_y$ direction. To measure the $Q_y$ dependence of the DXS the incident and outgoing beam angles are varied in a coordinated way so that their sum is to a good approximation fixed $Q_z$, namely $\alpha_i = \alpha \pm \Delta\alpha$; $\alpha_f = \alpha \mp \Delta\alpha$ thus, $Q_y = Q_z\sin\Delta\alpha$. Using Equations (64), (71), and (72) the DXS is given by,

$$R(\mathbf{Q}_{xy}, Q_z) = \frac{\left|t(k_z^i)\right|^2\left|t(k_z^f)\right|^2 Q_c^4}{128Q_z^2k_0^2\sin\alpha_i\sin\alpha_f Q_{\max}^\eta} Z(\eta)\left|\Phi(Q_z)\right|^2 D(Q_y) \tag{79}$$

It is instructive to calculate the ratio of the diffuse scattering to that of the reflectivity at the same $Q_z$ value as follows

$$\frac{R_{diff}(Q_y,Q_z)}{R(0,Q_z)}=\frac{\left|t(k_z^i)\right|^2\left|t(k_z^f)\right|^2 Q_c^4}{R_F(Q_z)128Q_z^2k_0^2\sin\alpha_i\sin\alpha_f Q_{\min}^{\eta}}D(Q_y) \qquad (80)$$

It should be noted that this ratio goes to 1 as $Q_y \rightarrow 0$ (for same resolution ($Q_{\min}$) in XR and DXS). The instrumental resolution function although complicating the calculations, is not the only considertaion when measuring XDS from a real liquid surface. Innate, unavoidable scattering from correlations in the bulk add incoherently to the scattering from the surface and need be accounted for. In the calculation above (of capillary wave contribution, for instance) a membrane with practically zero thickness at a finite surface tension is assumed. But, in practice there may be a relatively strong signal from the bulk structure-factor, $S_{Liq}(Q)$, that is superimposed on the surface scattering and cannot be ignored. The contribution of this bulk scattering as function of the angles $\alpha_i$ and $\alpha_f$ can be readily calculated as follows,

$$R_{Bulk}(\alpha_i,\alpha_f)=C(Q_z)\left|t(k_z^i)\right|^2\left|t(-k_z^f)\right|^2\frac{A}{\sin\alpha}\times \int S_{Liq}(Q)e^{-z/D(\alpha_i)}e^{-z/D(\alpha_f)}dz \qquad (81)$$

where $C$ is a constant (for a constant $Q_z$) and the incident beam cross-sectional area is $A$. The negative sign in the outgoing transmittance is added to indicate that this beam is in fact undergoing internal reflection, i.e., form the liquid to the vapor. This and Equation 72, show the DXS is asymmetric with respect to $Q_y$ being positive or negative. In the commonly used setup shown in Figure 8, the 2θ is practically fixed (2θ=0) and the value of $Q_y << Q_z$ so that total $Q \approx Q_z$, leading to $S_{Liq}(Q)$ being a constant. The contribution of the liquid bulk in Equation 81 can be written as

$$R_{Bulk}(\alpha_i,\alpha_f)=C(Q_z)\frac{\left|t(k_z^i)\right|^2\left|t(-k_z^f)\right|^2 AS_{Liq}(Q_z)}{\sin\alpha}\times\left(\frac{1}{D(\alpha_i)}+\frac{1}{D(\alpha_f)}\right) \qquad (82)$$

The measured diffuse scattering at a specific $Q_z$, can be modeled as follows,

$$\frac{I_{diff}(Q_y,Q_z)}{I(0,Q_z)}=\frac{R(Q_y,Q_z)+C(Q_z)R_{bulk}(\alpha_i,\alpha_f)}{R(0,Q_z)+C(Q_z)R_{bulk}(\alpha_i,\alpha_i)} \qquad (83)$$

So that one collects the actual bulk scattering count rate (minus room background), at $Q_y = 0$, experimentally of the reflectivity at a finite 2Θ angle

$$C(Q_z)=\frac{I_{BG}^{2\theta}(Q_z)}{I_0}=\frac{\left[I(Q_z)-I_{BG}^{2\theta}(Q_z)\right]}{\left[I(Q_z)-I_{BG}^{2\theta}(Q_z)\right]}\frac{I_{BG}^{2\theta}(Q_z)}{I_0}=R(0,Q_z)C'(Q_z)$$

$$C'(Q_z)\equiv\frac{I_{BG}^{2\theta}(Q_z)}{I(Q_z)-I_{BG}^{2\theta}(Q_z)} \qquad (84)$$

$$\frac{I_{diff}(Q_y,Q_z)}{I(0,Q_z)}=\frac{R(Q_y,Q_z)}{R(0,Q_z)[1+C'(Q_z)]}+\frac{C'(Q_z)R_{bulk}(\alpha_i,\alpha_f)}{[1+C'(Q_z)]} \qquad (85)$$

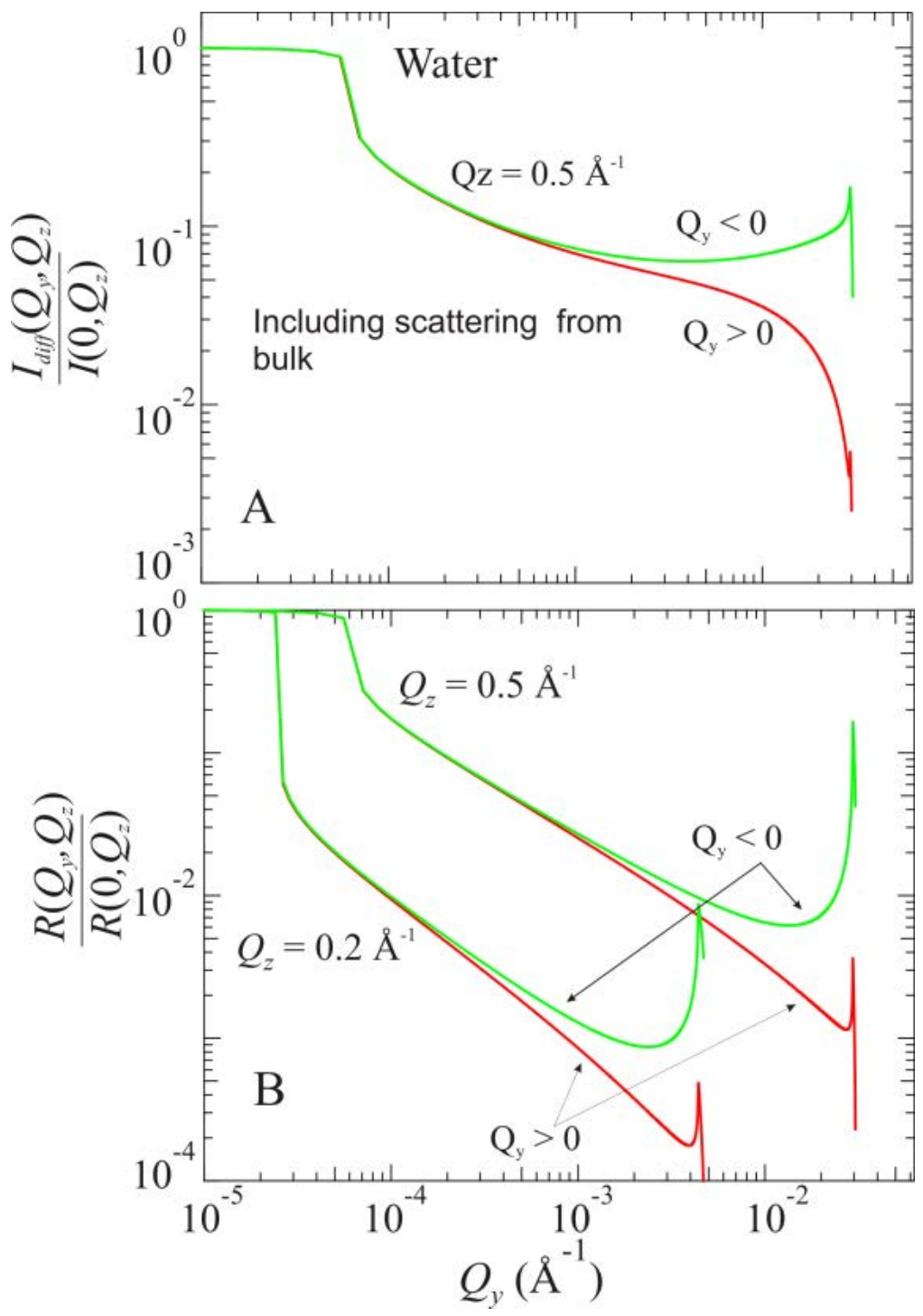

**Figure 10:** Calculated diffuse scattering from water surface at fixed $Q_z$ values (as indicated) normalized to the reflectivity at the same $Q_z$, including a $C'(Q_z) = 0.1$ due to bulk liquid scattering (using Equation 85). (B) DXS scattering calculations normalized to $R_F$ using Equation (80). Note the asymmetry with respect to the sign of $Q_y$. $\Delta\alpha_i = 0.0005$ was used in the calculations.

Figure 10 shows calculations of the DXS from capillary waves of water at various $Q_z$ values with (A) and without (B) accounting for bulk scattering using Equations 85, and 80, respectively. Figure 10 (B) shows that for this type of resolution, the normalized diffuse scattering (NDXS) has three prominent regions that can be estimated from Equation 72: 1) for $Q_y << \Delta Q_y$ T, NDXS → 1 and it falls of rapidly as $Q_y \geq \Delta Q_y$ 2) for $k_c > Q_y > \Delta Q_y$ the NDXS obeys a power law ~ $(Q_y)^{\eta-1}$3) for $Q_y \sim k_c$ it is dominated by multiple scattering from the surface as expressed by the Fresnel transmission function. In this intermediate region, the NDXS exhibits asymmetry with respect to the sign of $Q_y$ (it comes about from the illuminated area $A/\sin\alpha_i$ term in Equation 81). See, e.g., Fig. 7 of (Fukuto, 2006), which shows that $\sin(\alpha)*I(Q_y)/I_0$ is symmetric with respect to the sign of $Q_y$.

The behavior of the DXS gives insights on the origin and properties of surface roughness, in our case the $(Q_y)^{\eta-1}$ power law relates to the capillary waves of a surface [$\eta = k_B T Q_z^2/(2\pi\gamma)$]. The height-height correlation function due to capillary waves (Equation 67) is the most common contribution to the liquid surface roughness in XR and DXS. But, complex liquids and soft matter interfaces bring about different *h(r)*. Sinha considered a few of those cases where analytical solutions can be achieved (Sinha et al. 1988), in particular for two cases where $h(r_{xy}) \sim r_{xy}$, and $h(r_{xy}) \sim r_{xy}^2$ . Capillary wave fluctuations from coupled fluid-fluid interfaces have been studied by Fukuto et al. (2006) (see also Daillant 1992 and Li 2001.).

**Grazing Incidence Diffraction (GID), and Rod Scans:**

For simplicity, we assume a two dimensional (2D) lattice spanned by **a** and **b** vectors so that any in-plane lattice point, $\boldsymbol{\mu}_{nm}$ is given by

$$\boldsymbol{\mu}_{n,m} = n\mathbf{a} + m\mathbf{b} \qquad (86)$$

With a corresponding reciprocal lattice

$$\boldsymbol{\tau}_{h,k} = h\mathbf{a}^* + k\mathbf{b}^* \qquad (87)$$

Where *n, m, h,* and *k* are integers. The ED in each unit cell is then given as a sum over all atoms in the unit cell

$$\rho_{cell} = \sum \rho_j(r)\delta(\boldsymbol{\mu} - \boldsymbol{\mu}_j)\delta(z - z_j) \qquad (88)$$

The Fourier transform of the ED of the 2D-lattice yields the inplane Bragg peak for the condition $\mathbf{Q}_{xy} = \boldsymbol{\tau}_{h,k}$ and a unit cell structure factor

$$F_{h,k}(Q_z) = \sum_j f_j(Q) e^{2\pi i(hx_j + ky_j) + iQ_z z_j} \qquad (89)$$

Inserting Equation 91 in Equation 62 and integrating yields the differential cross-section for a quasi 2-D Bragg reflection

$$\frac{d\sigma}{d\Omega} \sim P(\mathbf{Q})\,|t(k_z^i)|^2 \left\langle |F_{h,k}(Q_z)|^2 \right\rangle |t(k_z^f)|^2 \qquad (90)$$

where $P(\mathbf{Q})$ is a polarization correction and the structure factor squared is averaged for multiplicity due to domains and weighted for orientation relative to the surface normal.. For s-wave (x-polarized) beam commonly available at an undulator beamline at a synchrotron source $P(\mathbf{Q};2\theta,\alpha_i,\alpha_f) \approx \cos^2 2\Theta$. For unpolarized beam $P(\mathbf{Q};2\theta,\alpha_i,\alpha_f) \approx (1+\cos^2 2\Theta)/2$. The ordering of monolayers at the air-water interface is usually in the form of 2D powder consisting of crystals with random orientation in the plane. If the film thickness is delta-function like, Equation 90 gives a form factor that is independent of $Q_Z$ . The expected scattering for such a system is a rod-like of scattering along $Q_Z$ that peaks at the critical angles due to multiple scattering, in the DWBA. The in-plane ordered entities giving rise to the Bragg reflection, have finite length *L*, for instance, the hydrocarbon chains of a Langmuir monolayer. The molecular form factor of a uniform long chain, with ED $\rho_{\text{chain}}(z)$ = constant, along $Q_Z$ can be readily evaluated by summing all the contributions of a dissected chain into *N* parts, to yield

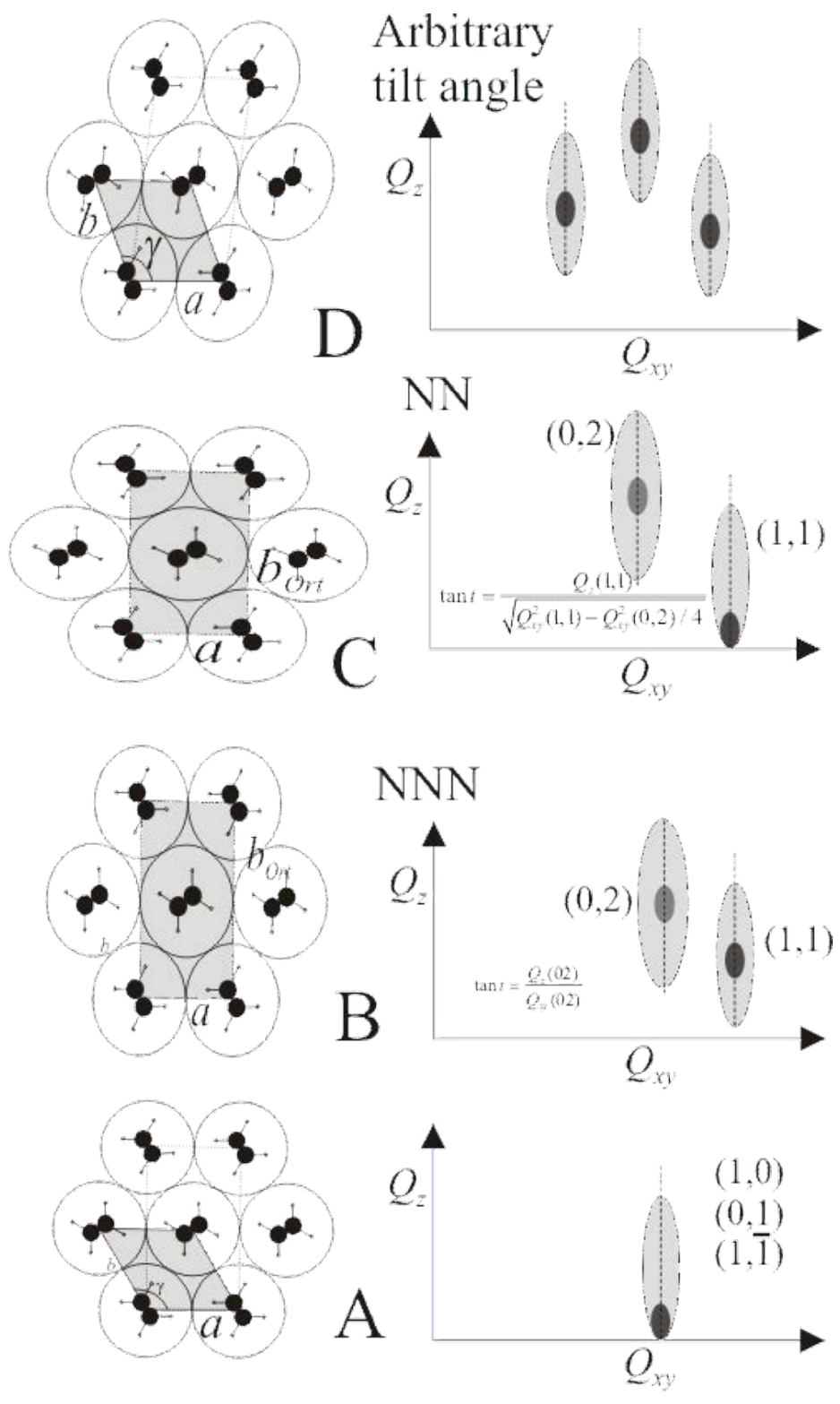

**Figure 11:** Various scenarios of hydrocarbon chain ordering induced primarily by their tilt angle with respect to the surface normal and with respect to their neighbors and the corresponding 2D contour plots of the GIXD in the ($Q_{xy}$, $Q_z$) plane; Black circles depict carbon atoms linked to two hydrogen atoms of a slice through the chains. (A) hexagonal, structure, (B) tilt towards NNN, (C) tilt to NN, (D) arbitrary tilt angle. Chain-chain interactions can induce more structures such as the herring-bone structure (See Kaganer 1999).

$$F_{h,k}(Q_z) = \frac{1 - e^{iQ_z Na}}{\left(1 - e^{iQ_z a}\right)N} F_{h,k}(0) \approx \frac{\sin Q_z L/2}{Q_z L/2} F_{h,k}(0) \qquad (91)$$

So, a scan at the Bragg reflection from a cylinder-like molecule forming a hexagonal lattice with a lattice constant gives a rod-scan as in Equation 91, and shown schematically in Figure 11. Experiments show that the chains can be tilted with respect to surface normal either towards nearest-neighbors (NN), towards next NN (NNN), or at an arbitrary direction as shown in Figure 11. Such tilt angles break the hexagonal symmetry and the predicted rod-scans peak at finite $Q_Z$ as schematically indicated in Figure 11 (Kaganer 1999). The chain tilt angle can be readily calculated from the peaks in the rod scans so that for a tilt towards NN the tilt angle is given by

$$\tan t = \frac{Q_z(1,1)}{\sqrt{Q_{xy}^2(1,1) - Q_{xy}^2(0,2)/4}} \qquad (92)$$

and if the tilt is towards NNN

$$\tan t = \frac{Q_z(02)}{Q_{xy}(02)} \qquad (93)$$

The area per chain is then given by

$$A_{chain} = \frac{2\pi}{Q_{xy}(0,2)} \frac{2\pi}{\sqrt{Q_{xy}^2(1,1) - Q_{xy}^2(0,2)/4}} \tag{94}$$

which gives the correct area for the hexagonal case as the (02) and (11) reflections (orthorhombic notation) are degenerate in the hexagonal with the other six reflections (for instance, (10) reflection in the hexagonal notation.)

**X-ray Spectroscopy at liquid interfaces**

Up to this point it was assumed that the electrons are free and uniformaly distributed across the liquid. However, we know that even though the electrons are highly concentrated at each atom, at large enough $Q$ values, in the BA gives rise to an atomic form factor $f^0(\mathbf{Q})$ that falls of rapidly at $Q$ values of order of the inverse of the average atomic radius $1/R$. Proper count of the effective number of electrons of an atom or (ion) with $Z$ electrons should then be

$$f^0(\mathbf{Q}) = \frac{1}{Z}\int \rho_e(\mathbf{r}) e^{i\mathbf{Q}\cdot\mathbf{r}} d\mathbf{r} \tag{95}$$

In a typical x-ray reflectivity measurement $Q_z \rightarrow 0$, and the form factor can be ignored, namely we take $f^0(Q_Z) = 1$. Furthermore, we have seen before that the generalized ED is a complex function, and that the absorption component is photon energy dependent, $E$. In fact, the energy dependence is much more dramatic when the photon energy is close to a binding energy level in the atom, and the energy dependence is incorporated into the form factor with a correction. The simplest way to treat the generalized ED near an absorption edge is by assuming that the electrons are not free but their motion obeys a forced and damped harmonic oscillator, with a resonant energy $E_0$. In analogy, to the forced harmonic oscillator we write the generalized form factor as follows

$$f(\mathbf{Q}, E) = f^0(\mathbf{Q}) + f'(E) + if''(E) \tag{96}$$

where $f'$ and $f''$ are the real and imaginary parts of the dispersion corrections. Because of its specificity, we deal with the contribution of an ion near one of its absorption edges, with core density that is highly localized, especially for K edge resonances, where the ED can be essentially treated as a delta function, so that the Fourier transform of core electrons is essentially 1 over the $Q$ ranges commonly used in x-ray experiments, (see discussion on the $Q$ dependence of $f'$ *and* $f''$ in the ITC page 215, edt. Wilson 1992). The generalized contribution due to each ion is

$$f' + if'' = \frac{\omega_0^2}{\omega^2 - \omega_0^2 - i\omega\Gamma} \tag{97}$$

Where it is assumed that $\omega_0 >> \Gamma$. The relation between the $f'$ and $f''$ is given by the Kramers-Kroning relations (Als-Nielsen and McMorrow 2001). Below, a few methods to monitor ions specifically at the liquid interfaces are described.

**Anomalous x-ray reflectivity:** To determine ion distribution at an interface by some mechanism, we perform x-ray reflectivity measurements at a few energies away from and at and around an absorption edge of the specific ion in the solution. Model profile EDs that fit the measurements at different energies are constructed and subsequently used to relate the differences to the ionic distribution. To achieve this goal, we introduce an effective number of electrons parameter

$$Z_{eff}^{ion} = Z^{ion}\left(1 + f'(E)\right) \tag{98}$$

At energies much higher than the absorption edge, $Z_{eff} \approx Z$, but at and around the resonance this number can be reduced significantly.

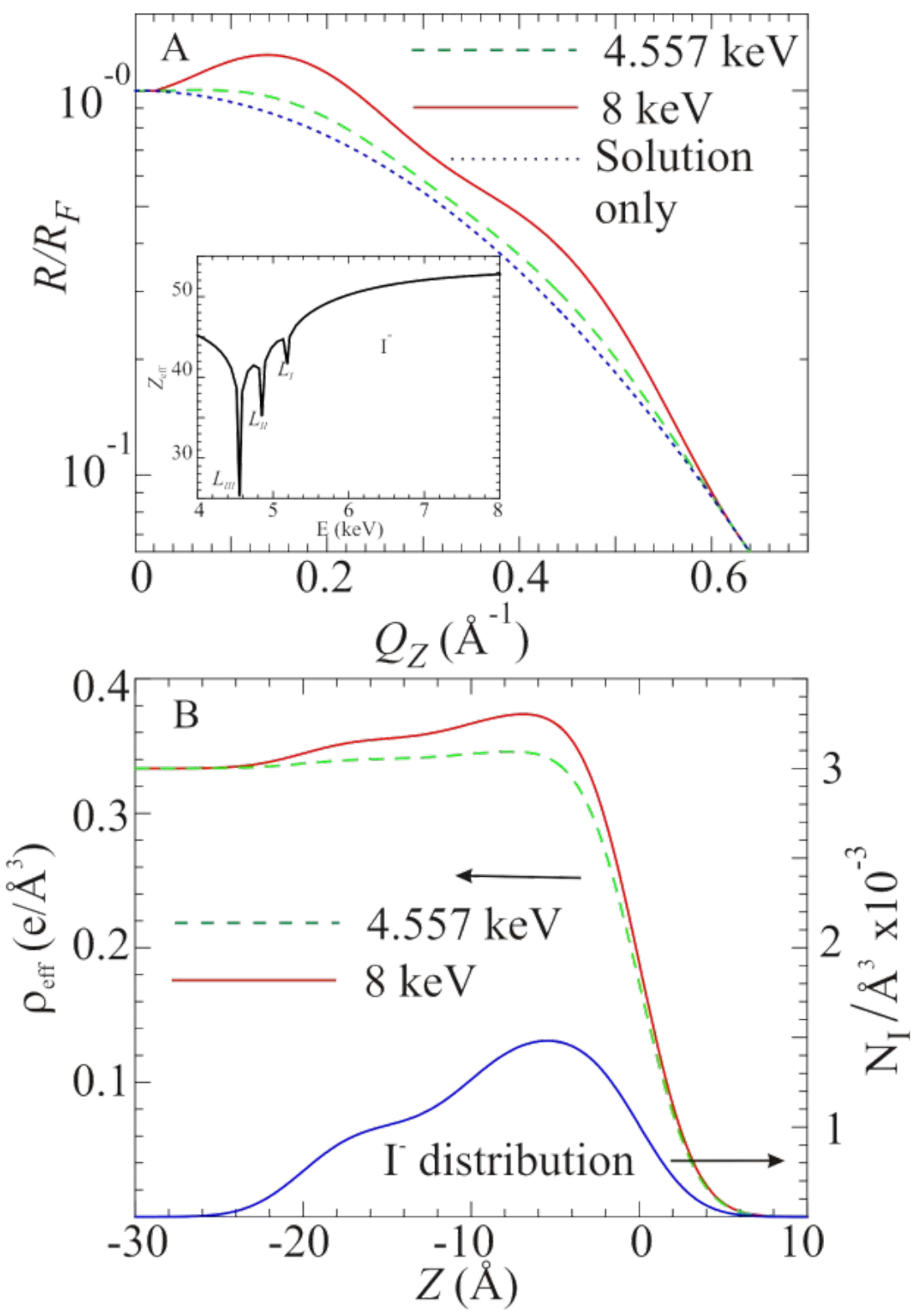

**Figure 12: (A)** XR calculations from a hypothetical water surface that is enriched with iodine ions at the interface. The calculations are performed at two different energies to demonstrate the effect of at and away from the $L_{III}$ iodine resonance. The inset above shows the effective number of electrons contributing to the reflectivity from each iodine ion versus photon energy. (B) The ED's that are used to calculate the reflectivities. Interfacial iodine distributions used is also shown.

As a hypothetical case we assume an iodine salt solution, and that the iodine spontaneously accumulates at the interface. We collect XR off resonance at $E_{off}$ = 8 keV and at the iodine $L_{III}$ edge, $E_0$ = 4.557.1 keV; At these energies the $Z_{eff}$ are 53 and 24.6 , respectively as shown in the inset of Figure 12. We then determine the ED's that best fit the data self consistently for both energies. The difference between the two EDs normalized to the "missing" electrons per ion yield the ion distribution as follows

$$N_{ion}(z) = \frac{\rho(E_{off}, z) - \rho(E_0, z)}{Z_{ion}\left|f'(E_0)\right|} \tag{99}$$

as shown in Figure 12(b). There are a few advantages to this method in determining ion distributions. First, it is highly specific, that is, in principle it can be used to determine relative surface enrichment of different ions. Second, it is to a large extent model independent of the details of the ED of the films at the interface. It should be noted that the reflectivities can be performed at various energies and not necessarily at exactly the edge. The exact energy of the edge may be chemically shifted and $Z_{\text{eff}}$ may change as well. It is therefore advantageous to measure the edge of the ion in the solution prior to these measurements.

### Energy scans at fixed momentum transfers

Another method to determine ion distributions at interfaces is by measuring the intensity of the reflected beam as a function of incident photon energy at fixed $Q_z$. The complex ED in this case can be written as,

$$\rho(z,E) = \sum_{j\neq ion} \rho_j + N_{ion}(z) Z_{ion}\left(1 + f'(E) + if''(E)\right) \qquad (100)$$

For this to be effective $E$-scans at a few $Q_z$ values are collected, which by self consistent modeling can provide not only the distributions of the ions $N_{\text{ion}}(z)$ but also the dispersion corrections of the ion in its environment at the interface. These scans can be performed at a limited range of energy near the absorption edge to yield the so called interfacial X-ray absorption near edge spectroscopy (XANES). Depending on signal strength such measurements can be extended to higher energies to yield interfacial EXAFS.

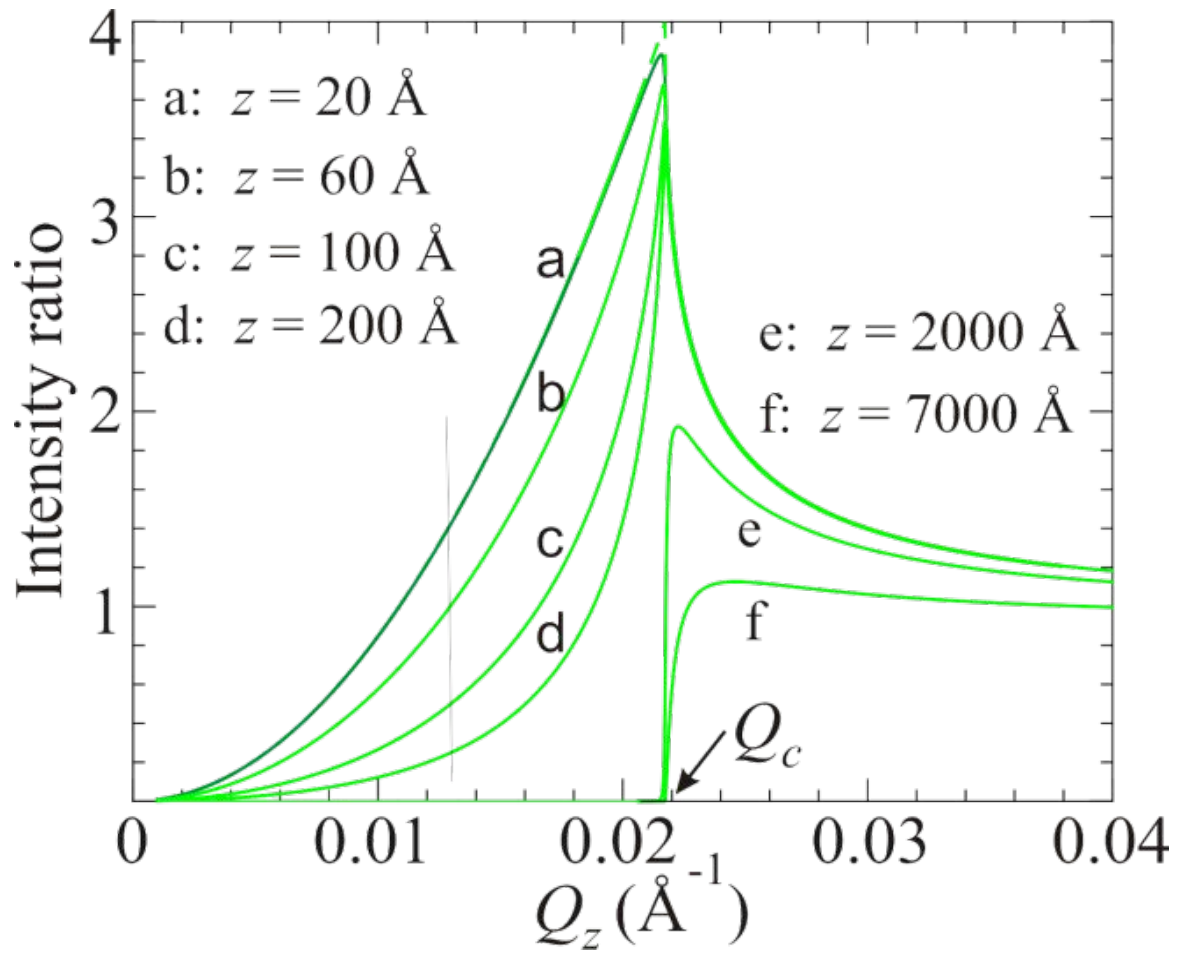

**Figure 13:** Intensity calculation based on Eq. 104 of a fluorescent signal from an object at various depth positions (z) with respect to the surface, assuming that the ED is that of water, and the absorption at 8 keV ($\xi = 10\,\text{cm}^{-1}$). The dashed line is calculated assuming the particle is at z = 0 and $\mu = 0$ where the enhancement at the critical angle is exactly 4. The intensity ratio at 0.015 compared to that at the $Q_c$ can be used to qualitatively determine the position of the emitting particle with respect to the surface.

### $Q_z$ and energy dependence of fluorescence at interfaces

As an x-ray beam travels through the liquid, in reflectivity experiment for instance, some or all of the transmitted beam is absorbed by elements that constitute the liquid. The absorbing elements are then excited to higher energy levels by the removal of core electrons that almost instantaneously are filled in by a cascade of electrons from outer shells, emitting photons that are characteristic to each element. Collecting and analyzing the emitted photons in a fluorescence experiment is therefore highly desirable to qualitatively determine the existence of a certain element in the liquid or at the interface. The fluorescence technique can be readily employed to quantitatively determine the ion density at the surface and to a certain extent also determine their spatial distribution with respect to the interface. Because of the incoherent character of the absorption/emission process, the intensity of an emission line is proportional to the density of the element in the liquid bulk or at the surface, and the effective illuminated volume. The relative intensity of the fluorescence beam is then given by

$$\frac{I(Q_z, z_{ion})}{I(Q_z, \infty)} = \left|t(k_z^i)\right|^2 e^{-z_{ion}/D(k_z^i)} \qquad (101)$$

Figure 13 shows the predicted relative fluorescence intensities characteristic to elements that reside at a specific position $z$ with respect to the surface. The actual fluorescence intensity from ion enriched-surface, with ion distribution $N_{ion}(z)$ can then be written as

$$I(k_z^i) = \left|t(k_z^i)\right|^2 \int N_{ion}(z) e^{-z/D(k_z^i)} dz \qquad (102)$$

Assuming, the ion distribution $N_{ion}(z)$ falls of exponentially with a characteristic length $\Lambda$ as is typical of ion distributions at charged aqueous interfaces, we can write it in the this form

$$N_{ion}(z) = N_{surf} e^{-z/\Lambda} + N_{bulk} \qquad (103)$$

and require that the surface density (two-dimensional density per unit area) of the ions is $N_{2D} = N_{surf}\Lambda$ is a constant, i.e., if $\Lambda = 0$, the distribution is a delta function like, and $N_{surf} = N_{2D}$ is located at exactly $z = 0$. Under these assumptions the following formula for the intensity from a distribution given by

$$I(k_z^i) = \left|t(k_z^i)\right|^2 \left[ N_{surf} \frac{\left(\Lambda D(k_z^i)\right)}{\Lambda + D(k_z^i)} + N_{bulk} D(k_z^i) \right] \qquad (104)$$

The fluorescence technique can be also used near an edge, where the intensity of a characteristic emission line is collected as a function of photon energy to yield directly the imaginary part of the dispersion correction $f''$ (Bu et al. 2009). Figure 14 shows calculations of relative fluorescence signals from a 1mM solution of ions that a few aggregated to the interface at exactly $z = 0$ and another with an exponentially decaying distribution of characteristic length $\Lambda = 20$ Å (total amount integrated over z for both is the same). One can clearly see two regions below and above the critical angle that practically distinguish the signal from the surface and from the bulk. The signal from an ion distribution over 20 Å is different from that of ions strictly at the interface only below the critical angle. Above the critical angle they are both practically indistinguishable. This is a general fact to consider; a peak at the critical angle does not necessarily mean the signal is due to an entity at the surface, i.e. $z = 0$. How the signal rises from $Q_z = 0$ to its maximum is more important in determining the spatial position of the object in fluorescence or in GIXD experiments.

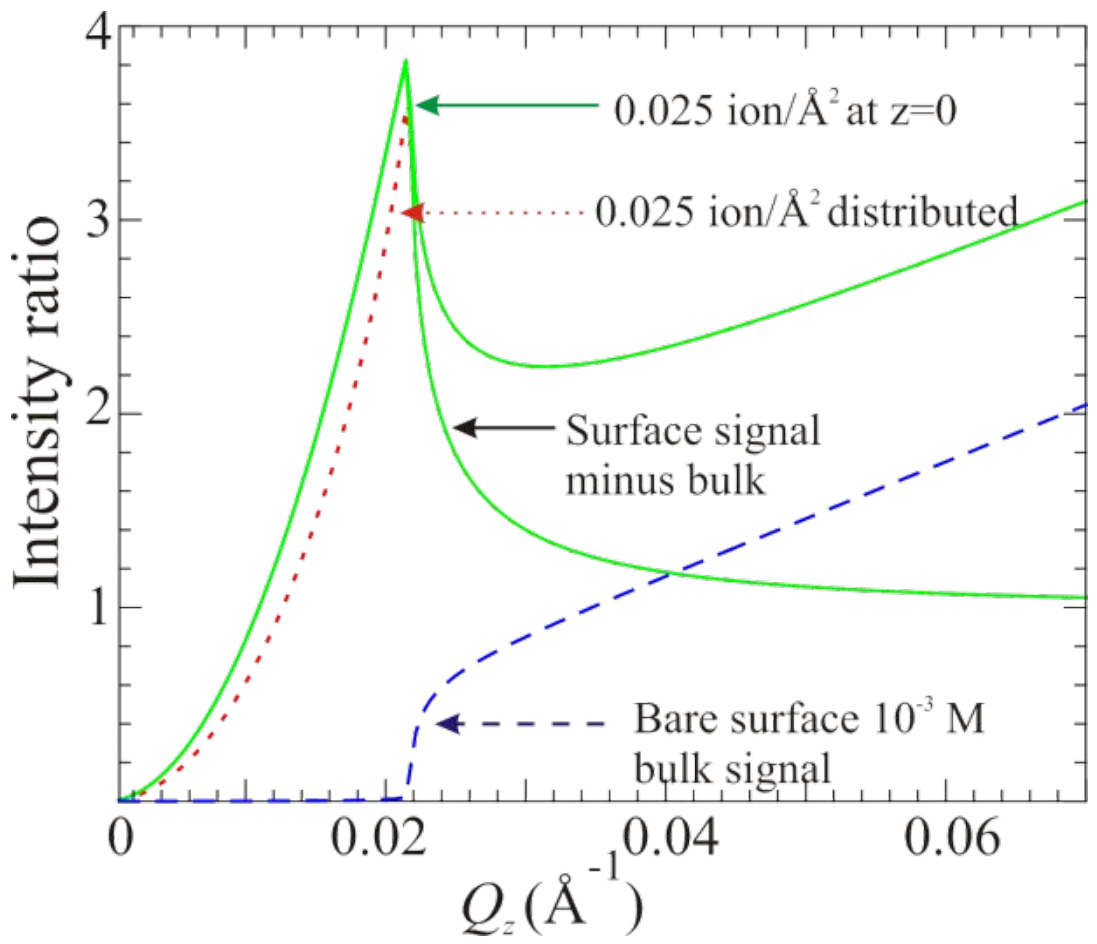

**Figure 14:** Calculated relative fluorescence signal from a 1mM solution of ions that a few aggregated at the interface at exactly $z = 0$ so (solid line) and with an exponentially decaying distribution with the characteristic length Λ = 20 Å (dotted line). One can clearly see two regions below and above the critical angle that practically distinguish the signal from the surface and from the bulk. Even for the case of a distribution of ions over 20 Å the difference from the case of ions at the interface is prominent only below the critical angle. Above the critical angle they are both practically indistinguishable. Dashed line is the expected fluorescence from the subphase.

## PRACTICAL ASPECTS OF THE METHOD

The minute sizes of interfacial samples on the sub-micro-gram level, combined with the weak interaction of x-rays with matter, result in very weak GIXD and reflectivity (at large $Q_Z$) signals that require highly intense incident beams, which are available at x-ray synchrotron sources (see SURFACE X-RAY DIFFRACTION). A well prepared incident beam for reflectivity experiments at a synchrotron (for example, on the Ames Laboratory Liquid Surface 6ID-B beamline spectrometer at the Advanced Photon Source at Argonne National Laboratory) has an intensity of $10^9$ to $10^{11}$ photons/sec (depending on photon energy, and for incident beam cross section of 0.1x1 mm$^2$ ), whereas, for a similar resolution, an 18-kW rotating anode generator produces $10^4$ to $10^5$ photons/sec. Although reflectivity measurements can be carried out with standard x-ray generators, the measurements are limited in $Q_z$ range compared to those accessible at synchrotron sources, and they take hours to complete compared to minutes at the synchrotron. The reflectivity from liquids at the synchrotron is practically limited by the strong scattering from the bulk of the liquid at large Q values, as discussed in relation to $Z(\eta)$ function (see Figure 9). GIXD experiments are practically impossible with x-ray generators, since the expected signals (2-D Bragg reflections, for example) normalized to the incident beam are on the order of $10^{-7}$ to $10^{-10}$.

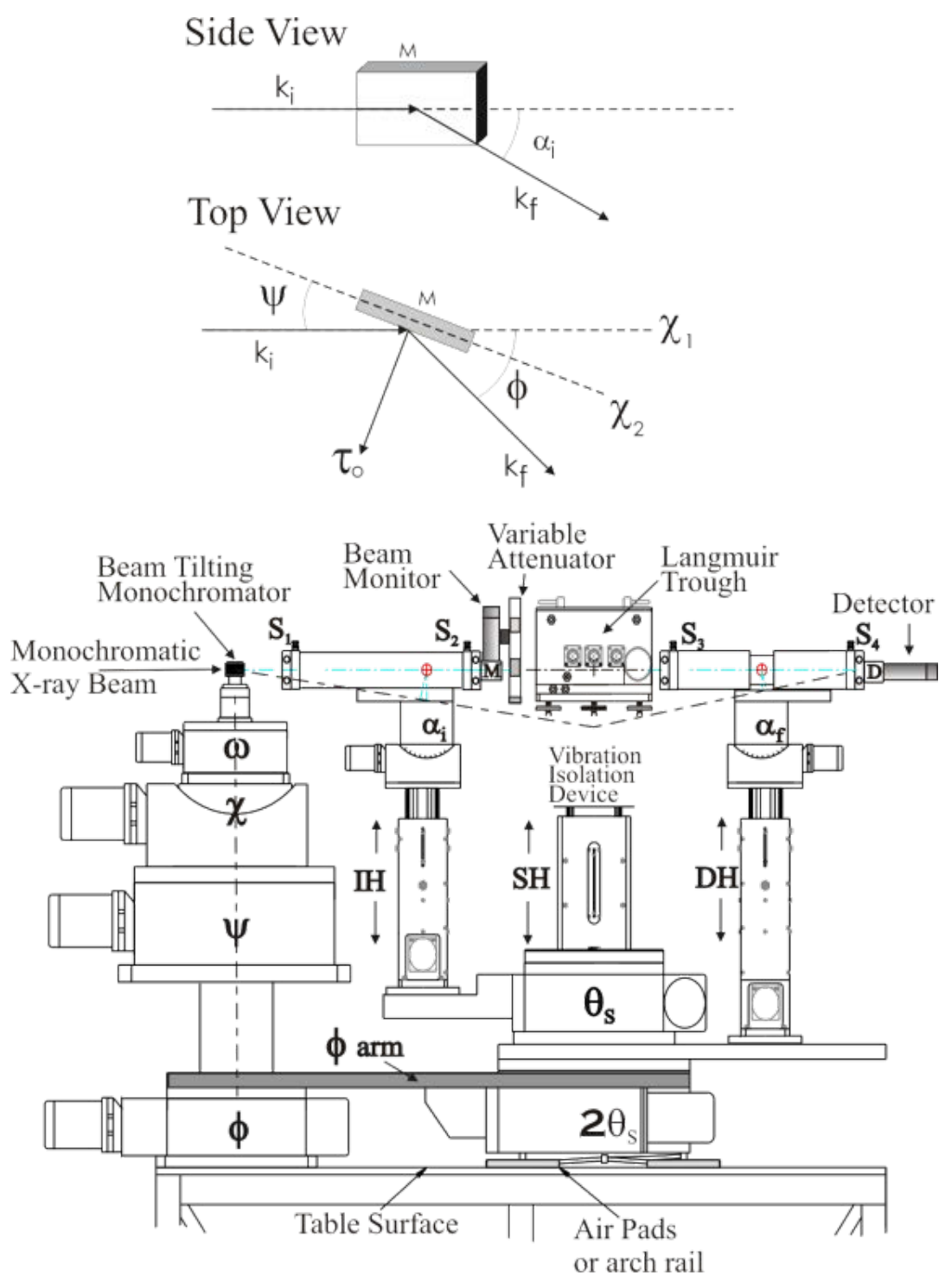

**Figure 15:** (**top**) Monochromator geometry to tilt a Bragg reflected beam from the horizon on a liquid surface. Two possible tilting configurations one about the primary beam axis and another about an axis along the surface of the reflecting planes ($\chi_1$ and $\chi_2$, respectively) are shown. A side view diagram of the Ames Laboratory Liquid Surfaces Diffractometer at the 6-ID beam line at the Advanced Photon Source at Argonne National Laboratory.

### Spectrometer

X-ray reflectivity, GIXD, and spectroscopy measurements of liquid surfaces are carried out on special spectrometers that enable manipulation of the incident as well as the outgoing beam with respect to a horizontal liquid surface. A prototype liquid surface reflectometer was described by Als-Nielsen and Pershan (1983) (see also, Schlossman 1997). In order to bring the beam to an angle of incidence $\alpha_i$ with respect to the liquid surface, a secondary monochromator crystal (referred to as mirror) is placed on the spectrometer itself. The monochromator crystal is tilted by an angle $\chi$ either about the axis of the incident beam (indicated by $\chi_1$ in Fig. 14) or about the axis normal to the reciprocal lattice wave vector of the monochromator, $\tau_0$ ($\chi_2$). Figure 15 shows the geometry that is used to deflect the beam from the horizontal onto the liquid surface at an angle, $\alpha_i$, by tilting the monochromator. At the Bragg condition, the surface of the monochromator crystal is at an angle $\psi$ with respect to the incoming beam. Tilting over the incident beam axis is like tracing the Bragg reflection on the Debye-Scherer cone so that the $\psi$ axis remains fixed, with a constant wavelength at different tilting angles. The rotated reciprocal lattice vector and the final wave vector in this frame are given by

$$\begin{aligned} \tau_0 &= \tau_0(-\sin\psi, \cos\psi\cos\chi_1, \cos\psi\sin\chi_1) \\ \mathbf{k}_f &= k_0(\cos\alpha_i, \cos\phi, \cos\alpha_i\sin\phi, \sin\alpha_i) \end{aligned} \tag{105}$$

where $\phi$ is the horizontal scattering angle. The Bragg conditions for scattering are given by

$$\mathbf{k}_i + \tau_0 = \mathbf{k}_f; \qquad |\mathbf{k}_f| = k_0 \tag{106}$$

Using Equations 105 and 106, the following relations for the monochromator axes are obtained

$$\begin{aligned} \sin\psi &= \frac{\tau_0}{2k_0} \\ \sin\chi_1 &= \frac{k_0}{t_0}\cos\psi\sin\alpha_i \\ \cos\phi &= \left(1-\frac{\tau_0^2}{2k_0^2}\right)\Big/\cos\alpha_i \end{aligned} \tag{107}$$

and we notice that the monochromator angle $\psi$ is independent of $\alpha_i$. However, the scattering angle $\phi$ has to be modified as $\alpha_i$ is varied. This means that the whole reflectometer arm has to be rotated. Similarly, for the configuration where the monochromator is tilted over the axis normal to $\tau_0$ we get

$$\begin{aligned} \sin\psi &= \frac{\tau_0}{2k_0\cos\chi_2} \\ \sin\chi_2 &= \frac{k_0}{t_0}\sin\alpha_i \\ \cos\phi &= \left(1-\frac{\tau_0^2}{2k_0^2}\right)\Big/\cos\alpha_i \end{aligned} \tag{108}$$

From these relations, the conditions for a constant wavelength operation for any angle of incidence, $\alpha_i$, can be calculated and applied to the reflectometer. Here, unlike the previous mode, beam deflection to different angles of incidence requires both the adjustment of $\psi$ and of $\phi$ in order to maintain a constant wavelength. If $\psi$ is not corrected in this mode of operation, the wavelength varies as $\chi$ is varied. This mode is sometimes desirable, especially when the incident beam hitting the monochromator consists of a continuous distribution of wavelengths around the wavelength at horizontal scattering, $\chi_2 = 0$. Such continuous wavelength distribution exists when operating with x-ray tubes or when the tilting monochromator is facing the white beam of a synchrotron. Although the variation in the wavelength is negligible as $\chi_2$ is varied, without the correction of $\psi$, the exact wavelength and the momentum transfer can be computed using the relations in Equation 109 and Equation 110. In both modes of monochromator tilting, the surface height as well as the height of the slits are adjusted with vertical translations.

Figure 15, panel B, shows a side-view diagram of the Ames Laboratory Liquid Surfaces Diffractometer at the 6-ID beam line at the Advanced Photon Source at Argonne National Laboratory. A downstream Si double-crystal monochromatcr selects a very narrow-bandwidth energy (> 1 eV) beam (in the range 4 to 40 keV) from an undulator-generated beam. The highly monochromatic beam is deflected onto the liquid surface to any desired angle of incidence, $\alpha_i$, by the beam-tilting monochromator; typically Ge(111), Ge(220), or x-cut quartz crystals are used (for various photon energy ranges). To simplify its description, the diffractometer can be divided into two main stages, with components that may vary slightly depending on the details of the experiment. In the first stage, the incident beam on the liquid surface is optimized. This part consists of the axes that adjust the beam tilting monochromator ($\psi$, $\phi$, $\chi$), incident beam

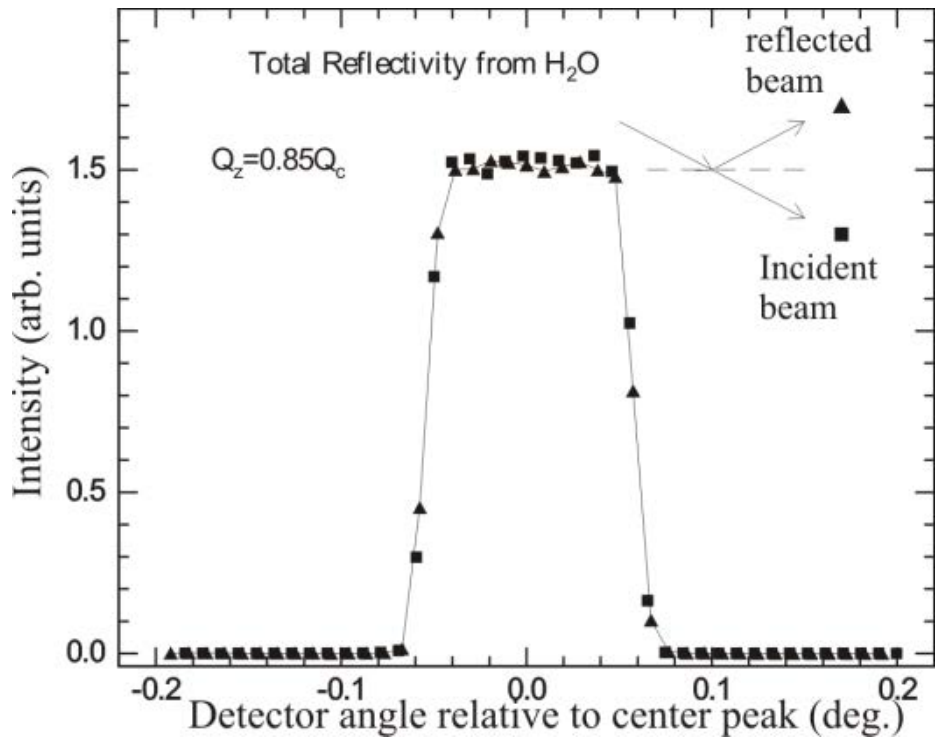

**Figure 16:** Superposition of the reflected beam intensity below the critical angle (triangles) and direct beam intensity (squares), demonstrating near-total reflectivity of x-rays from the surface of water. Severe surface roughness reduces the intensity and broadens the reflected signal. A reduction from total reflectivity can also occur if the slits of the incident beam are too wide, so that the beam-footprint is larger than the surface sample.

slits ($S_i$), beam monitor, and attenuator. The $\omega$ axis, below the monochromator crystal, is adjusted in the initial alignment process to ensure that the tilting axis $\chi$ is well defined. For each angle of incidence $\alpha_i$, the $\psi$, $\phi$, $\chi$ and the height of the incident beam arm (IH) carrying the $S_1$, $S_2$ slits, are adjusted (note that the $\chi$ degree of freedom in Figure 15 is rotated so it is seen). In addition, the liquid surface height (SH) is brought to the intersecting point between the incident beam and the detector arm axis of rotation $2\theta_s$. In the second stage of the diffractometer, the intensity of the scattered beam from the surface is mapped out. In this section, the angles $\alpha_f$ and $2\theta_s$ and the detector height (DH) are adjusted to select and detect the outgoing beam. The two stages are coupled through the $\phi$-arm of the diffractometer. In general, $\theta_s$ is kept idle because of the polycrystalline nature of monolayers at liquid surfaces. By contrast surface crystallization of alkanes proceeds with the formation of a few large single crystals (Wu et al 1994), and the use of $\theta_s$ is essential to orienting one of these crystals with respect to the incoming beam. The scattering from liquid metals is complicated by the fact that their surfaces are not uniformly flat. For details on how to scatter from liquids with curved surfaces see Regan et al. (1997).

For aqueous surfaces, a trough (approximate dimensions ~120 × 270 × 5 $mm^3$) is placed in an airtight aluminum enclosure that allows for the exchange of the gaseous environment around the liquid surface. To get total reflectivity from water below the critical angle (e.g., $\alpha_c = 0.1538°$ and $0.07688°$ at 8 keV and 16 keV, respectively), the footprint of the beam has to be smaller than the specimen surface. A typical cross-section of the incident beam is $0.5 \times 0.1$ mm$^2$ (at 8 keV) with approximately $10^{10}$ to $10^{11}$ photons per second (6-ID beamline). At about $0.8\alpha_c$, the footprint of the beam with a vertical size (slit $S_2$) of about 0.1 mm is ~47 mm at 8 keV and about the same size for a 0.05 mm slit at 16 keV, compared to 120 mm liquid-surface dimension in the direction of the primary beam. Whereas a 0.1 mm beam size is adequate for getting total reflectivity at 8 keV (exploiting about half of the liquid surface in the beam direction), a 0.05-mm beam size is more appropriate at 16 keV. This vertical beam size (slit $S_2$) can be increased at larger incident angles to maintain a relatively constant beam footprint on the surface.

The alignment of the diffractometer encompasses two interconnected iterated processes. First, the angles of the first stage ($\alpha_i$, $\psi$, $\phi$, $\chi$, and $\omega$) are optimized so that the x-ray flux at the monitor position is preserved upon deflection of the beam (tracking procedure). Second, the beam is steered so that it is parallel to the liquid surface. It should be emphasized that the beam, after the tracking procedure, is not necessarily parallel to the 1iquid surface. In this process, reflectivities from the liquid surface at various incident angles are employed to define the parallel beam, by adjustment of the angles and heights of the diffractometer. Then, repeatedly, the first and second processes are iterated until convergence is achieved (i.e., until corrections to the initial positions of motors are smaller than the mechanical accuracies).

The divergence of the monochromatic incident beam on the surface is determined by at least two horizontal slits located between the sample and the source. One of these slits is usually located as close as possible to the sample, and the other as close as possible to the source. These two slits determine the resolution of the incident beam. By deflecting the beam from the horizontal, the shape of the beam changes, and that may change the incident beam intensity going through the slits; therefore the use of a monitor right after the slit in front of the sample is essential for the absolute determination of the reflectivity. The size of the two slits defining the incident beam is chosen in such a way that the footprint of the beam is much smaller than the width of the reflecting surface, so that total reflectivity is observed. Figure 16 shows the reflected beam and the direct beam from a flat surface of water, demonstrating total reflectivity at $Q_z = 0.85Q_c$. In this experiment the detector slit is wide open at ~10 times the opening of the one of the incident beam. As is demonstrated, the effect of absorption is negligible for water, and roughness is significantly reduced by damping surface waves. Damping of mechanical vibrations is achieved by reducing the height of the water film to below ~0.3 mm by placing a passive as well as an active antivibration unit underneath the liquid sample holder, suppressing mechanical vibrations (Kjaer et al., 1994).
The adjustment of the sample height and other spectrometer components with the variation of beam incident angle in this type of spectrometers introduces mechanical vibrations that can persist for some time (seconds to even minutes) and interfere with consistent data collection. To rapidly dampen such mechanical vibrations, a thin liquid film, thick enough to maintain the liquid phase integrity with no influence from the flat solid holder (a few to 300 μm thick) is desirable. In addition, a waiting time between spectrometer adjustment and starting photon count is routinely used to mitigate the effect of these self inflicted mechanical vibrations. However, even these two remedies may not be practical (very long waiting time before counting) when dealing with liquid/liquid and liquid-metal/vapor interfaces and a completely stationary sample holder that is completely decoupled from the rest of the spectrometer is desired. Indeed, a novel design that allows manipulation of the incoming and outgoing beams and produces a stationary scattering volume at the same spot on the sample surface has been introduced at the high-energy diffraction beamline ID15A at the European Synchrotron Radiation Facility (ESRF, Grenoble, France). This has been achieved by using a pair of deflecting crystals that maintain the incident X-ray beam centered at a stationary sample position while the incident angle with respect to the surface is varied (Honkimäki et al., 2006). A similar design, using a double crystal beam-tilter in Bragg geometry, has also been implemented on the PETRA III High Resolution Diffraction Beamline at the DESY synchrotron in Hamburg, Germany (Murphy et al., 2010).

**Spectrometer resolution**
The resolution of the spectrometer is a function of beam incident and outgoing beam divergence $\Delta\alpha_i$ , $\Delta\alpha_f$ , and $\Delta 2\theta$ and the photon energy bandwidth expressed by $\Delta k_0$. These Δ−values correspond to the full-width at half maximum (FWHM) values of a resolution function that could be modeled as Gaussian or a square function, depending on the specifics of the instrument and beam source. Based on the geometry and Equation (77) the most general FWHM of the Q-vector components are

$$\begin{aligned}
\Delta Q_x &= \Delta k_0 \cos\alpha_f \sin 2\theta + k_0(\sin\alpha_f \Delta\alpha_f + \cos\alpha_f \cos 2\theta \Delta 2\theta) \\
\Delta Q_y &= \Delta k_0 (\cos\alpha_f \cos 2\theta - \cos\alpha_i) + \\
&\quad k_0(\sin\alpha_f \Delta\alpha_f + \cos\alpha_f \sin 2\theta \Delta 2\theta + \sin\alpha_i \Delta\alpha_i) \\
\Delta Q_z &= \Delta k_0 (\sin\alpha_i + \sin\alpha_f) + k_0(\cos\alpha_f \Delta\alpha_f + \cos\alpha_i \Delta\alpha_i)
\end{aligned} \tag{109}$$

Including all these terms to the resolution function can be complicated and is not always necessary. For different types of measurements and setups (reflectivity, GIXD, DXS, etc .), most of these terms may be insignificant with respect to the dominating ones. In particular, when performing any of these measurements at a synchrotron the energy bandwidth ($\Delta k_0$) term can be completely neglected as it is determined by the Darwin widths of the first stage monochromator and the mirror that make $\Delta k_0 \sim 10^{-4}$ Å$^{-1}$. This low value coupled with the fact that in reflectivity and in DXS experiments the angles are small, make other terms negligible. Also, to get any significant intensity in DXS, and reflectivity measurements, the divergence in $2\theta\,(\Delta Q_x)$ and $\Delta\alpha_f$ have to be relaxed, and integrations such as the one we performed in Equation (71) have to be performed to account for the total scattering in those directions. In addition the divergence of the incoming beam, in a synchrotron setup is orders of magnitude smaller than that of the outgoing beam, and it can be neglected (i.e., the resolution function is considered a $\delta(\alpha_i)$). All these considerations, in reflectivity and DXS experiments, simplify Equation 109 to

$$\Delta Q_y = k_0 \sin\alpha_f \Delta\alpha_f$$
$$\Delta Q_z = k_0 \cos\alpha_f \Delta\alpha_f \qquad (110)$$

For GIXD measurements at relatively large angles from a 2D polycrystalline ordering one gets a simple dominating term for the resolution

$$\Delta Q_{xy} = k_0 \cos\alpha_f \Delta 2\theta \qquad (111)$$

The bottom line is that for each configuration all the terms in Equation (111) need be considered keeping the most dominating ones at the end of the process (for details see Tolan 1999, Bouwman 1996, Pershan 2009).

To determine the in-plane correlations, the horizontal resolution of the diffractometer can be adjusted with a Soller-slit collimator consisting of vertical absorbing foils stacked together between the surface and the detector. The area that is probed at each scattering angle 2θ is proportional to $S_0$/sin 2θ, where $S_0$ is the area probed at $2\theta = \pi/2$. The probed area must be taken into account in the analysis of a GIXD scan that is performed over a wide range of angles.

Position-sensitive detectors (PSD) are commonly used to measure the intensity along the 2-D rods. It should be kept in mind that the intensity along the PSD is not a true rod scan of a Bragg reflection at a nominal $Q_{xy}$ because its variation with $\alpha_f$, as is seen in Equation 69.

**Fluorescence Spectroscopy Setup**

Here, we differentiate two types of experimental setups. In one, the same set up as in reflectivity and GIXD experiments, without additional instruments, is used. So, for anomalous reflectivity (GIXD) experiments the x-ray energy is changed to various energies at and away from resonance. At resonance, for transition metal ions photon energy of the absorption edge associated with K-1s binding is selected, whereas for the heavier elements (Cs, Ba, La) the $L_2 2p_{3/2}$ is used, as it gives a stronger signal than at other $L$ binding energies. In the second setup an additional energy dispersive detector is placed at about 2 cm above the liquid surface as shown schematically in Figure 17. The detector itself is situated in a well outside the trough with a thin Kapton window. Calibration of the detector with radioactive elements is performed prior to the experiment. To monitor the data qualitatively as the experiment is performed regions of interest (ROI) on the MCA are set up for a specific emission line of the ions. The performance of silicon-drift diode EDD is not consistent and may vary significantly within a period of an experiment therefore, monitoring the scattered direct beam (which includes Compton processes) as a control of the detector consistency is advisable.

## DATA ANALYSIS AND INITIAL INTERPRETATION

The task of finding the ED from a reflectivity curve is similar to that of finding an effective potential for Equation 7 from the modulus of the wave-function. Direct inversion of the scattering amplitude to ED is not possible except for special cases (Sacks, 1993, and references therein). If the modulus and the phase are known, they can be converted to ED by the method of Gelfand-Levitan-Marchenko (Sacks, 1993) (GLM method). However, in reflectivity experiments, the intensity of the scattered beam alone is measured, and phase information is lost. The main issue here is not that a model cannot be constructed so it fits the data satisfactorily, but that there may be more than one solution, i.e., the solution is not unique.

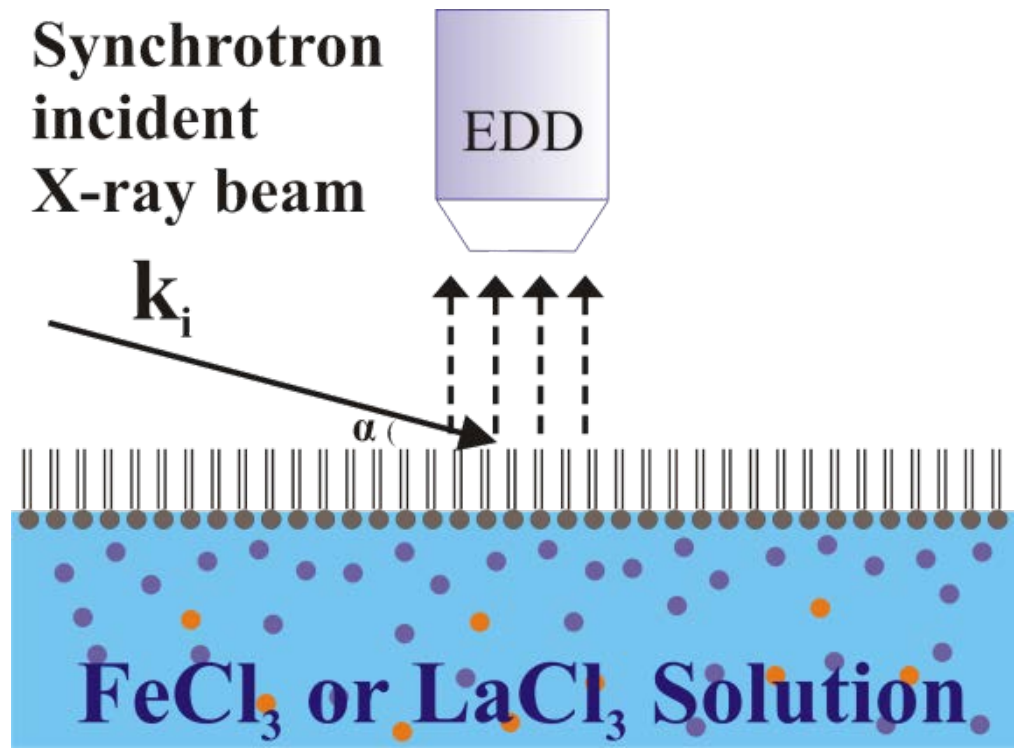

**Figure 17**: The setup used to measure fluorescence with an energy dispersive detector. The detector is placed about 1-2 cm above the liquid surface in front a Kapton window that keeps the sample under the desired environment, .i.e, water saturated helium, for aqueous interfaces.

Step-like potentials have been directly reconstructed by retrieving the phase from the modulus—i.e., reflectivity—and then using the GLM method (Clinton, 1993; also Sacks, 1993,). Model-independent methods which are based on optimization of a model to reflectivity, without requiring any knowledge of the chemical composition of the SLD at the interface, were also developed recently (Pedersen, 1992; Zhou and Chen, 1995). Such models incorporate a certain degree of objectivity. These methods are based on the kinematical and the dynamical approaches for calculating the reflectivity. One method (Pedersen, 1992) uses indirect Fourier transformation to calculate the correlation function of $d\rho/dz$, which is subsequently used in a square-root deconvolution model to construct the ED model. Zhou and Chen (1995), on the other hand, developed a groove tracking method that is based on an optimization algorithm to reconstruct an ED using the dynamical approach to calculate the reflectivity at each step.

The most common procedure to extract structural information from reflectivity is by use of standard nonlinear least squares refinement of an initial ED model defined in by P-dimensional set of independent parameters, **p**, using all the experimental information available in guessing $\rho(z, \mathbf{p})$. The parameters are then refined by calculating the reflectivity ($R[\boldsymbol{Q}^i_z\mathbf{p}]$) with the tools described earlier, and by minimizing the $\chi^2(\mathbf{p})$ quantity

$$\chi^2(\mathbf{p}) = \frac{1}{N-P}\sum_{i=1}\left[\frac{R_{\exp}(Q_z^i) - R(Q_z^i,\mathbf{p})}{\varepsilon(Q_z^i)}\right]^2 \qquad (112)$$

where $\varepsilon(Q^i_z)$ is the uncertainty of the measured reflectivity, $R_{\exp}(Q^i_z)$ and $N$ is the number of measured points. The criteria for a good fit can be found in the book by Bevington (1968). A good rule of thumb is to use the minimum set of parameters that are necessary to obtain the best fit to the data, and that also make physical sense considering the experimental conditions. It is well known that the height and thickness of a box ($d_i$, $\rho_i$, as shown in Figure 5) are strongly coupled, and their individual uncertainty can be large. However their product (or area) is highly sensitive to the fitting parameters. Thus, if the thickness of the box is such that the ED cannot be achieved with light elements but only under the forces of a "black hole", constraints can be made on certain values of the ED to confine them to physical values. Uncertainties of a certain parameter can be obtained by fixing this parameters at various values and for each value refining the rest of the parameters until $\chi^2$ is increased to achieve a desired confidence limit.

The direct methods and model-independent procedures of reconstruction ED do not guarantee uniqueness of the potential—i.e., there can be multiple profiles that essentially yield the same reflectivity curve, as discussed with regard to Figure 7, for example. Near-uniqueness can be achieved by introducing physical constraints that are incorporated into the parameters of the model. Volume, in-plane density of electrons, etc., are among such constraints that can be used. Applying such constraints is discussed briefly below; see Examples (also see Vaknin et al., 1991a,b; Gregory et al., 1997). These constraints reduce the uncertainties and make the relationship of the ED to the actual molecular arrangement apparent. In the dynamical approach, no two potentials yield exactly the same reflectivity, although the differences between two models might be too small to be detected in an experiment.

An experimental method for solving this problem was demonstrated by Sanyal et al. (1992) using anomalous x-ray reflectivity methods for a film on the surface of Ge crystal. Two reflectivity curves from the same sample are measured with two different x-ray energies, one below and one above an absorption edge of the substrate atoms, thereby varying the scattering length density of the substrate. Subsequently the two reflectivity curves can be used to perform a direct Fourier reconstruction (Sanyal et al., 1992), or refinement methods can be used to remove ambiguities. This method is not practical when dealing with liquids that consist of light atoms, because of the very low energy of the absorption edges with respect to standard x-ray energies. Another way to overcome the problem of uniqueness is by performing reflectivity experiments on similar samples with x-rays and with neutrons. In neutron scattering experiments the scattering length density (SLD) across the interface can be changed significantly, by chemical exchange of isotopes that change the SLD, but maintain the chemical integrity of the sample (Vaknin et al., 1991b; Penfold and Thomas, 1990). The reflectivities (x-ray as well as neutrons) can be modeled to one parameterized structure that is defined in terms of geometrical parameters only, and calculating the EDs and SLDs from scattering lengths of the constituents and the geometrical parameters (Vaknin et al., 1991b,c). One can of course add heavy ions to a solution to utilize the anomalous technique, but in general such additions affect both the interfacial solution and films of the aqueous surface. If such an addition is not already part of the protocol of the preparation of the studied specimen no gain is achieved in trying to understand the pristine unperturbed system.

**Examples**

Since the pioneering work of Als-Nielsen and Pershan (1983), x-ray reflectivity, GIXD, and DXS became standard tools for determining structure and phenomena at liquid surfaces on atomic length scales. The techniques have been exploited in studies of the physical properties of simple liquids (Braslau et al., 1988; Sanyal et al., 1991; Ocko et al., 1994), Langmuir monolayers (Dutta et al., 1987; Kjaer et al., 1987, 1989, 1994; Als-Nielsen and Kjaer, 1989; Vaknin et al., 1991b, Lin 1999), liquid metals (Rice et al., 1986; Magnussen et al., 1995; Regan et al., 1995, Shpyrko et al. 2006), surface crystallization (Wu et al., 1993a,b, 1995; Ocko et al., 1997), liquid crystals (Pershan et al., 1987), surface properties of quantum liquids (Lurio et al., 1992), protein recognition processes and orientation at liquid surfaces (Vaknin et al., 1991a, 1993; Lösche et al., 1993, Strzalka et al., 2006), ionic liquids (Sloutskin et al. 2005, Jeon et al. 2008) and many other applications. They have also been used in liquid-liquid interfaces (McClain et al. 1994, Schlossman and Tikhonov, 2008). Subsequently they have been extended to measure fluorescence from ion enriched surfaces (Bloch et al. 1985, Daillant et al. 1991, Bu et al. 2006). Here, a few examples are briefly described in order to demonstrate the strengths and the limitations of the technique. In presenting the examples, there is no intention of giving a full theoretical background of the systems.

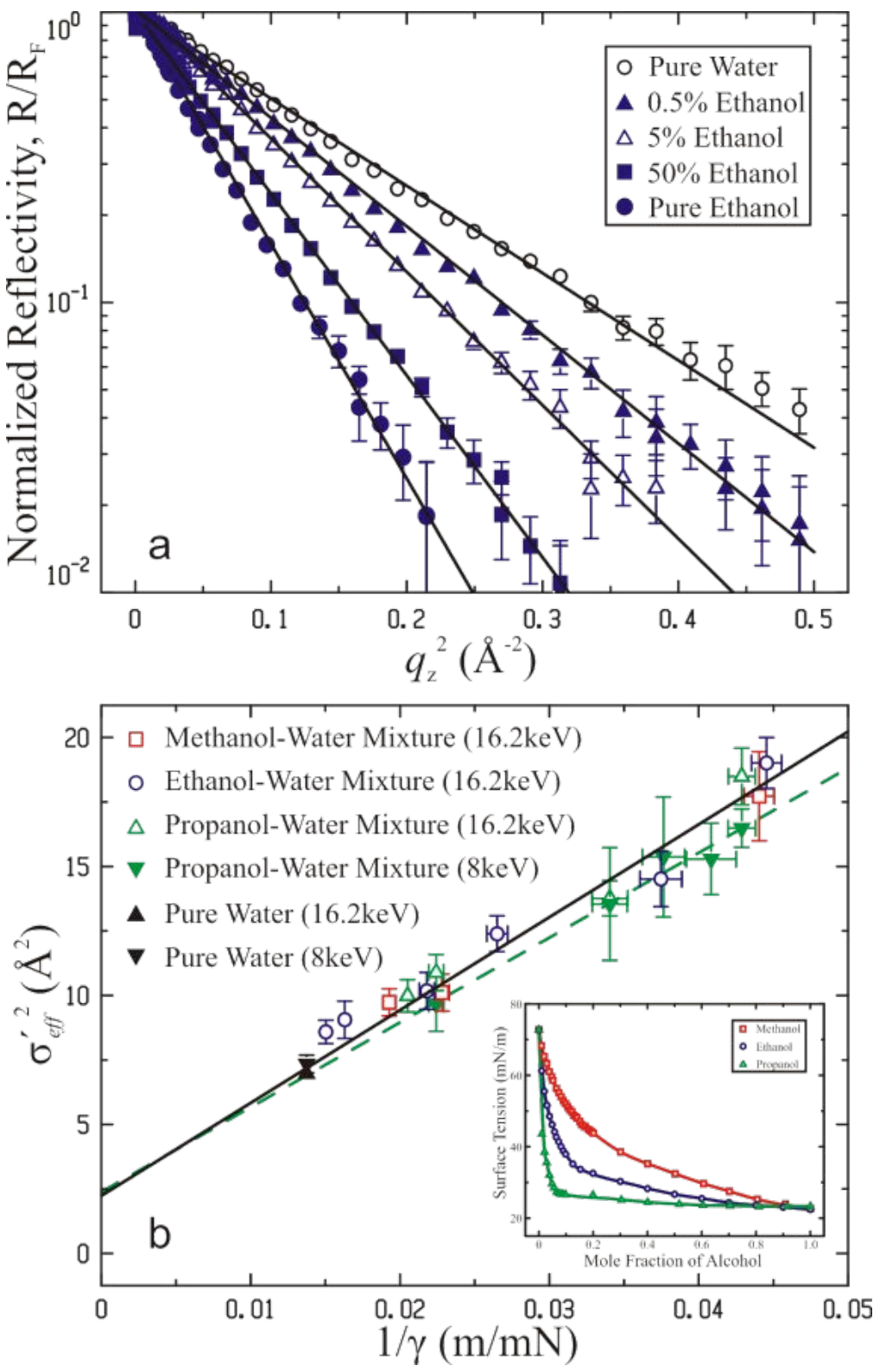

**Figure 18: (a)** Normalized X-ray reflectivity of ethanol-water mixtures versus $Q_z$. The solid lines are linear fits to the (logarithm of) data yielding $\sigma_{eff}$. (b) Compilation of $(\sigma_{eff})^2$ values obtained from the analysis of the reflectivities water (measured at 8 and 16.2 keV) and mixtures with three alcohols (as indicated) and *versus* the inverse of surface tension of these mixtures. The solid line (16.2 keV) and dashed line (8 keV) are the best linear fits to the data, differ because different setups, i.e., $Q_{min}$. The inset in (b) shows the surface tension of water-alcohol mixtures with varying bulk concentrations (Vaknin 2009).

**Capillary waves and interfacial properties of simple liquids.**

Although, X-ray reflectivity, DXS, and wide angle GIXD have been extensively used to unravel the structure and properties of liquid/vapor interfaces, they may not be sufficient to provide definitive answers with regard to the exact structure of the top most layer of liquids, because of low signals and the inherent dynamics of the liquid. The capillary wave model (Buff et al. 1965, Evans et al. 1979), based on the phenomenological surface-tension seems to capture the essential interfacial properties of the liquid/vapor interface on the intermediate length scales, i.e., between the macroscopic and short-range correlations of the liquid. To examine the nature of capillary-waves one has to measure the reflectivity and the DXS from a liquid surface, as has been done from the outset (Braslau et al., 1988; Sanyal et al., 1991; Ocko et al., 1994). Ocko et al. (1994) have measured the reflectivity of liquid alkanes over a wide range of temperatures, verifying that the surface roughness is of the form given in Equations 75 and 76. Numerous studies have been conducted subsequently on simple and more complex liquids and a detailed account of these is provided in Tolan's book (1999). Wide angle GIXD of liquid vapor interfaces has been less explored, but it is the one technique necessary to provide insight into the short range order correlations at the liquid-vapor interface (Fradin et el., Vaknin et al. 2009)

Here, we describe an XR and DXS study of water-alcohol mixtures at various surface tensions (Vaknin 2009). Figure 18(a) shows the normalized reflectivities from water/ethanol mixtures versus the square of $Q_z$ on a logarithmic scale. The addition of ethanol decreases the surface tension of the water and enables systematic examination of the capillary wave theory. The solid lines are the best fits to the data using Equation 75 by a single parameter $\sigma_{\rm eff}$. Similar measurements were conducted on mixtures of methanol and propanol with water, and on the pure liquids. For all liquids, the region of the reflectivity near the critical angle was examined, and found to be in agreement with the average electron density of each mixture. Figure 18(b) shows a compilation of the $(\sigma_{\rm eff})^2$ values for all three mixtures versus the inverse of surface tension, $\gamma$, for the corresponding mixture, as indicated. Measured surface tension values of the mixtures are shown in the inset of Figure 18(b) as a function of the molar fraction of alcohol in water. To a good approximation, all the measured roughness values squared fall on a straight line (solid line 8 keV dashed line 16.2 keV) that within experimental error confirms Equation 75. This behavior strongly suggests the profile of the pure and liquid mixtures of small molecules is predominantly determined by surface tension (at a given temperature). Since $Q_{\rm min}$ can be calculated and also determined experimentally, the only unknown here are $\sigma_{\rm I}$ and $Q_{\rm max}$. The intercept at $1/\gamma=0$ yields the intrinsic roughness $\sigma_{\rm I}$ = 1.5 □0.2 Å. This value is very close to that of bond and coordination lengths in these systems (e.g., C-C and C-O with bond lengths 1.54 and 1.43 Å respectively). The theory for X-ray reflectivity from capillary waves above assumes the electron density is a continuum, but physically, the electrons are concentrated around discrete nuclei thereby giving rise to intrinsic roughness on the scale of atomic separations. Hypothetically, capillary wave theory predict zero roughness at $T = 0$ at but we know random uncorrelated discreteness of the atomic constituents should give rise to some intrinsic roughness. Although in this study of simple liquids $\sigma_{\rm I}$ is of order of average inter-atomic distances, this is only a lower limit and for more complex liquids, the intrinsic roughness may be larger depending on the degree that molecules are uncorrelated at the liquid interface. A few reported studies ignore the contribution from $\sigma_{\rm I}$. From the slope of the data in Figure 18(b), $S = k_BT\ln(Q_{\rm max}/Q_{\rm min})$, and knowledge of $Q_{\rm min}$ (confirmed in DXS measurements on the same samples as shown in Figure 19), we can obtain $Q_{\rm max}$. The values reported for $Q_{\rm max}$ (Vaknin2009) are significantly smaller than the commonly used value of $\pi/a$ where $a$ is typical molecular size. The importance of this result is that it indicates that capillary wave theory breaks down at distances that are on length scales comparable to bulk correlations lengths in the liquid (Vaknin 2009).

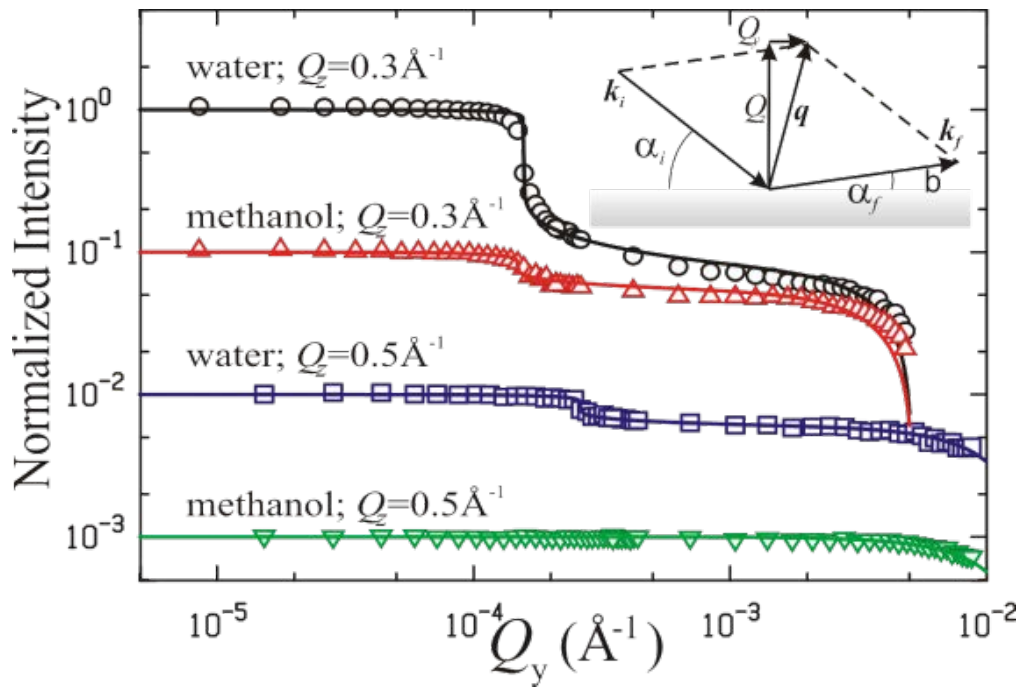

**Figure 19:** The DXS from pure water and pure methanol at constant $Q_z$ values as indicated (shifted by a decade each for clarity). The solid lines are fits to the data using Eq. (86) with no varied parameters. Note that for methanol at $Q_z$ = 0.5 **Å$^{-1}$** the signal is purely *background* (i.e., not from capillary waves) from the bulk structure factor as the function $Z(\eta)$ diverges at lower values $Q_z$. The inset shows the experimental beam geometry (Vaknin 2009).

**GIXD of long-chain molecules**

**Langmuir Monolayers.** A Langmuir monolayer (LM) is a monomolecular amphiphilic film spread at the air-water interface. Each amphiphilic molecule consists of a polar head group (hydrophilic moiety) and a nonpolar tail, typically hydrocarbon (hydrophobic) chain (Gaines, 1966; 1987; Möhwald, 1990). Typical examples are fatty acids, lipids, alcohols, and others. The length of the hydrocarbon chain can be chemically varied, affecting the hydrophobic character of the molecule, and the head group can be ionic, dipolar, or it may possess a certain shape and electrical properties that attract a specific compound (protein, for instance). LMs have been used extensively as model systems of functional lipid and protein membranes.

In this example, XR and GIXD results from a simple lipid, dihexadecyl hydrogen phosphate (DHDP), consisting of a phosphate head group ( $PO_4^-$ ) and its two attached hydrocarbon chains, are presented. Figure 20A shows the normalized reflectivity of a DHDP monolayer at the air-water interface at a lateral pressure of 40 mN/m. The corresponding electron density profile is shown in the inset as a solid line. The profile in the absence of surface roughness ($\sigma = 0$) is displayed as a dashed line. The bulk water sub-phase corresponds to $z < 0$, the phosphate headgroup region is at $0 \geq z \geq 3.4$ Å, and the hydrocarbon tails are at the $3.4$ Å $\geq z \geq 23.1$ Å region. As a first-stage analysis of the reflectivity, a model ED with minimum number of boxes, $i = 1, 2, 3\ldots$, is constructed. Each box is characterized by a thickness $d_i$ and an electron density $N_{e,i}$, and one surface roughness, $\sigma$ for all interfaces. Refinement of the reflectivity with Equation 61 shows that the two-box model is sufficient. In order to get more insight into the structure, we can take advantage of information we know of the monolayer, i.e., the constituents used and the molecular area determined from the lateral-pressure versus molecular area isotherm and the GIXD. If the monolayer is homogeneous (not necessarily ordered), we can assume an average area per molecule at the interface $A$, and calculate the electron density of the tail region as follows

$$\rho_{\text{tail}} = N_{\text{e,tail}}\, r_0 / (A d_{\text{tail}}) \tag{113}$$

where $N_{e,\text{tail}}$ is the number of electrons in the hydrocarbon tail and $d_{\text{tail}}$ is the length of the tail in the monolayer. The advantage of this description is two-fold: the number of independent parameters can be reduced, and constraints on the total number of electrons can be introduced. However, in this case, the simple relation $\rho_{\text{head}} = N_{e,\text{phosphate}}/(Ad_{\text{head}})$ is not satisfactory, and in order to get a reasonable fit, additional electrons are necessary in the head-group region. These additional electrons are due to water molecules that interpenetrate the head-group region, which is not densely packed, by itself. The cross-section of the phosphate head group is smaller than the area occupied by the two hydrocarbon tails, allowing for water molecules to penetrate the head group region. We therefore introduce an extra parameter $N_{\text{H}_2\text{O}}$, the number of water molecules with ten electrons each. The electron density of the head group region is given by

$$\rho_{\text{head}} = (N_{\text{e,phosphate}} + 10N_{\text{H}_2\text{O}})/(Ad_{\text{head}}) \tag{114}$$

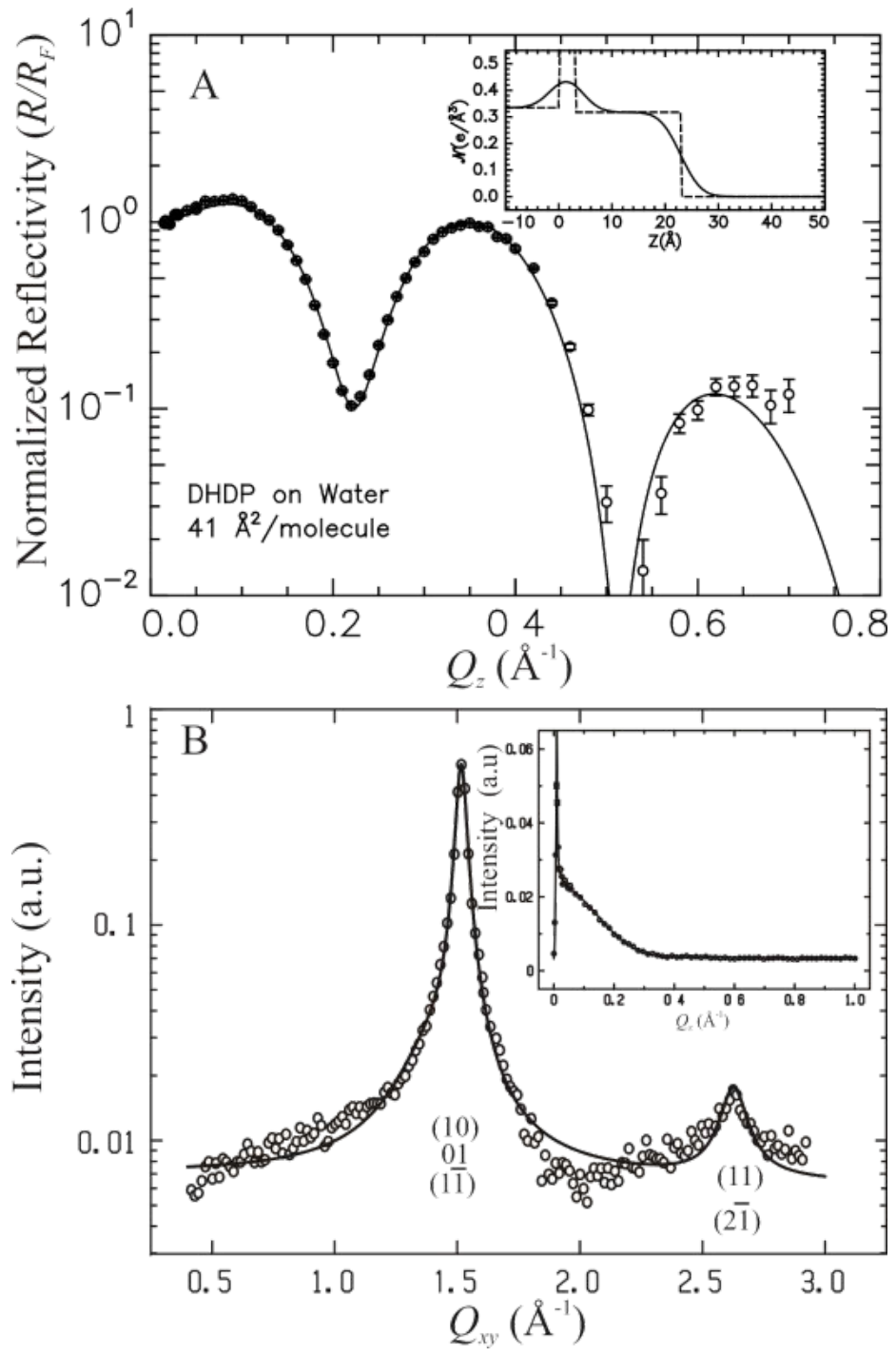

**Figure 20:** (**A**) Normalized x-ray reflectivity from dihexadecyl phosphate (DHDP) monolayer at the air-water interface with best-fit electron density, $N_e$, shown with solid line in the inset. The calculated reflectivity from the best model is shown with a solid line, and dashed line is the ED with no roughness, $\sigma = 0$ (**B**) Diffraction pattern from the same monolayer showing prominent 2-D Bragg reflections corresponding to the hexagonal ordering of individual hydrocarbon chains at $Q^B_{xy} = 1.516$ Å$^{-1}$ and at $\sqrt{3}Q^B_{xy} = 2.626$ Å$^{-1}$. The inset shows a rod scan from the quasi-2D Bragg reflection at $Q^B_{xy}$, with a calculated model for tilted chains denoted by solid line.

This approach gives a physical insight into the chemical constituents at the interface. In modeling the reflectivity with the above assumptions, we can either apply volume constraints or, equivalently, examine the consistency of the model with the literature values of closely packed moieties. In this case the following volume constraint can be applied

$$V_{\text{headgroup}} = Ad_{\text{head}} = N_{\text{H}_2\text{O}} V_{\text{H}_2\text{O}} + V_{\text{phosphate}} \tag{115}$$

where $V_{\text{H}_2\text{O}} \approx 30$ Å$^3$ is known from the density of water. The value of $V_{\text{phosphate}}$ determined from the refinement should be consistent within error with known values extracted from crystal structures of salt phosphate (Gregory et al., 1997).

Another parameter that can be deduced from the analysis is the average tilt angle, $t$, of the tails with respect to the surface. For this the following relation is used

$$d_{\text{tail}} / l_{\text{tail}} = \cos t \tag{116}$$

where $l_{\text{tail}}$ is the full length of the extended alkyl chain evaluated from the crystal data for alkanes (Gregory et al., 1997). Such a relation is valid under the condition that the electron density of the tails when tilted is about the same as that of closely packed hydrocarbon chains in a crystal, $\rho_{\text{tail}} \approx 0.32e/$Å$^3 r_0$, as observed by Kjaer et al. (1989). Such a tilt of the hydrocarbon tails would lead to an average increase in the molecular area compared to the cross-section of the hydrocarbon tails ($A_0$)

$$A_0 / A = \cos t \tag{117}$$

Gregory et al. (1997) found that at lateral pressure $\pi = 40$ mN/m, the average tilt angle is very close to zero ($\approx 7 \pm 7°$), and extracted an $A_0 \approx 40.7$Å$^2$ compared with a value of 39.8 Å$^2$ for closely packed crystalline hydrocarbon chains. The small discrepancy can be attributed to defects at domain boundaries.

The GIXD for the same monolayer is shown in Figure 20B, where a lowest-order Bragg reflection at $Q_{(1,0)} = 1.516$ Å$^{-1}$ and a second order one at $Q_{(1,1)} = (3)^{1/2}\, Q_{(1,0)} = 2.626$ Å$^{-1}$ are observed. These reflections corresponds to the hexagonal ordering of the individual hydrocarbon chains (Kjaer et al., 1987, 1989) with lattice constant $d = 4.1144$ Å, and molecular area per chain $A_{\text{chain}} = 19.83$ Å$^2$ (using Equation 96). Note that in DHDP the phosphate group is anchored to a pair of hydrocarbon chains with molecular area $A = 39.66$ Å$^2$, and it is surprising that ordering of the head group with a larger unit cell (twice that of the hydrocarbon unit cell) is not observed, at $Q_{xy}$ values below the first order reflection of the hexagonal structure, as is evident in Figure 20 B. Also shown in the inset of Figure 20B is a rod scan of the Bragg reflection. No extra Bragg reflections were found in the vicinity of this reflection in the ($Q_{xy}$, $Q_z$) plane indicating that at 40 mN/m the hydrocarbon chains are practically normal to the surface as explained in regard to Figure 11. Fitting the rod using Equation 93 yields a chain length that within error is stretched confirming the orientation of the chain with respect to the surface.

Modified methods have been applied to monolayers that do not possess long chain molecules. GIXD measurements of $C_{60}$-propylamine spread at the air-water interface (see more details on fullerene films; Vaknin, 1996) show a broad in-plane GIXD signal (Fukuto et al., 1997). The diffraction pattern indicated that the local packing of molecules on water is hexagonal, forming a 2-D amorphous solid. The GIXD signal was analyzed with a 2-D radial distribution function that implied short-range positional correlations extending to only few molecular distances.

**Surface Crystallization of Liquid Alkanes**. Normal alkanes are linear hydrocarbon chains $(CH_2)_n$ terminating with $CH_3$ groups on both ends of the chain. Extensive x-ray studies of pure and mixed liquid alkanes (Wu et al., 1993a, b, 1995; Ocko et al., 1997) revealed rich and remarkable properties near their melting temperature, $T_f$. In particular, a single crystal monolayer is formed at the surface of an isotropic liquid bulk up to ≈3°C above $T_f$ for a range of hydrocarbon number $n$. The surface freezing phenomenon exists for a wide range of chain lengths, $16 \geq n \geq 50$. The molecules in the ordered layer are hexagonally packed and show three distinct ordered phases: two rotator phases, one with the molecules oriented vertically ($16 \geq n \geq 30$), and the other tilted toward nearest neighbors ($30 \geq n \geq 44$). The third phase ($44 \geq n$) orders with the molecules tilted towards next-nearest neighbors. In addition to the 2-D Bragg reflections observed in the GIXD studies, reflectivity curves from the same monolayers were found to be consistent with a one-box model of densely packed hydrocarbon chains, and a thickness that corresponds to slightly tilted chains. This is an excellent demonstration of a case where no other technique but synchrotron x-ray experiments reveal the structure of the monolayers. Neutron scattering from this system would have yielded similar information; however, the intensities available today from reactors and spallation sources are smaller by at least a factor of $10^5$ counts/sec for similar resolutions, and will not allow observation of any GIXD signals above background levels.

**Liquid Metals**. Liquid metals, unlike dielectric liquids, consist of the classical ionic liquid and quantum free-electron gas. Scattering of conduction electrons at a step-like potential (representing the metal-vacuum interface) gives rise to quantum interference effects and leads to oscillations of the electron density across the interface (Lang and Kohn, 1970). This effect is similar to the Friedel oscillations in the screened potential arising from the scattering of conduction electrons by an isolated charge in a metal. By virtue of their mobility, the ions in a liquid metal can in turn rearrange and conform to these oscillations to form layers at the interface, not necessarily commensurate with the conduction electron density (Rice et al., 1986). Such theoretical predictions of atomic layering at surfaces of liquid metals have been known for some time, and with the advent of synchrotron radiation have been confirmed by x-ray reflectivity studies for liquid gallium and liquid mercury (Magnussen et al., 1995; Regan et al., 1995, Shpyrko 2006). X-ray reflectivities of these liquids were extended to $Q_z \sim 3\text{Å}^{-1}$, showing a single peak that indicates layering with spacing on the order of atomic diameters. The exponential decay for layer penetration into the bulk of Ga (6.5 Å) was found to be larger than that of Hg (~3 Å). The normalized reflectivities from such interfaces were modeled to oscillating and exponentially decaying ED profile (Regan et al., 1995) of the form

$$\rho(z)/\rho_s = \text{erf}\,[(z-z_0)/\sigma] + \theta(z) A \sin(2\pi z/d) e^{-z/\xi} \qquad (118)$$

where $\theta(z)$ is a step function, $d$ is the inter-layer spacing, $\xi$ is the exponential decay length, and $A$ is an amplitude. The layering phenomena in Ga showed a strong temperature dependence. Although liquid Hg exhibits layering with a different decay length, the reflectivity at small momentum transfers, $Q_z$, is significantly different than that of liquid Ga, indicating fundamental differences in the surface structures of the two metals. The layering phenomena suggest in-plane correlations that might be different than those of the bulk, but had not been observed yet with GID studies.

**Anomalous Reflectivity**

This and subsequent examples deal with the interfacial Poisson-Boltzmann problem, with the objective to evaluate ion distributions at a charged aqueous interfaces (Bu et al. 2006). Here, results on the distribution of monovalent ions near a charged interface, for which the PB theory seems to work well, are presented. Multivalent ions are complicated by the fact that, in addition to electrostatic interactions assumed in PB theory, quantum effects (chemical bonds) play a role in forming the distributions. The DHDP lipid, mentioned above, is spread as a monolayer at the air-water interface to control surface charge density. And by using anomalous reflectivity off and at the Cs $L_{III}$ resonance (at 5.012 keV), the spatial distributions of $Cs^+$ ions are determined. Figure 21A shows the variation of the effective number of electrons $Z_{eff}$ and the absorption coefficient β with photon energy. Figure 21 also shows an illustration of the monolayer and ions in solution (B), a model ED (C) and its correspondence to the moleculcar arrangements in the film (D).

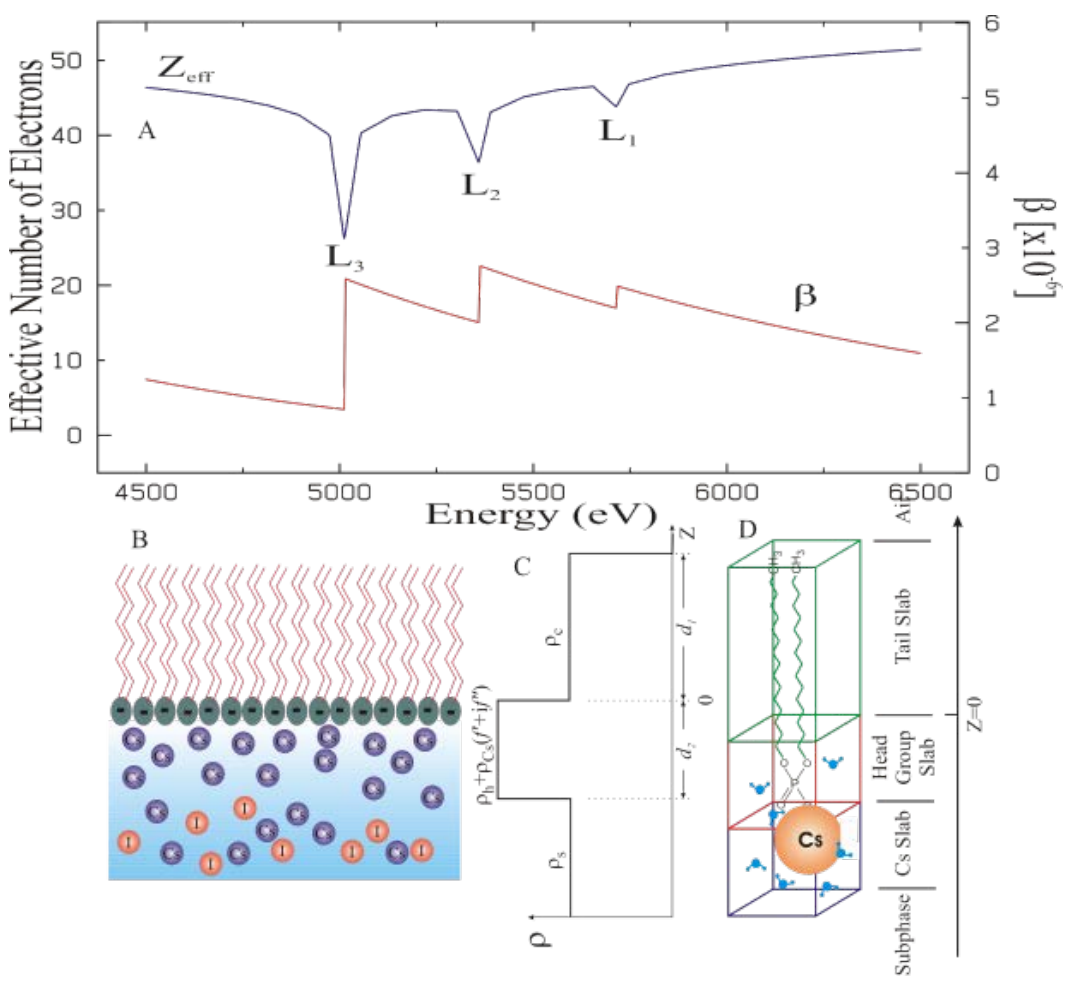

Figure 21: (A) The effective number of electrons per Cs ion and the corresponding absorption coefficient as a function of photon energy. (B) Illustration of the model monolayer on the solution and (C) the expected ED profile (without surface roughness). (D) Boxes show molecular association with the monolayer- and ions.

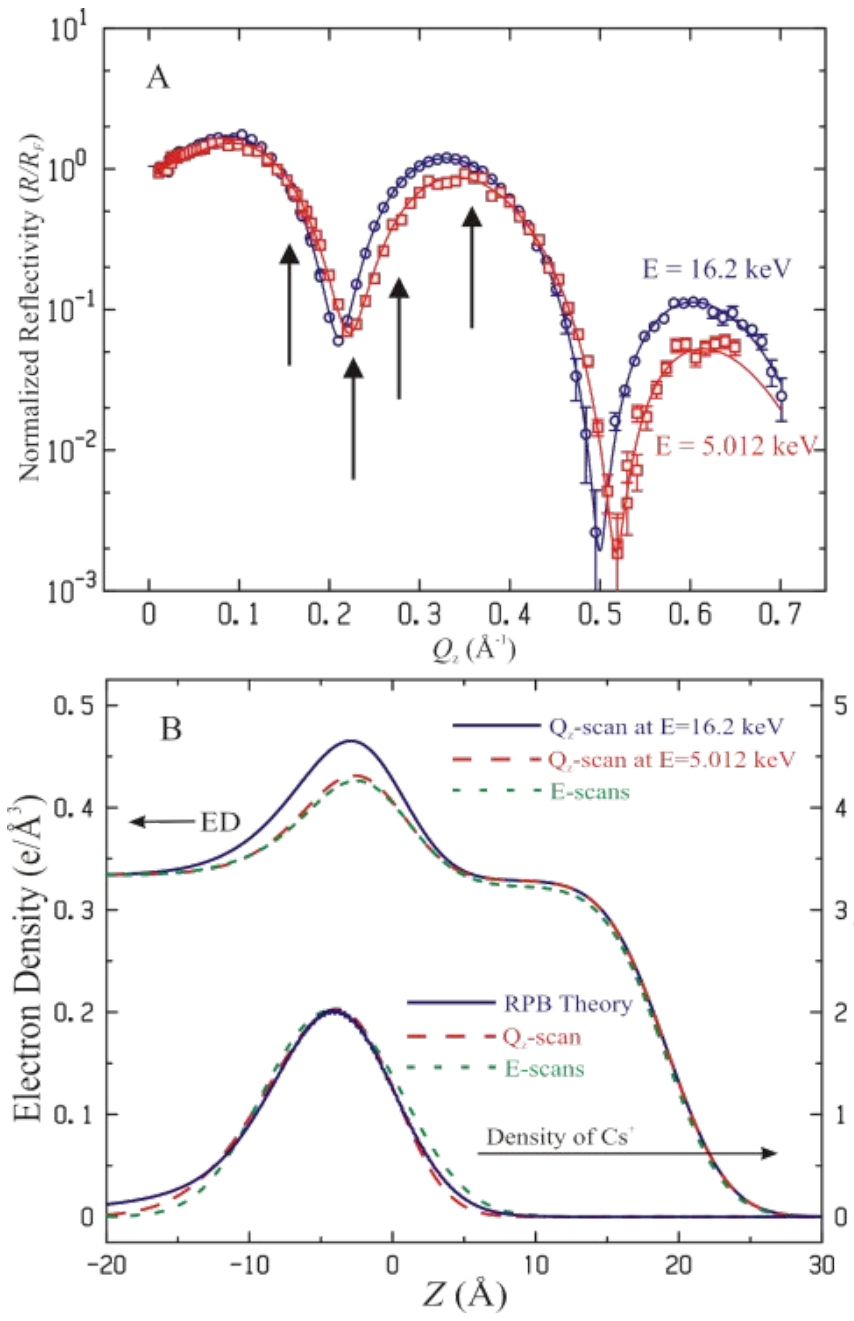

**Figure 22:** (A) Normalized x-ray reflectivities measured at 16.2 and 5.012 keV for DHDP monolayer spread on $10^{-3}$ M CsI solution (π = 40 mN/m). The solid lines are calculated reflectivities using the ED profiles shown in (B) The arrows indicate points where E-scans at fixed $Q_z$ were performed (see Figure 25). (B)The two data sets were combined and refined to a model with common structural adjustable parameters. Also shown is the profile of absorption factor β which at 5.012 keV is dominated by the presence of $Cs^+$ close to the interface. The solid smooth line is the distribution of $Cs^+$ determined from the modified Poissson-Boltzmann theory. The dashed line is the ion distribution extracted from the XR data shown in (A). The dotted line is obtained from the analysis of the energy scans at fixed shown Figure 24 (Bu et al. 2006).

Figure 22A shows reflectivities of DHDP spread on $10^{-3}$ M CsI at 16.2 keV and at the $L_{III}$ edge of Cs at 5.012 keV. The reflectivity taken at 16.2 keV has sharper and deeper minima that are slightly shifted to smaller $Q_z$, compared to the reflectivity taken at reosnance. This is due to the dependence of ρ' and ρ'' on the energy of the x-ray energy, as shown in Figure 21A. At the two energies that the measurements were conducted (16.2 and 5.012 keV), the XR are different for cesium (as shown in Figure 21 and only slightly for the phosphorous ion, whereas for the remaining constituents binding energies are smaller than 5.012 keV and therefore their electrons are effectively free. Bu et al (2006) applied analysis similar to the one described for DHDP monolayer on water (above) to obtain the EDs shown in Figure 22B. The difference between the ED at the two different energies is shown in Figure 22B and normalized (see Equation 101), to yield the distribution of ions at the interface. Bu et al (2006) compared the experimental distribution to the theoretical model (solid line) at various ion concentrations and found excellent agreement with the Poisson Boltzmann theory, that needed to be modified to account for the presence of protons in the subphase (i.e., finite pH values). More complex distributions and strong binding have been also reported for multivalent ions (Wang 2011).

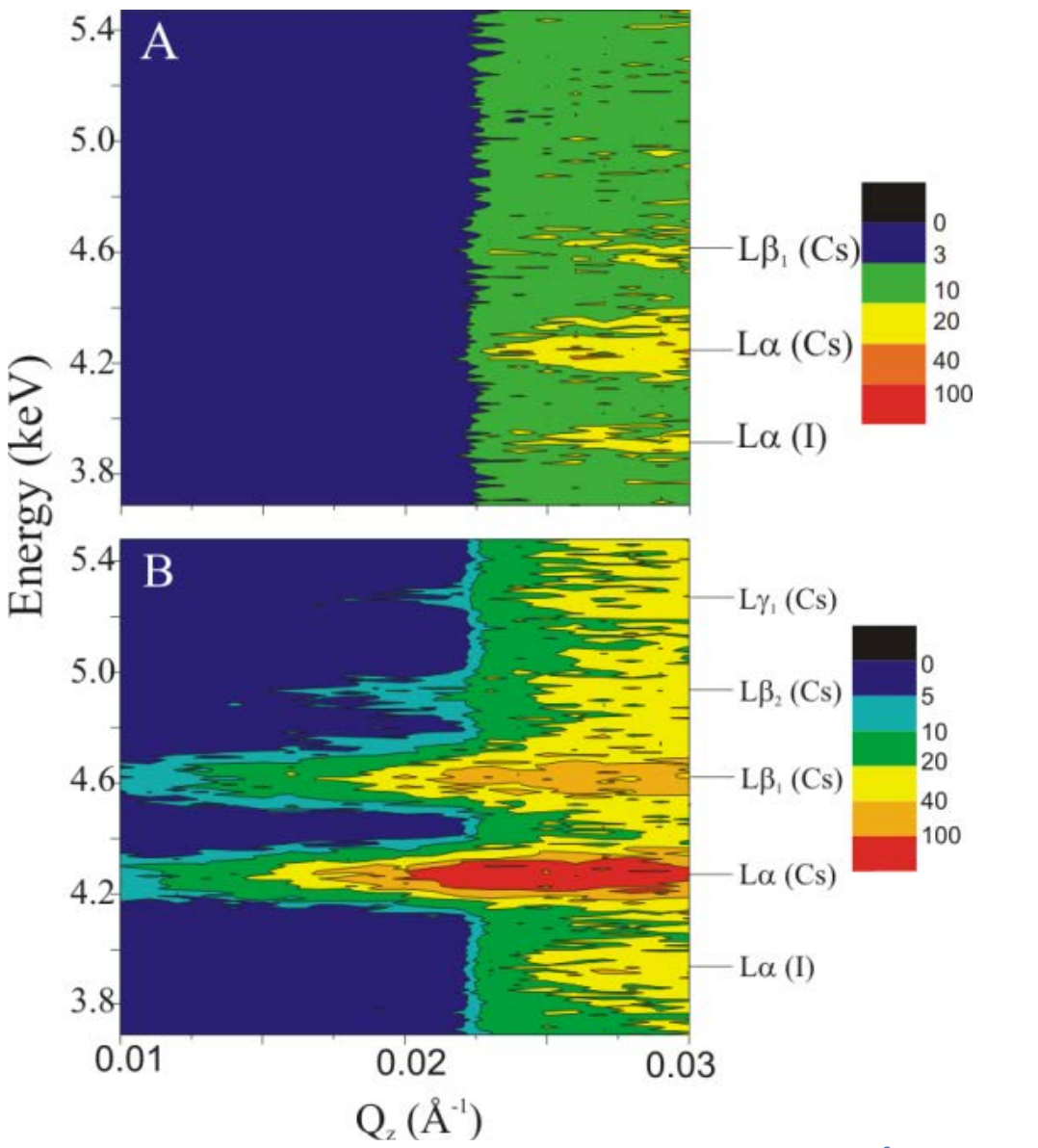

**Figure 23:** Contour plots of fluorescence intensity for $10^{-3}$ M CsI without (A) and with monolayer DHDP (B). Emission lines are labeled on the right side. The surface density of Cs ions at the interface is ~ 0.0125 per Å$^2$. Incident x-ray beam energy is 8 keV. (Bu and Vaknin 2009).

## Spectroscopy at Liquid Interfaces

**Fluorescence Spectra** Figure 23 shows contour plots of fluorescence intensity as functions of emitted photon energy, $E$, and momentum transfer, $Q_z$, for $10^{-3}$ M CsI with and without DHDP monolayer present at the surface (Bu and Vaknin, 2009). Without a monolayer on the surface 23A, the fluorescence pattern is relatively simple. Below the critical angle, no significant fluorescence intensity is observed, consistent with the fact that ions ($Cs^+$ and $I^-$) in the bulk are not concentrated enough to generate any detectable signal over the very short penetration depth of the evanescent beam. So at this concentration, bulk ions do not have any contribution to the signal below $Q_c$ showing a clear line at $Q_c$ that effectively separates surface signal from bulk signal (at higher concentrations under the same conditions one can get a signal below the critical angle). Above the critical angle the x-ray beam penetrates much deeper, sufficient to generate fluorescence signals that show a few main emission lines from Cs and I as indicated. The fluorescence pattern with the DHDP monolayer on (Figure 23B) is dramatically different from that of the bare surface. Now, below the critical angle strong emission lines from Cs (as labeled) are observed, but none from I are. This is qualitative evidence that $Cs^+$ exclusively adsorb at the negatively charged surface that is spanned by the $PO_4^-$ of DHDP. This kind experiment can be used to monitor specifically and selectively ion binding from a mixed ion solution to a charged interfaces. Energy and $Q_z$ cuts obtained from similar plot as shown in Figure 23 (incident photon energy 5.015 keV) are shown in Figure 24. Using Equation (106) to simultaneously fit the bare surface data and that with DHDP monolayer data, yields the density and location of the ions with respect to the interface (Bu and Vaknin 2009).

## Energy scans at fixed momentum Transfer – Interfacial XANES

Figure 25 shows normalized reflectivities ($R/R_F$) at fixed momentum transfer $Q_z$ as a function of photon energy near the Cs $L_{III}$ resonance. Similar measurements on the CsI solution, in the absence of a monolayer, did not reveal significant anomalies in the reflected beam as a function of photon energy, as is shown in Fig. 24 at $Q_z$. = 0.15 Å$^{-1}$ (triangles). By contrast with the monolayer a strong signal is observed. Bu et al. (2006) analyzed the data using the BA method using the generalized complex ED (Equation 21) and obtained good agreement with the variation of the spectra at different $Q_z$ values and the average thickness of the film. They also used the Parratt recursive formalism with the complex ED and a model *f'* and *f''* to obtain the ion distributions and also the XANES of the ions at the interface as shown in Figure 25. The model *f'* and *f''* where coupled by the Kramers-Kroning relation to minimize the number of variable parameters. No signal was detected in energy scans around the iodine $L_{III}$ absorption edge (4.557 keV) at a few fixed $Q_z$ values, demonstrating that the concentration of $I^-$ is in fact depleted with respect to that of Cs at the interface. Figure 26 shows the *f'* and *f''* for a $Cs^+$ at the interface in comparison to their values in different and related environments.

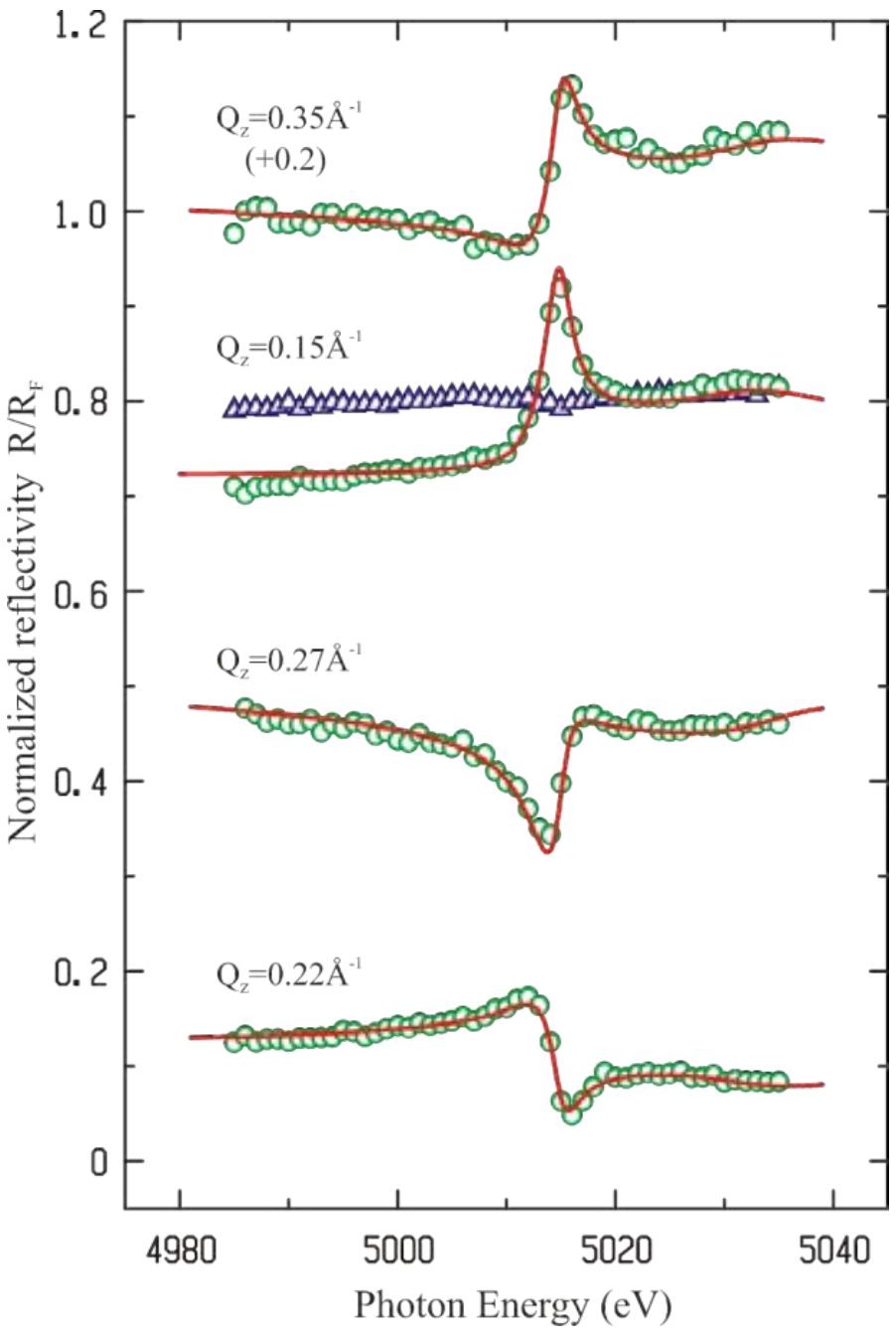

**Figure 25:** Energy scans at fixed momentum transfers $Q_z$ (as indicated in Figure 22) near the Cs $L_{III}$ edge of a DHDP monolayer spread on $10^{-3}$ M CsI solution ( π = 40 mN/m). Solid lines are best fits to the data as described in detail by Bu et al (2006) (energy scan at $Q_z$ = 0.35 Å$^{-1}$ is offset vertically for clarification). Energy scan of the same CsI solution at $Q_z$ = 0.15 Å$^{-1}$ (triangles) do not show any anomalies in the absence of interfacial charges, i.e., with no monolayer. (Bu et al., 2006).

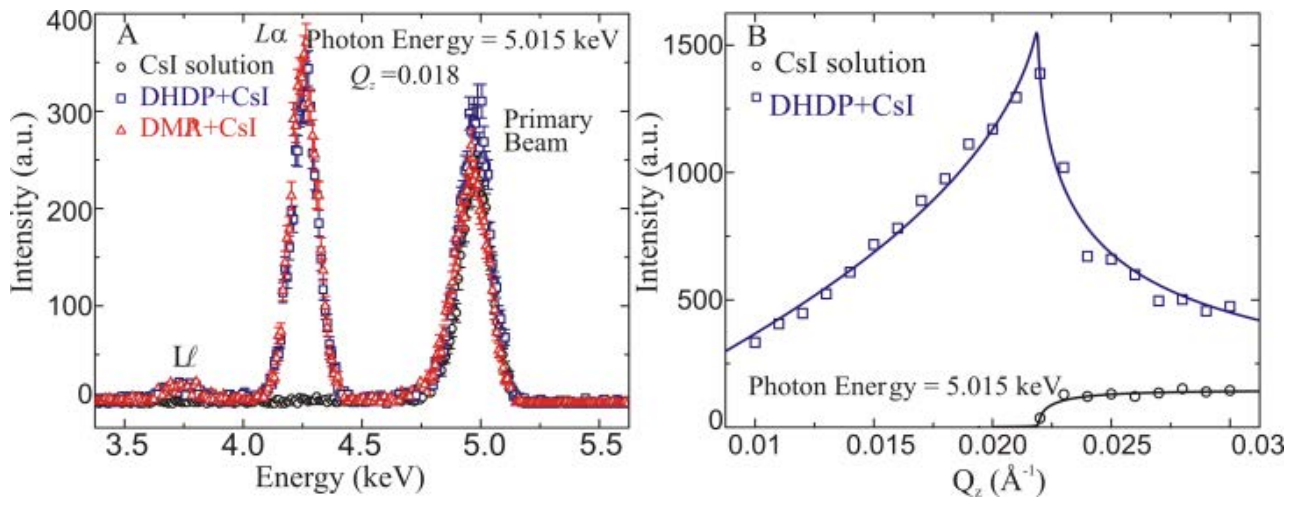

**Figure 24: (A)** Energy cuts below the critical angle for a bare solution and for a similar solution covered with a charged monolayer (DHDP or DMPA) (B) $Q_z$ dependence of fluorescence form a bare CsI solution surface (circles) and from the same surface after covering it with a charged monolayer (squares). Measurements conducted near resonance at 5.015 keV (Bu and Vaknin 2009).

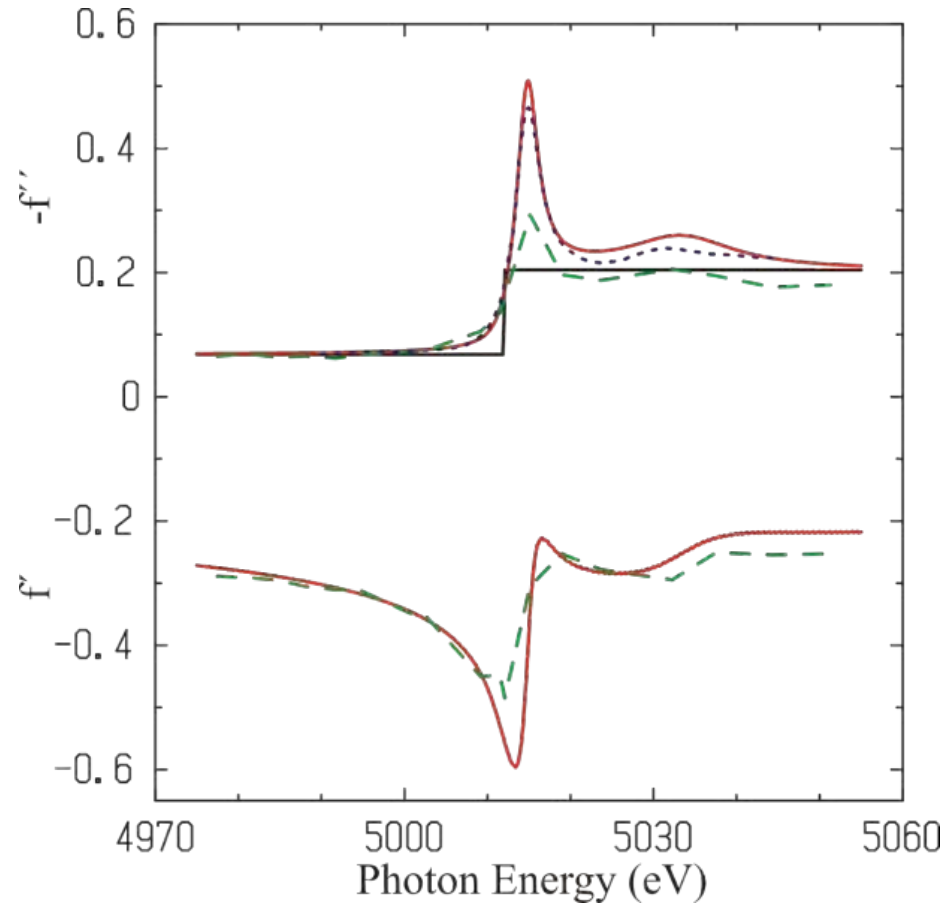

**Figure 26:** Dispersion corrections near the Cs $L_{III}$ edge. The solid lines are derived from the best fit and Karmers-Kronig relations to the spectra shown in Fig. 24. The dashed line was obtained from the Bijvoet-pairs measurements of $CsHC_4O_6$ single crystal (Templeton 1980), the dotted line was extracted from absorption measurements of $CsNO_3H_2O$ solution (Gao 2005). (Bu et al., 2006).

## SPECIMEN MODIFICATION AND PROBLEMS

In conducting experiments from liquid surfaces, the experimenter faces problems that are commonly encountered in other x-ray techniques. A common nuisance in an undulator beam line is the presence of high harmonic components that accompany the desired monochromatic beam. These high harmonics can skew the data as they affect the counting efficiency of the detector, can damage the specimen or the detector, and can increase undesirable background. In photon energy below an edge the higher harmonic can give rise to weak emission line signals that may be lead to wrong interpretation of data. The higher harmonic components can be reduced significantly (but not totally eliminated) by inserting a mirror below the critical angle for total reflection, or by detuning the second crystal of the double-crystal monochromator (XAFS SPECTROSCOPY). However, there are also problems that are more specific to reflectivity and GIXD in general, and to liquid surfaces in particular. Such problems can arise at any stage of the study, initially during the alignment of the diffractometer, or subsequently during data collection, and also in the final data analysis.

Poor initial alignment of the diffractometer will eventually result in poor data. Accurate tracking, namely small variation (5% to 15%) in the count rate of the monitor as the incident angle $\alpha_i$ is varied (workable range of $\alpha_i$ is 0° to 10°) is a key to getting reliable data. Although seemingly simple, one of the most important factors in getting a good alignment is the accurate determination of relevant distances in the diffractometer. For the specific design in Figure 7B, these include the distances from the center of the tilting monochromator to the center of the IH elevator and to the sample SH, or the distances from the center of the liquid surface to the detector ($S_4$) and to the elevator (DH). These distances are used to calculate the translations of the three elevators (IH, SH, and DH) for each angle. Although initial values measured directly with a ruler can be used, effective values based on the use of the x-ray beam yield the best numbers of the diffractometer.

Radiation damage to the specimen is a common nuisance when dealing with liquid surfaces. Many of the studies of liquid surfaces and monolayers involve investigations of organic or biomaterials that are susceptible to chemical transformations in general and more so in the presence of the intense synchrotron beam. Radiation damage is of course not unique to monolayers on liquid surfaces; other x-ray techniques that involve organic materials (protein, polymer, liquid crystals, and others) face similar problems. Radiation damage to a specimen proceeds in two steps. First, the specimen or a molecule in its surroundings is ionized or excited to higher energy levels (creating transient radicals). Subsequently, the ionized/excited product can react with a nearby site of the same molecule or with a neighboring molecule to form a new species, altering the chemistry as well the structure at the surface. In principle, recovery without damage is also possible, but there will be a recovery time involved. The remedies that are proposed here are in part specific to liquid surfaces and cannot be always fulfilled in view of the specific requirements of an experiment. To minimize the primary effects (photoelectric, Auger, and Compton scattering) one can employ one or all of the following remedies. First, the exposure time to the x-ray beam can be minimized to signal counting only. For instance, while motors are still moving to their final positions, the beam can be blocked. Reduced exposure can be also achieved by attenuating the flux on the sample to roughly match it to the expected signal, and desirable uncertainty, so that the full intense beam is used only for signals with cross-sections for scattering that require it. That, of course, requires rough knowledge of signal intensity, which is usually the case in x-ray reflectivity experiments from surfaces. Monolayers on liquid surfaces are inevitably disposable and it takes a few of them to complete a study, so that in the advanced stage of a study many peak intensities are known. Another approach to reducing the effect of the primary stage is by operating at high x-ray energies. It is well known that the cross-section for all the primary effects is significantly reduced with the increase of x-ray energy. If the experiment does not require a specific energy, such as in resonance studies (anomalous scattering), it is advantageous to operate at high x-ray energies. However, operation at higher x-ray energies introduces technical difficulties of its own. Higher mechanical angular resolutions, and smaller slits, are required in order to achieve reciprocal space resolutions comparable to those at lower energies. As discussed above, slit $S_2$ at 16 keV has to be set at about 50 μm width, which cannot be reliably determined at high photon energies because of leakages through slits edges. Finally, reducing the high harmonic component in the x-ray beam will also reduce the primary effect of the radiation.

Reducing the effects of secondary processes also depends on the requirements of a specific experiment. Air surrounding the sample has probably the worst effect on the integrity of an organic film at the liquid interface. The x-ray radiation readily creates potent radicals in air (such as monatomic oxygen) that are highly diffusive and can interact with almost any site of an organic molecule. Working under helium environment or under vacuum, if vapor pressure allows that, can significantly reduce this source of radiation damage. Another approach to consider for reducing the secondary effect of radiation is scattering at low temperatures. It is well documented in protein crystallography that radiation damage is significantly reduced at liquid nitrogen temperatures. However, such low temperatures are not a choice when dealing with aqueous surfaces, and variations in temperature can also lead to dramatic structural transition in the films which may alter the objectives of a study.

The liquid substrate, even water, can create temporary radicals that can damage the monolayer: in particular, the head group region of lipids. Water under intense x-ray radiation can produce many reactive products such as $H_2O_2$ or monatomic oxygen that can readily interact with the monolayer. Thus, some radiation damage, with an extent that may vary from sample to sample, is inevitable, and fresh samples are required to complete a study. Moving the sample underneath the footprint is a quick fix in that regard, assuming that the radiation damage is mostly localized around the illuminated area. To accomplish that, one can introduce a translation of the trough (perpendicular to the incident beam direction at $\alpha_i = 0$) to probe different parts of the surface. It should be noted that for lipid monolayers, a common observation suggests that radiation damage is much more severe at lower in-plane densities than for closely packed monolayers.

Another serious problem concerning scattering from surfaces is background radiation that can give count rates comparable to those expected from GIXD or rod scan signals. Background can be classified into two groups, one due to room background and the other due to the sample and its immediate surroundings. Although it is very hard to locate the sources of room background, it is important to trace and block them, as they reduce the capability of the diffractometer. The specimen can give unavoidable background signal due to diffuse or incoherent scattering from the bulk liquid substrate, and this needs to be accounted for in the interpretation of the data (as discussed in regard to DXS). An important source of background is the very immediate environment of the liquid surface that does not include the sample but is included in the scattering volume of the beam: worst of all is air. Working under vacuum is not an option with monolayers, and therefore such samples are kept under water saturated helium. Air scattering in the trough can give rise to background levels that are at least two or three orders of magnitude higher than the expected signal from a typical 2-D Bragg reflection in the GIXD.

As discussed previously (see Data Analysis and Initial Interpretation), reflectivity data can give ambiguous ED values, although that rarely happens. More often, however, the reflectivity is overinterpreted, with ED profiles with details that cannot be supported by the data. The reflectivity from aqueous surfaces can be at best measured up to a momentum transfer $Q_z$ of ~1 Å$^{-1}$, which, roughly speaking, in an objective reconstruction of the ED, should give uncertainties on the order of $2\pi/Q_z \sim 6$ Å. It is only by virtue of complementary knowledge on the constituents of the monolayer that these uncertainties can be lowered to about 1 to 2 Å. Another potential pitfall for overinterpretation lies in the fact that in GIXD experiments only a few peaks are observed, that without extra knowledge on the 3-D packing of the constituents it is difficult to interpret the data unequivocally.

## ACKNOWLEDGMENTS

Work at the Ames Laboratory was supported by the Office of Basic Energy Sciences, U.S. Department of Energy under Contract No. DE-AC02-07CH11358.

## KEY REFERENCES